\documentclass[%
 reprint,
 amsmath,amssymb,
 aps,
floatfix, 
]{revtex4-2} 

\usepackage{graphicx}
\usepackage{dcolumn}
\usepackage{bm}
\usepackage{afterpage, longtable} 
\usepackage{upgreek} 
\usepackage{placeins} 
\usepackage{float}

\begin{document}

\preprint{APS/123-QED}

\title{Tailoring the Acidity of Liquid Media with Ionizing Radiation \\-- Rethinking the Acid-Base Correlation Beyond pH}  

\author{Birk Fritsch}
\email{birk.fritsch@fau.de}
\affiliation{Friedrich-Alexander-Universität Erlangen-Nürnberg \\ Department of Electrical, Electronic and Communication Engineering,\\
Electron Devices (LEB),
Cauerstraße 6, 91058 Erlangen, Germany\\ }
 \affiliation{Friedrich-Alexander-Universität Erlangen-Nürnberg, \\
Department of Materials Science and Engineering, \\
Institute of Micro- and Nanostructure Research (IMN) and Center for Nanoanalysis and Electron Microscopy (CENEM) \\
Cauerstraße 3, 91058 Erlangen, Germany}

\author{Andreas Körner}%
\email{a.koerner@fz-jeulich.de}
\affiliation{%
 Forschungszentrum Jülich GmbH\\
 Helmholtz Institute Erlangen-Nürnberg for Renewable Energy (IEK-11), Cauerstraße 1, 91058 Erlangen, Germany \\}
 \affiliation{Friedrich-Alexander-Universität Erlangen-Nürnberg, \\
Department of Chemical and Biological Engineering, \\
Immerwahrstraße 2a, 91058 Erlangen, Germany}

\author{Tha\"{i}s Couasnon}
\affiliation{
Deutsches GeoForschungsZentrum,
Helmholtz-Zentrum Potsdam,\\
Telegrafenberg, 14473 Potsdam, Germany}%

\author{Roberts Blukis}
\affiliation{
Deutsches GeoForschungsZentrum,
Helmholtz-Zentrum Potsdam,\\
Telegrafenberg, 14473 Potsdam, Germany}%

\author{Mehran Taherkhani}
\affiliation{Friedrich-Alexander-Universität Erlangen-Nürnberg, Department of Electrical, Electronic and Communication Engineering, \\
Electron Devices (LEB),
Cauerstraße 6, 91058 Erlangen, Germany}%

\author{Liane G. Benning}
\affiliation{
Deutsches GeoForschungsZentrum,
Helmholtz-Zentrum Potsdam,\\
Deutsches GeoForschungsZentrum (GFZ), Telegrafenberg,
14473 Potsdam, Germany}%
\affiliation{
Department of Earth Sciences,\\
Free University of Berlin, 12249 Berlin, Germany}%

\author{Michael P. M. Jank}
\affiliation{Fraunhofer Institute for Integrated Systems and Device Technology IISB,
Schottkystraße 10, 91058 Erlangen, Germany}
\affiliation{Friedrich-Alexander-Universität Erlangen-Nürnberg, Department of Electrical, Electronic and Communication Engineering, \\
Electron Devices (LEB),
Cauerstraße 6, 91058 Erlangen, Germany}%

\author{Erdmann Spiecker}
\affiliation{Friedrich-Alexander-Universität Erlangen-Nürnberg, \\
Department of Materials Science and Engineering, \\
Institute of Micro- and Nanostructure Research (IMN) and Center for Nanoanalysis and Electron Microscopy (CENEM), \\
Cauerstraße 3, 91058 Erlangen, Germany}

\author{Andreas Hutzler}%
\email{a.hutzler@fz-juelich.de}
\affiliation{%
    Forschungszentrum Jülich GmbH\\
 Helmholtz Institute Erlangen-Nürnberg for Renewable Energy (IEK-11), Cauerstraße 1, 91058 Erlangen, Germany}
\date{\today}

\begin{abstract}
Advanced \textit{in situ} techniques based on electrons and X-rays are increasingly used to gain insights into fundamental materials dynamics in liquid media. Yet, ionizing radiation changes the solution chemistry. In this work, we show that ionizing radiation decouples the acidity from autoprotolysis. Consequently, pH is insufficient to capture the acidity of water-based systems under irradiation. \textit{Via} radiolysis simulations, we provide a more conclusive description of the acid-base interplay. Finally, we demonstrate that acidity can be tailored by adjusting the dose rate and adding pH-irrelevant species. This opens up a huge parameter landscape for studies involving ionizing radiation.
\end{abstract}

\maketitle


\section{\label{sec:level1}Introduction}

\textit{In situ} studies employing ionizing radiation enable unique insights into fundamental dynamics in liquid \cite{LaForge.2019, Jahnke.2021, Wang.2022}.
Yet, performing reliable cutting-edge research demands precise knowledge of radiation -- matter interaction and the related parameters during the experiment \cite{Signorell.2020, Yesibolati.2020, Yesibolati.2020b}. Particularly when studying chemical phenomena in liquid using electrons (e.g. during liquid-phase transmission electron microscopy (LP-TEM)) or X-rays (e.g. in X-ray diffraction (XRD)) it must be ensured that the effect of radiation on the observation is accounted for \cite{Woehl.2020, Steinruck.2020, Bras.2021, Bras.2022, Fritsch.2022b}.

One of the main parameters characterizing the physicochemical properties is the acidity of the liquid phase, generally described by the negative decadic logarithm of the concentration $c\rm (H^+)$ of hydrogen ions, known as pH. Simulations show that electron irradiation of pure water cause a dose-rate dependent increase of $c\rm (H^+)$, thus lowering pH \cite{Schneider.2014, Gupta.2018}.

In contrast, precipitation phenomena observed in aqueous solutions \cite{Abellan.2017}
and analyses of growth kinetics \cite{Su.2019b}
also suggest an elevated concentration of $c\rm (OH^-)$ under irradiation. Nevertheless, simultaneous electron-beam induced changes of $c\rm (H^+)$ and $c\rm (OH^-)$ were not yet discussed in literature. Moreover, the interpretation of pH in irradiated liquid must be evaluated in general.

Highly reactive radiolysis products and their subsequent deactivation reactions enable diverse reaction pathways which drastically depend on the chemical environment \cite{Schneider.2014, Hutzler.2019, Korpanty.2021, Fritsch.2022b}. 
In this sense, the impact of additives to pure water on the acidity has not been discussed to date.

In this letter, we reconsider the interpretation of pH in irradiated liquids by modeling electron beam and X-ray radiation chemistry in pure water.
We show that current models of irradiation-induced acidification are insufficient and introduce a more conclusive description of the acidity in irradiated solutions.
Furthermore, the impact of different supposedly pH-irrelevant ionic species typically present in LP-TEM like chloride \cite{Aliyah.2020, Hermannsdorfer.2015, Dong.2021, Fritsch.2022b, Fritsch.2021},  bromide \cite{Wang.2022, Hutzler.2018, Bae.2020, Crook.2021, Dang.2021},
and nitrate \cite{Aliyah.2020, Loh.2017, Hutzler.2018, Woehl.2012, Wang.2018, Abellan.2014, Dong.2019} on the acidity are investigated. 

\section{\label{sec:level2}Theory and experimental procedures}

During autoprotolysis, water molecules dissociate into protons $\mathrm{H^+}$ and hydroxide ions $\mathrm{OH^-}$ changing the respective concentrations until an equilibrium is reached.

\begin{eqnarray}
    \rm H_2 O \rightleftharpoons H^+ + OH^-
\label{eq:Chem_WaterAutoprotolysis}
\end{eqnarray}

Considering the law of mass action, an equilibrium constant $K$ can be formulated, which depends on the activity $\alpha$ of the respective species:

\begin{eqnarray}
    K = \frac{\alpha(\rm{H^+})\cdot \alpha(\rm{OH^-})}{\alpha(\rm{H_2O})}
\label{eq:LawOfMAss_WaterAutoprotolysis}
\end{eqnarray}

Due to its normally low dissociation degree, the activity of the solvent ($\rm H_2O$) can be assumed to be unity. Therefore it can be incorporated into the ion product $K_{\rm W}$:

\begin{eqnarray}
\label{eq:Kw}
    K_{\mathrm W} = c(\mathrm{H^+})\cdot c(\mathrm{OH^-})
\end{eqnarray}

pH and complementary pOH are defined as the negative decadic logarithms of the $\rm{H}^+$ and $\rm OH^-$ concentration normalized to unit molar concentration $c_{\rm unit} = 1\,\rm{M}$:

\begin{align}
    \rm{pH} &= - \lg\left(\frac{c(\rm{H}^+)}{c_{\rm{unit}}}\right), \hspace{0.2cm}
    \rm{pOH} &= - \lg\left(\frac{c(\rm{OH}^-)}{c_{\rm{unit}}}\right)
\label{eq:DecadicLog}
\end{align}

At standard conditions the $\rm H^+$ concentration of $\mathrm{0.1\,\upmu M}$ corresponds to a neutral pH value of 7 in pure water. According to equation (\ref{eq:Chem_WaterAutoprotolysis}) also the $\rm OH^-$ concentration equals $\mathrm{0.1\,\upmu M}$ as $c\rm (H^+)$ and $c\rm (OH^-)$ are coupled. Adding acids or bases manipulates $c\rm (H^+)$ and $c\rm (OH^-)$ for the solution to become more acidic or basic, respectively,
while maintaining $K_{\rm W} = 1\cdot10^{-14}\,\rm{M}^2$.

Due to the generation of several primary species via radiolytic fission and subsequent reactions \cite{LeCaer.2011}, the inverse proportionality of $c\rm (H^+)$ and $c\rm (OH^-)$ is decoupled under irradiation which is in contrast to classical chemical conditions:

\begin{eqnarray}
    \rm{H_2O \xrightarrow[{radiation}]{{ionizing}} 
        \substack{\rm e_h^-, HO^\bullet, H^\bullet, HO_2^\bullet,\\ \rm H^+, OH^-, H_2O_2, H_2}}
\label{eq:ionizing_radiation}
\end{eqnarray}

\begin{figure*}
\includegraphics[scale=0.6]{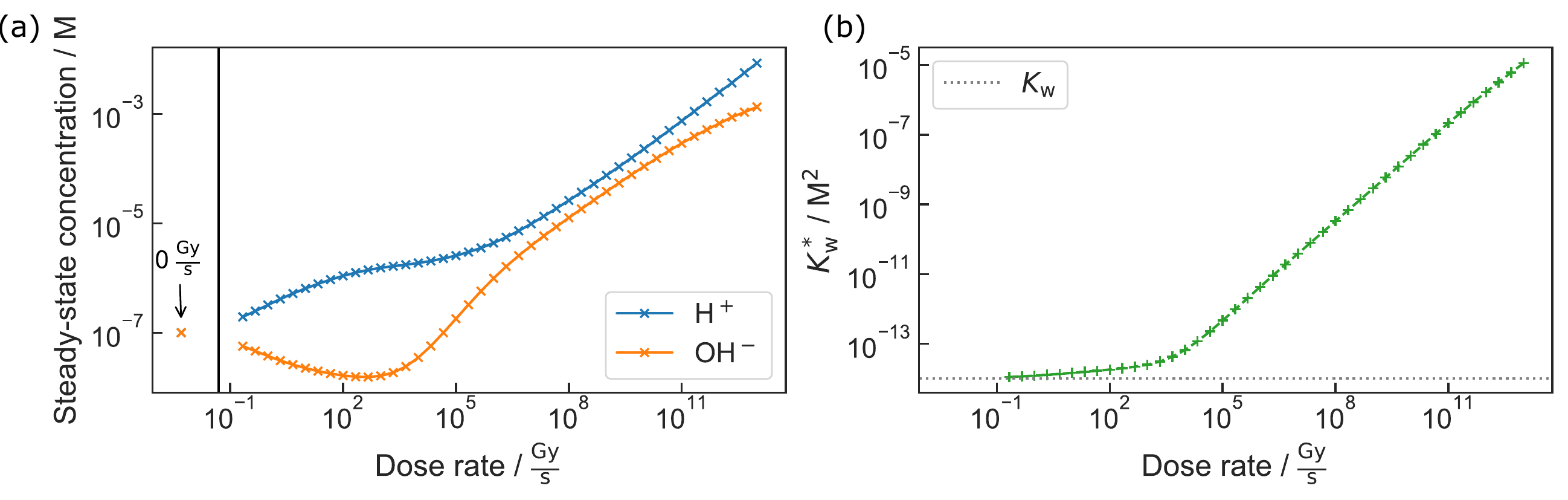}
\caption{\label{fig:SteadyStateWater2}(a) Steady-state concentrations of $\rm H^+$ and $\rm OH^-$ in pure, aerated $\rm H_2O$ and (b) respective ion product, both as a function of the dose rate of electron irradiation.}
\end{figure*}

Consequently, $K_{\mathrm W}$ does not necessarily remain constant in irradiated solutions. Access to the kinetic reaction constants and $G$-values (number of molecules created by energy unit, see Table\,\ref{tab:GenerationValues} of the supplementary information) typical for specific types of radiation allows to simulate these reaction pathways \cite{Schneider.2014, Fritsch.2022b} (see Figure\,\ref{fig:Network1}(b) in the supplementary information).

For pure, aerated ($c_{\rm sat} \rm (O_2) = 0.255\,mol\cdot L^{-1}$, \cite{Schneider.2014}) water exposed to electron irradiation the steady state concentrations of $\rm H^+$ and $\rm OH^-$ are plotted as function of dose rate (Figure\,\ref{fig:SteadyStateWater2}(a)). Evidently, the concentrations of $\rm H^+$ and $\rm OH^-$ are strongly influenced by reactions with such radiolysis products.

Furthermore, $\rm H^+$ and $\rm OH^-$ themselves are primarily generated (eq.\,\ref{eq:ionizing_radiation}) so that the ion product is remarkably changed when the solution is exposed to ionizing radiation.
As illustrated in Figure\,\ref{fig:SteadyStateWater2}(b), the ion product under irradiation does not denote the equilibrium constant but the product of $c(\mathrm{H+})$ and $c(\mathrm{OH^-})$ instead.
To emphasize this fundamental difference, the radiolytic ion product $K_\mathrm{W}^*$ is introduced:
\begin{equation}
K_\mathrm{W} \xrightarrow[\mathrm{radiation}]{\mathrm{ionizing}} K_\mathrm{W}^* = \left(c(\mathrm{H^+})\cdot c(\mathrm{OH^-})\right)_\mathrm{irradiated}
\label{eq:KW*}    
\end{equation}
A direct proportionality (power law with an exponent of unity)
of $K_{\rm W}^*$ to the dose rate is observed for values above $1\,\rm kGy\cdot\,s^{-1}$ (Figure\,\ref{fig:SteadyStateWater2}(b)). 

Consequently, a more conclusive interpretation of the acidification of irradiated solutions is required that accounts for the drastically different interplay of both species within the solution. 

To predict whether $c \rm (H^+)$ or $c (\rm OH^-)$ is dominating and, thus, if an irradiated solution constitutes an acidic or basic environment, we introduce the logarithmic ratio of $c(\rm H^+)$ and $c(\rm OH^-)$ as a new measure.
This is denoted as radiolytic acidity $\pi^*$:

\begin{eqnarray}
\label{eq:pi*}
    \pi^* = \lg\left(\frac{c(\rm H^+)}{c(\rm OH^-)}\right)
\label{eq:radiolytic_acidity}
\end{eqnarray}

A $\pi^*$ of zero represents a neutral environment, whereas positive and negative values describe acidic or basic solutions, respectively. 

\begin{figure*}
\includegraphics[
scale=0.75]{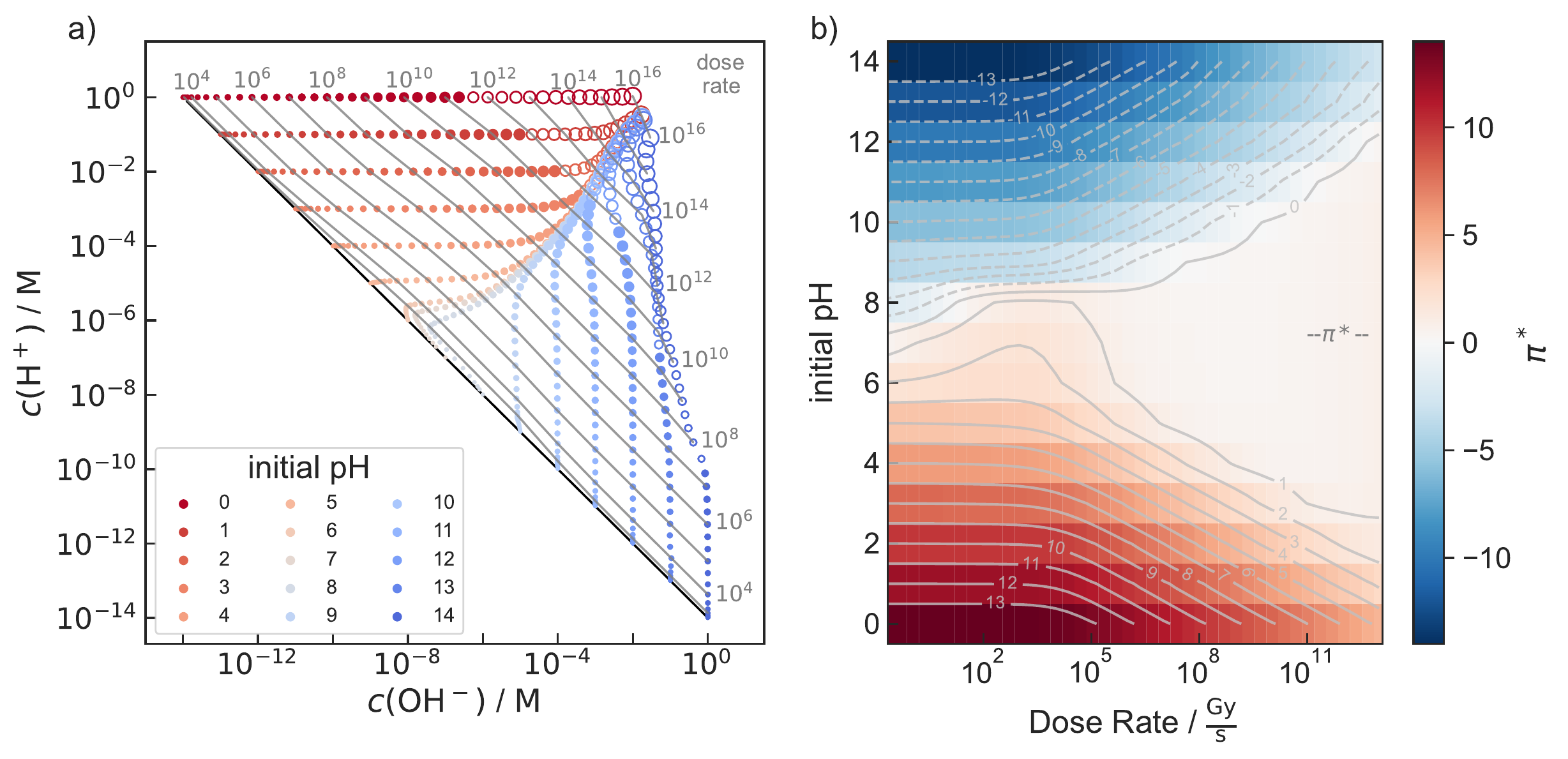}
\caption{\label{fig:pHMap_RadAcidity}Acid/base chemistry of neat, aerated water as a function of dose rate of an electron beam and the initial pH value. (a) Concentrations of $\rm H^+$ and $\rm OH^-$ in the steady state. Each dot represents a simulation, while its size is a measure of the dose rate. Dose rate (grey numbers) is given in $\rm Gy\cdot s^{-1}$ and indicated by contour lines. The black diagonal line corresponds to water under equilibrium conditions ($K_{\rm W} = 10^{-14}\,\rm M^2$) without irradiation. Empty dots represent simulation results, in which the concentration of water in the steady state drops below 99\% of that of non-irradiated solution (see section \ref{sec:level5} for further information). (b) $\pi^*$ (color map and grey contour lines) as function of initial pH and dose rate. The equivalent plots for X-ray irradiation are shown in Figure\,\ref{fig:pHMap_RadAcidity_Xray} in the supplementary information.}
\end{figure*}

For neat 
water, $c \rm (H^+)$ and $c \rm (OH^-)$ depend on the initial pH and especially on the type and dose rate of irradiation. For electron exposure the result for a wide range of initial pH and dose rate is visualized in Figure\,\ref{fig:pHMap_RadAcidity}. Figure\,\ref{fig:pHMap_RadAcidity}\,(a) indicates the decoupling of $\rm H^+$ and $\rm OH^-$ concentrations  under irradiation, while $\pi^*$ is depicted in \ref{fig:pHMap_RadAcidity}\,(b). Equivalent plots for X-ray exposure are shown in Figure\,\ref{fig:pHMap_RadAcidity_Xray} in the supplementary information. Steady-states with a water concentration dropping below 99\% of the non-irradiated solution are indicated by unfilled markers, as discussed in Section V. 

Remarkably, independent of the initial pH value of the specimen solution prior to irradiation, $\pi^*$ converges towards neutral conditions for increasing dose rates. This becomes prominent above $\sim 1\rm\,MGy\cdot\,s^{-1}$. 

A slight asymmetry favoring acidic conditions is assumed to be related to the (slow) decay of $\rm H_2O_2$ and $\rm O_3$ yielding para-oxygen ($\rm O$, \textsuperscript{3}\textit{P}). This in turn triggers a reaction cascade in which, beside others, $\rm OH^-$ is consumed (see Supplementary Table\,\ref{tab:model_Water_S1}).

This interplay of radiation chemistry products with acidity highlights the necessity of elucidating the complete reaction chemistry network, which becomes even more pronounced in systems more complex than water. Hence, in the following, the influence of additives on $\pi^*$ is exemplarily demonstrated with chloride, bromide and nitrate ions. 
All simulations are based on radiation chemistry of pure water (17 species, 83 coupled reactions, Table\,\ref{tab:model_Water_S1}, Figure\,\ref{fig:Network1} in the supplementary information). 
Additional reactions and species are considered for chlorine, bromine and nitrate-containing solutions (see supplementary information).

\begin{figure*}
\includegraphics[width=1\textwidth]{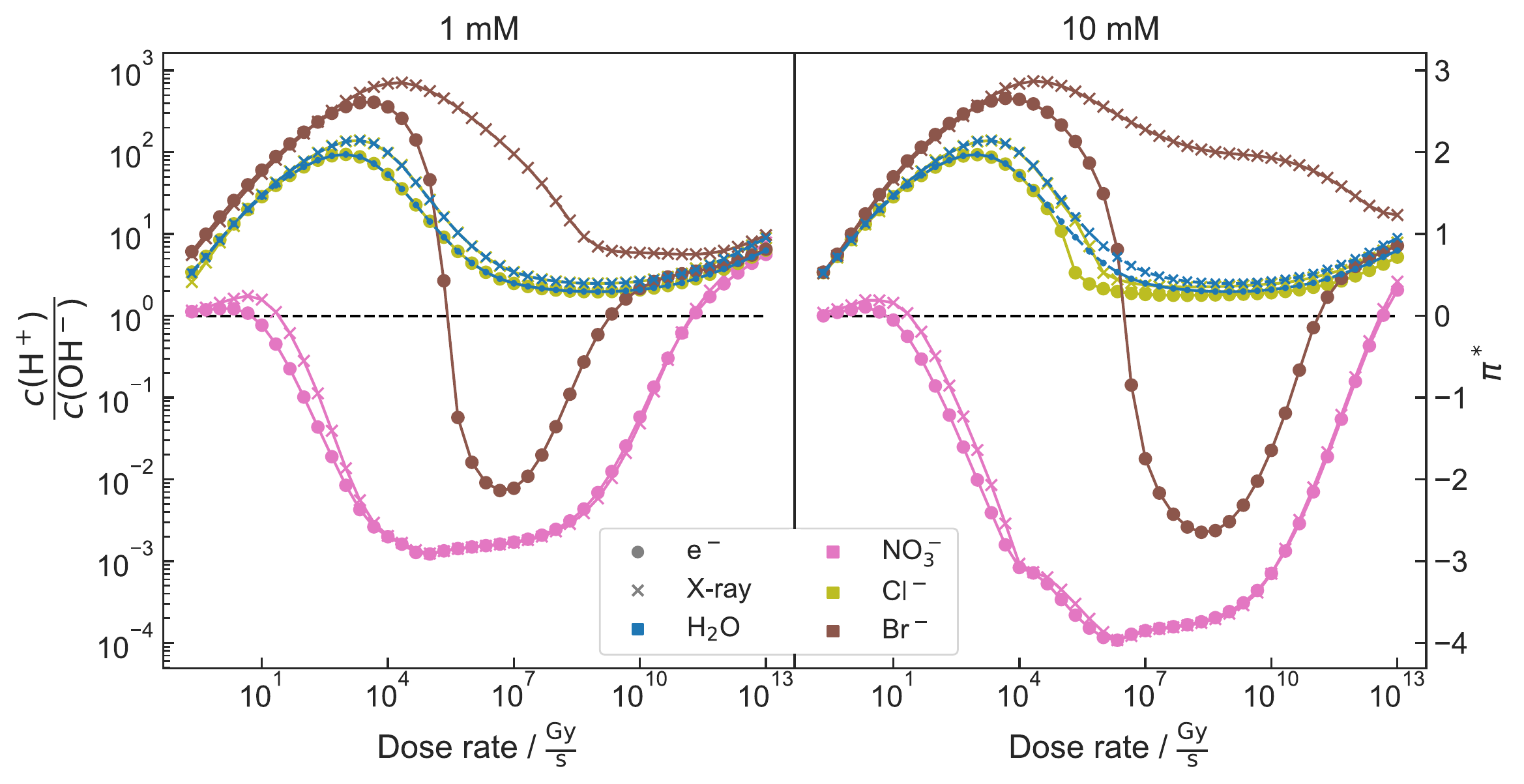}
\caption{\label{fig:ClBrNO3}$\pi^*$ of $\rm Cl^-$, $\rm Br^-$, and $\rm NO_3^-$ solutions and pure water as a function of dose rate. Anion concentrations of $1\,\rm mM$ (left) and $10\,\rm mM$ (right) are considered. `X'-markers correspond to the simulation performed based on $G$-values of X-rays.
The dotted line corresponds to a balance of $c(\rm H^+)$ / $c(\rm OH^-)$ = 1 ($\pi^* = 0$). Note that for $\rm 1\,mM$, $\pi^*$ for $\rm H_2O$ and $\rm Cl^-$ overlap.
Individual simulations and $K_\mathrm{W}^*$ are denoted in Figures\,\ref{fig:SteadyStates4} and \ref{fig:AnionsKW_Relation}.} 
\end{figure*}

The evolution of $\pi^*$ for a solution of $\rm pH = 7$ containing these anions at concentrations of $\rm 1\,mM$ and $\rm 10\,mM$ is shown in Figure\,\ref{fig:ClBrNO3}. The individual concentrations of $c \rm{ (H^+)}$ and $c \rm (OH^-)$ are separately plotted in the supplementary information Figure\,\ref{fig:SteadyStates4}.

As all mentioned anions represent conjugated bases of strong acids ($\rm HCl$, $\rm HBr$, and $\rm HNO_3$), their basic strength is generally negligible.
Consequently, solutions containing these anions can have a neutral pH prior to irradiation.
In combination with thermodynamically stable cations such as Li$^+$,
the anion-impact on radiation chemistry can be investigated experimentally.
As the standard electrode potential of Li$^+$ ($E^\circ(\mathrm{Li/Li^+})=-3.0401\,\mathrm{V}$)
\cite{Rumble.2022} required for reduction exceed $E^\circ$ of the strongest reductant present ($E^\circ(\mathrm{H_2O/e_h^-})=-2.9\,\mathrm{V}$ \cite{LeCaer.2011}), kinetics focusing on the anion -- water interplay were performed exclusively.

Differences in kinetics influence the radiation chemistry, which is evident when comparing dissociation rate constants of conjugated acids ($1.46\cdot 10^{10}\, \rm s^{-1}: \rm HNO_3 \rightarrow H^+ + NO_3^-$\,\cite{Horne.2016}, $5\cdot 10^5\, \rm s^{-1}: HCl \rightarrow H^+ + Cl^-$\,\cite{Kelm.2004}, $1\cdot 10^{13}\, \rm s^{-1}: HBr \rightarrow H^+ + Br^-$\,\cite{Williams.2002}; backward reactions are about 3, 6, and 9 orders of magnitude slower).

Nevertheless, simulation results shown in Figure\,\ref{fig:ClBrNO3} suggest that chloride ions barely influence the evolution of $c(\rm H^+)$ and $c(\rm OH^-)$. However, bromide and nitrate ions strongly alter the acidity of the irradiated solution: 

For both radiation types, nitrate mitigates the impact on $c(\rm H^+)$ and $c(\rm OH^-)$ at low dose rates. Above  ~$1\rm\, kGy\cdot\,s^{-1}$ the solution becomes more basic (negative $\pi^*$) compared to pure water, becomes more neutral for higher dose rates and finally turns acidic beyond $\sim 100\rm\, GGy\cdot\,s^{-1}$.

While the type of radiation - electrons or X-ray photons - appears to have no qualitative influence on the evolution of $c(\rm H^+)$ and $c(\rm OH^-)$ for aqueous solutions of $\rm Cl^-$ or $\rm NO_3^-$, the simulation indicates a difference for $\rm Br^-$. When irradiated with X-rays, the solution remains acidic for the entire simulated dose rate range. When considering electron irradiation, however, the solution shows acidic behavior for low and high dose rates, while a basic behavior evolves for dose rates between $500\rm\,kGy\cdot\,s^{-1}$ and $5\rm\,GGy\cdot\,s^{-1}$ for an initial concentration of $1\,\rm mM$. 

A change of the concentration from $1\,\rm mM$ to $10\,\rm mM$ does not qualitatively change the shape of the curves, but enhances tendencies and therefore shifts intersection points by about one to two orders of magnitude. 

For bromide ion concentrations of $10\,\rm mM$ this shift causes the intersection point to come close to parameters accessible in standard LP-TEM, indicating that $\rm Br^-$ could be a promising candidate for \textit{in situ} studies of acidity-dependent precipitation reactions, using only dose rate adjustments. 
The different impact of bromide and chloride is remarkable. Although in general both halides show comparable chemical properties, the larger bromine radical is lower in energy than the chlorine equivalent. This is reflected by $E^\circ$ of $+2.43\,\mathrm{V}$ for $\rm Cl/Cl^-$ and $+1.96\,\mathrm{V}$ for $\rm Br/Br^-$ \cite{Armstrong.2015} which impacts the radical chemistry that dominates kinetic models. Consequently, reaction kinetics of bromide ions exhibit a stronger involvement of acidity-mediating pathways. 

\section{\label{sec:level3}Discussion}
By referring to $\pi^*$ in non-irradiated solutions, it can be mapped for an initial pH value (see supplement for details). Nevertheless, this should not be carelessly translated to irradiated solutions, as there, pH does not provide a holistic picture of the acid-base interplay.

Moreover, $\pi^*$ assumes an equivalent reactivity with $\rm H^+$ and $\rm OH^-$. This might be misleading in situations where this prerequisite is not fulfilled. In addition, identical $\pi^*$ values can be obtained by different absolute concentrations. Therefore, both, $\pi^*$ and $K_\mathrm{W}^*$ should be considered in combination as this fully describes the acidity under irradiation.

Albeit neat, aerated water is the basis of many experiments, this work emphasises once more that any extrapolation of these findings to different settings must be treated with caution. Although multiple scenarios have already been elucidated here, additional changes in experimental conditions may significantly alter steady-state concentrations of $\rm H^+$, $\rm OH^-$.

Experimental conditions may deviate significantly from the described simulations. In particular, the simulations shown here consider neither diffusion nor phase boundaries and are therefore only accurate when an isotropic volume element is irradiated homogeneously. Thus, it is only a guidance for experiments using scanning probes in large non-irradiated liquid reservoirs or flow setups.
Mind also that consumption of the solvent - water - limits the validity of the assumption that the radiation only interacts with water for high dose rates (usually above $\mathrm{10^{13}\,Gy\cdot s^{-1}}$ \cite{Schneider.2014}).
Furthermore, as demonstrated, even additives considered as non-reactive, can drastically change the chemistry at hand. Thus, any extrapolation should be performed cautiously.

Especially relevant for LP-TEM is electron beam induced heating, which can significantly affect the redox chemistry. However, this effect is simulated to have a negligible influence on $c(\mathrm{H^+})$ and $c(\mathrm{OH^-})$ in pure water \cite{Fritsch.2021}, suggesting that $\pi^*$ is not affected by beam-heating.

Nevertheless, the herein presented work provides a good approximation for liquid cell architectures with small, static volumes irradiated completely by X-rays (e.g. in synchrotron beam line end stations) and/or electron beams in TEM (e.g. graphene liquid cells \cite{Yuk.2012} and derivatives \cite{Yin.2019, Hutzler.2018, Hutzler.2019b, Bae.2020, Lim.2020, Kelly.2018, Liu.2021, Yang.2019d}).

The large parameter space comprising types of radiation, dose rate, additives, initial concentrations etc. allows for tailoring specific conditions. In this letter, we merely scratch the surface to illustrate the observable effects.
However, experimental verification of the model is necessary.
Hence, suitable marker reactions, showing structural changes, precipitation or dissolution in the accessible $c \mathrm{(H^+)} / c \mathrm{ (OH^-)}$  range could be employed.

\section{\label{sec:level4}Conclusion and Outlook}

While our work seconds the finding that irradiation increases $c(\mathrm{ H^+})$ \cite{Schneider.2014} we show that $\mathrm{ H^+}$ alone is insufficient to quantify the acidity of aqueous solutions interacting with ionizing radiation. Hence, by introducing $\pi^*$ and $K_{\rm W}^*$ as more adequate measures that consider the relation of $c \mathrm{ (H^+)}$ and $c (\mathrm{ OH^-})$ we unveil that in pure water, electron beam and X-ray irradiation drives the acidity towards a balanced environment, even for high or low initial pH values. Moreover, we show that adding $\mathrm{ Cl^-}$, $\mathrm{ Br^-}$, and $\mathrm{ NO_3^-}$ ions significantly impacts $\pi^*$. This allows for tailoring $\pi^*$ during LP-TEM experiments by means of initial concentration and dose rate for quantitative \textit{in situ} studies. Here, $\mathrm{ Br^-}$ ions are promising candidates for validating the predictions made.
Finally, our simulations provide valuable insights for radiation chemistry even beyond LP-TEM and XRD techniques even towards astrochemical physics \cite{Shingledecker.2018, Arumainayagam.2019}. 

\section{\label{sec:level5}Experimental/Computational Methods}

Radiolysis simulations were performed utilizing AuRaCh, a custom-build algorithm which has been described in our previous work \cite{Fritsch.2022b}. Coupled ordinary differential equations (ODEs) are used to simulate the concentration $c$ of species $i$ over time $t$, depending on the concentration of reactants $l$ and $n$. With the liquid density $\rho$, Faraday constant $F$, dose rate $\Psi$ in $[\mathrm{ Gy s^{-1}}]$, generation value $G_i$ of species $i$, and the kinetic constant $k_j$ and $k_m$ for reaction $j$ and $m$, respectively, it can be expressed as:

\begin{align}
    \frac{\partial c_i}{\partial t} = \frac{\rho}{F}\Psi G_i + \sum\limits_j k_j\left(\prod\limits_l c_l\right) - \sum\limits_{m\neq j} k_m \left( \prod\limits_n c_n\right)
\end{align}

Here we assume sole interaction of radiation with water, for which $G$-values are well-known. For electron beam-irradiation the herein used $G$ values are valid for an energy of $\rm 200\,- 300\,keV$, while the $G$-values for X-ray irradiation are generally valid for low linear energy transfer (low-LET) radiation \cite{Gupta.2018, Hill.1994}.
Note that other acceleration voltages (e.g. that typical for SEM) can have deviating $G$ values and thus would yield a different chemistry.
This can easily be simulated with the herein presented tools. Simulation results where the amount of radiolytic products in steady state exceeds $1\%$ of the water concentration are indicated by hollow markers in Figure\,\ref{fig:pHMap_RadAcidity} (a) and \ref{fig:pHMap_RadAcidity_Xray} (a), because the assumption of radiation only interacting with $\rm H_2O$ becomes questionable. \\

\begin{acknowledgments}
Financial support by the German Research Foundation \textit{via} the Research Training Group GRK 1896 "In situ microscopy with electrons, X-rays and scanning probes", by the Cluster of Excellence "Engineering of Advanced Materials (EAM)" is gratefully acknowledged. AH and AK acknowledge the financial support by the Federal Ministry of Education and Research (BMBF) of Germany in the programme H\textsubscript{2}Giga - StacIE. (Project identification number: 03HY103H). TC, RB and LGB, furthermore, acknowledge support by the Helmholtz Recruiting Initiative grant (no. I-044-16-01).
\end{acknowledgments}

\section*{Author contribution}

B. F. -- Writing - original draft (equal), formal analysis (equal) - electron simulations (lead), data curation (equal), investigation (equal) - electron simulations (lead), conceptualization (equal), methodology (lead), review and editing (supporting), 
A. K. -- Writing - original draft (equal), methodology (supporting), formal analysis (equal) - X-ray simulations (lead), data curation (equal), investigation (equal) - X-ray simulations (lead), validation (lead), Review and editing (lead),
M. T. -- validation (supporting), methodology (supporting), review and editing (supporting),
T. C. -- methodology (supporting), review and editing (supporting),
R. B. -- methodology (supporting), review and editing (supporting),
L. B. -- methodology (supporting), review and editing (supporting), resources (supporting),
M. J. -- methodology (supporting), review and editing (supporting), supervision (supporting), resources (equal),
E. S. -- methodology (supporting), review and editing (supporting), supervision (supporting), resources (equal),
A. H. -- methodology (supporting), review and editing (supporting), supervision (lead), resources (equal), Writing - original draft (supporting).

B. Fritsch and A. Körner contributed equally to this work. 

\bibliography{Bibliography}

\begin{thebibliography}{72}%
\makeatletter
\providecommand \@ifxundefined [1]{%
 \@ifx{#1\undefined}
}%
\providecommand \@ifnum [1]{%
 \ifnum #1\expandafter \@firstoftwo
 \else \expandafter \@secondoftwo
 \fi
}%
\providecommand \@ifx [1]{%
 \ifx #1\expandafter \@firstoftwo
 \else \expandafter \@secondoftwo
 \fi
}%
\providecommand \natexlab [1]{#1}%
\providecommand \enquote  [1]{``#1''}%
\providecommand \bibnamefont  [1]{#1}%
\providecommand \bibfnamefont [1]{#1}%
\providecommand \citenamefont [1]{#1}%
\providecommand \href@noop [0]{\@secondoftwo}%
\providecommand \href [0]{\begingroup \@sanitize@url \@href}%
\providecommand \@href[1]{\@@startlink{#1}\@@href}%
\providecommand \@@href[1]{\endgroup#1\@@endlink}%
\providecommand \@sanitize@url [0]{\catcode `\\12\catcode `\$12\catcode
  `\&12\catcode `\#12\catcode `\^12\catcode `\_12\catcode `\%12\relax}%
\providecommand \@@startlink[1]{}%
\providecommand \@@endlink[0]{}%
\providecommand \url  [0]{\begingroup\@sanitize@url \@url }%
\providecommand \@url [1]{\endgroup\@href {#1}{\urlprefix }}%
\providecommand \urlprefix  [0]{URL }%
\providecommand \Eprint [0]{\href }%
\providecommand \doibase [0]{https://doi.org/}%
\providecommand \selectlanguage [0]{\@gobble}%
\providecommand \bibinfo  [0]{\@secondoftwo}%
\providecommand \bibfield  [0]{\@secondoftwo}%
\providecommand \translation [1]{[#1]}%
\providecommand \BibitemOpen [0]{}%
\providecommand \bibitemStop [0]{}%
\providecommand \bibitemNoStop [0]{.\EOS\space}%
\providecommand \EOS [0]{\spacefactor3000\relax}%
\providecommand \BibitemShut  [1]{\csname bibitem#1\endcsname}%
\let\auto@bib@innerbib\@empty
\bibitem [{\citenamefont {LaForge}\ \emph {et~al.}(2019)\citenamefont
  {LaForge}, \citenamefont {Michiels}, \citenamefont {Bohlen}, \citenamefont
  {Callegari}, \citenamefont {Clark}, \citenamefont {von Conta}, \citenamefont
  {Coreno}, \citenamefont {{Di Fraia}}, \citenamefont {Drabbels}, \citenamefont
  {Huppert} \emph {et~al.}}]{LaForge.2019}%
  \BibitemOpen
  \bibfield  {author} {\bibinfo {author} {\bibfnamefont {A.~C.}\ \bibnamefont
  {LaForge}}, \bibinfo {author} {\bibfnamefont {R.}~\bibnamefont {Michiels}},
  \bibinfo {author} {\bibfnamefont {M.}~\bibnamefont {Bohlen}}, \bibinfo
  {author} {\bibfnamefont {C.}~\bibnamefont {Callegari}}, \bibinfo {author}
  {\bibfnamefont {A.}~\bibnamefont {Clark}}, \bibinfo {author} {\bibfnamefont
  {A.}~\bibnamefont {von Conta}}, \bibinfo {author} {\bibfnamefont
  {M.}~\bibnamefont {Coreno}}, \bibinfo {author} {\bibfnamefont
  {M.}~\bibnamefont {{Di Fraia}}}, \bibinfo {author} {\bibfnamefont
  {M.}~\bibnamefont {Drabbels}}, \bibinfo {author} {\bibfnamefont
  {M.}~\bibnamefont {Huppert}}, \emph {et~al.},\ }\href
  {https://doi.org/10.1103/PhysRevLett.122.133001} {\bibfield  {journal}
  {\bibinfo  {journal} {Physical Review Letters}\ }\textbf {\bibinfo {volume}
  {122}},\ \bibinfo {pages} {133001} (\bibinfo {year} {2019})},\ \Eprint
  {https://arxiv.org/abs/31012607} {31012607} \BibitemShut {NoStop}%
\bibitem [{\citenamefont {Jahnke}\ \emph {et~al.}(2021)\citenamefont {Jahnke},
  \citenamefont {Guillemin}, \citenamefont {Inhester}, \citenamefont {Son},
  \citenamefont {Kastirke}, \citenamefont {Ilchen}, \citenamefont {Rist},
  \citenamefont {Trabert}, \citenamefont {Melzer}, \citenamefont {Anders} \emph
  {et~al.}}]{Jahnke.2021}%
  \BibitemOpen
  \bibfield  {author} {\bibinfo {author} {\bibfnamefont {T.}~\bibnamefont
  {Jahnke}}, \bibinfo {author} {\bibfnamefont {R.}~\bibnamefont {Guillemin}},
  \bibinfo {author} {\bibfnamefont {L.}~\bibnamefont {Inhester}}, \bibinfo
  {author} {\bibfnamefont {S.-K.}\ \bibnamefont {Son}}, \bibinfo {author}
  {\bibfnamefont {G.}~\bibnamefont {Kastirke}}, \bibinfo {author}
  {\bibfnamefont {M.}~\bibnamefont {Ilchen}}, \bibinfo {author} {\bibfnamefont
  {J.}~\bibnamefont {Rist}}, \bibinfo {author} {\bibfnamefont {D.}~\bibnamefont
  {Trabert}}, \bibinfo {author} {\bibfnamefont {N.}~\bibnamefont {Melzer}},
  \bibinfo {author} {\bibfnamefont {N.}~\bibnamefont {Anders}}, \emph
  {et~al.},\ }\bibfield  {journal} {\bibinfo  {journal} {Physical Review X}\
  }\textbf {\bibinfo {volume} {11}},\ \href
  {https://doi.org/10.1103/PhysRevX.11.041044} {10.1103/PhysRevX.11.041044}
  (\bibinfo {year} {2021})\BibitemShut {NoStop}%
\bibitem [{\citenamefont {Wang}\ \emph {et~al.}(2022)\citenamefont {Wang},
  \citenamefont {Xu}, \citenamefont {Chen}, \citenamefont {Shangguan},
  \citenamefont {Dong}, \citenamefont {Ma}, \citenamefont {Zhang},
  \citenamefont {Yang}, \citenamefont {Bai}, \citenamefont {Guo}, \citenamefont
  {Fang}, \citenamefont {Zheng},\ and\ \citenamefont {Sun}}]{Wang.2022}%
  \BibitemOpen
  \bibfield  {author} {\bibinfo {author} {\bibfnamefont {W.}~\bibnamefont
  {Wang}}, \bibinfo {author} {\bibfnamefont {T.}~\bibnamefont {Xu}}, \bibinfo
  {author} {\bibfnamefont {J.}~\bibnamefont {Chen}}, \bibinfo {author}
  {\bibfnamefont {J.}~\bibnamefont {Shangguan}}, \bibinfo {author}
  {\bibfnamefont {H.}~\bibnamefont {Dong}}, \bibinfo {author} {\bibfnamefont
  {H.}~\bibnamefont {Ma}}, \bibinfo {author} {\bibfnamefont {Q.}~\bibnamefont
  {Zhang}}, \bibinfo {author} {\bibfnamefont {J.}~\bibnamefont {Yang}},
  \bibinfo {author} {\bibfnamefont {T.}~\bibnamefont {Bai}}, \bibinfo {author}
  {\bibfnamefont {Z.}~\bibnamefont {Guo}}, \bibinfo {author} {\bibfnamefont
  {H.}~\bibnamefont {Fang}}, \bibinfo {author} {\bibfnamefont {H.}~\bibnamefont
  {Zheng}},\ and\ \bibinfo {author} {\bibfnamefont {L.}~\bibnamefont {Sun}},\
  }\href {https://doi.org/10.1038/s41563-022-01261-x} {\bibfield  {journal}
  {\bibinfo  {journal} {Nature Materials}\ ,\ \bibinfo {pages} {1}} (\bibinfo
  {year} {2022})},\ \Eprint {https://arxiv.org/abs/35618827} {35618827}
  \BibitemShut {NoStop}%
\bibitem [{\citenamefont {Signorell}(2020)}]{Signorell.2020}%
  \BibitemOpen
  \bibfield  {author} {\bibinfo {author} {\bibfnamefont {R.}~\bibnamefont
  {Signorell}},\ }\href {https://doi.org/10.1103/PhysRevLett.124.205501}
  {\bibfield  {journal} {\bibinfo  {journal} {Physical Review Letters}\
  }\textbf {\bibinfo {volume} {124}},\ \bibinfo {pages} {205501} (\bibinfo
  {year} {2020})},\ \Eprint {https://arxiv.org/abs/32501058} {32501058}
  \BibitemShut {NoStop}%
\bibitem [{\citenamefont {Yesibolati}\ \emph
  {et~al.}(2020{\natexlab{a}})\citenamefont {Yesibolati}, \citenamefont
  {Lagan{\'a}}, \citenamefont {Kadkhodazadeh}, \citenamefont {Mikkelsen},
  \citenamefont {Sun}, \citenamefont {Kasama}, \citenamefont {Hansen},
  \citenamefont {Zaluzec},\ and\ \citenamefont {M{\o}lhave}}]{Yesibolati.2020}%
  \BibitemOpen
  \bibfield  {author} {\bibinfo {author} {\bibfnamefont {M.~N.}\ \bibnamefont
  {Yesibolati}}, \bibinfo {author} {\bibfnamefont {S.}~\bibnamefont
  {Lagan{\'a}}}, \bibinfo {author} {\bibfnamefont {S.}~\bibnamefont
  {Kadkhodazadeh}}, \bibinfo {author} {\bibfnamefont {E.~K.}\ \bibnamefont
  {Mikkelsen}}, \bibinfo {author} {\bibfnamefont {H.}~\bibnamefont {Sun}},
  \bibinfo {author} {\bibfnamefont {T.}~\bibnamefont {Kasama}}, \bibinfo
  {author} {\bibfnamefont {O.}~\bibnamefont {Hansen}}, \bibinfo {author}
  {\bibfnamefont {N.~J.}\ \bibnamefont {Zaluzec}},\ and\ \bibinfo {author}
  {\bibfnamefont {K.}~\bibnamefont {M{\o}lhave}},\ }\bibfield  {journal}
  {\bibinfo  {journal} {Nanoscale}\ }\href {https://doi.org/10.1039/D0NR04352D}
  {10.1039/D0NR04352D} (\bibinfo {year} {2020}{\natexlab{a}})\BibitemShut
  {NoStop}%
\bibitem [{\citenamefont {Yesibolati}\ \emph
  {et~al.}(2020{\natexlab{b}})\citenamefont {Yesibolati}, \citenamefont
  {Lagan{\`a}}, \citenamefont {Sun}, \citenamefont {Beleggia}, \citenamefont
  {Kathmann}, \citenamefont {Kasama},\ and\ \citenamefont
  {M{\o}lhave}}]{Yesibolati.2020b}%
  \BibitemOpen
  \bibfield  {author} {\bibinfo {author} {\bibfnamefont {M.~N.}\ \bibnamefont
  {Yesibolati}}, \bibinfo {author} {\bibfnamefont {S.}~\bibnamefont
  {Lagan{\`a}}}, \bibinfo {author} {\bibfnamefont {H.}~\bibnamefont {Sun}},
  \bibinfo {author} {\bibfnamefont {M.}~\bibnamefont {Beleggia}}, \bibinfo
  {author} {\bibfnamefont {S.~M.}\ \bibnamefont {Kathmann}}, \bibinfo {author}
  {\bibfnamefont {T.}~\bibnamefont {Kasama}},\ and\ \bibinfo {author}
  {\bibfnamefont {K.}~\bibnamefont {M{\o}lhave}},\ }\href
  {https://doi.org/10.1103/PhysRevLett.124.065502} {\bibfield  {journal}
  {\bibinfo  {journal} {Physical Review Letters}\ }\textbf {\bibinfo {volume}
  {124}},\ \bibinfo {pages} {065502} (\bibinfo {year} {2020}{\natexlab{b}})},\
  \Eprint {https://arxiv.org/abs/32109081} {32109081} \BibitemShut {NoStop}%
\bibitem [{\citenamefont {Woehl}(2020)}]{Woehl.2020}%
  \BibitemOpen
  \bibfield  {author} {\bibinfo {author} {\bibfnamefont {T.~J.}\ \bibnamefont
  {Woehl}},\ }\bibfield  {journal} {\bibinfo  {journal} {Chemistry of
  Materials}\ }\href {https://doi.org/10.1021/acs.chemmater.0c01360}
  {10.1021/acs.chemmater.0c01360} (\bibinfo {year} {2020})\BibitemShut
  {NoStop}%
\bibitem [{\citenamefont {Steinr{\"u}ck}\ \emph {et~al.}(2020)\citenamefont
  {Steinr{\"u}ck}, \citenamefont {Cao}, \citenamefont {Lukatskaya},
  \citenamefont {Takacs}, \citenamefont {Wan}, \citenamefont {Mackanic},
  \citenamefont {Tsao}, \citenamefont {Zhao}, \citenamefont {Helms},
  \citenamefont {Xu}, \citenamefont {Borodin}, \citenamefont {Wishart},\ and\
  \citenamefont {Toney}}]{Steinruck.2020}%
  \BibitemOpen
  \bibfield  {author} {\bibinfo {author} {\bibfnamefont {H.-G.}\ \bibnamefont
  {Steinr{\"u}ck}}, \bibinfo {author} {\bibfnamefont {C.}~\bibnamefont {Cao}},
  \bibinfo {author} {\bibfnamefont {M.~R.}\ \bibnamefont {Lukatskaya}},
  \bibinfo {author} {\bibfnamefont {C.~J.}\ \bibnamefont {Takacs}}, \bibinfo
  {author} {\bibfnamefont {G.}~\bibnamefont {Wan}}, \bibinfo {author}
  {\bibfnamefont {D.~G.}\ \bibnamefont {Mackanic}}, \bibinfo {author}
  {\bibfnamefont {Y.}~\bibnamefont {Tsao}}, \bibinfo {author} {\bibfnamefont
  {J.}~\bibnamefont {Zhao}}, \bibinfo {author} {\bibfnamefont {B.~A.}\
  \bibnamefont {Helms}}, \bibinfo {author} {\bibfnamefont {K.}~\bibnamefont
  {Xu}}, \bibinfo {author} {\bibfnamefont {O.}~\bibnamefont {Borodin}},
  \bibinfo {author} {\bibfnamefont {J.~F.}\ \bibnamefont {Wishart}},\ and\
  \bibinfo {author} {\bibfnamefont {M.~F.}\ \bibnamefont {Toney}},\ }\href
  {https://doi.org/10.1002/anie.202007745} {\bibfield  {journal} {\bibinfo
  {journal} {Angewandte Chemie International Edition}\ }\textbf {\bibinfo
  {volume} {59}},\ \bibinfo {pages} {23180} (\bibinfo {year}
  {2020})}\BibitemShut {NoStop}%
\bibitem [{\citenamefont {Bras}\ \emph {et~al.}(2021)\citenamefont {Bras},
  \citenamefont {Myles},\ and\ \citenamefont {Felici}}]{Bras.2021}%
  \BibitemOpen
  \bibfield  {author} {\bibinfo {author} {\bibfnamefont {W.}~\bibnamefont
  {Bras}}, \bibinfo {author} {\bibfnamefont {D.~A.~A.}\ \bibnamefont {Myles}},\
  and\ \bibinfo {author} {\bibfnamefont {R.}~\bibnamefont {Felici}},\ }\href
  {https://doi.org/10.1088/1361-648X/ac1767} {\bibfield  {journal} {\bibinfo
  {journal} {Journal of physics. Condensed matter : an Institute of Physics
  journal}\ }\textbf {\bibinfo {volume} {33}},\ \bibinfo {pages} {423002}
  (\bibinfo {year} {2021})},\ \Eprint {https://arxiv.org/abs/34298526}
  {34298526} \BibitemShut {NoStop}%
\bibitem [{\citenamefont {Bras}\ \emph {et~al.}(2022)\citenamefont {Bras},
  \citenamefont {Newton}, \citenamefont {Myles},\ and\ \citenamefont
  {Felici}}]{Bras.2022}%
  \BibitemOpen
  \bibfield  {author} {\bibinfo {author} {\bibfnamefont {W.}~\bibnamefont
  {Bras}}, \bibinfo {author} {\bibfnamefont {M.~A.}\ \bibnamefont {Newton}},
  \bibinfo {author} {\bibfnamefont {D.~A.~A.}\ \bibnamefont {Myles}},\ and\
  \bibinfo {author} {\bibfnamefont {R.}~\bibnamefont {Felici}},\ }\bibfield
  {journal} {\bibinfo  {journal} {Nature Reviews Methods Primers}\ }\textbf
  {\bibinfo {volume} {2}},\ \href {https://doi.org/10.1038/s43586-022-00112-y}
  {10.1038/s43586-022-00112-y} (\bibinfo {year} {2022}),\ \bibinfo {note} {pII:
  112}\BibitemShut {NoStop}%
\bibitem [{\citenamefont {Fritsch}\ \emph {et~al.}(2022)\citenamefont
  {Fritsch}, \citenamefont {Zech}, \citenamefont {Bruns}, \citenamefont
  {Körner}, \citenamefont {Khadivianazar}, \citenamefont {Wu}, \citenamefont
  {Zargar~Talebi}, \citenamefont {Virtanen}, \citenamefont {Unruh},
  \citenamefont {Jank}, \citenamefont {Spiecker},\ and\ \citenamefont
  {Hutzler}}]{Fritsch.2022b}%
  \BibitemOpen
  \bibfield  {author} {\bibinfo {author} {\bibfnamefont {B.}~\bibnamefont
  {Fritsch}}, \bibinfo {author} {\bibfnamefont {T.~S.}\ \bibnamefont {Zech}},
  \bibinfo {author} {\bibfnamefont {M.~P.}\ \bibnamefont {Bruns}}, \bibinfo
  {author} {\bibfnamefont {A.}~\bibnamefont {Körner}}, \bibinfo {author}
  {\bibfnamefont {S.}~\bibnamefont {Khadivianazar}}, \bibinfo {author}
  {\bibfnamefont {M.}~\bibnamefont {Wu}}, \bibinfo {author} {\bibfnamefont
  {N.}~\bibnamefont {Zargar~Talebi}}, \bibinfo {author} {\bibfnamefont
  {S.}~\bibnamefont {Virtanen}}, \bibinfo {author} {\bibfnamefont
  {T.}~\bibnamefont {Unruh}}, \bibinfo {author} {\bibfnamefont {M.~P.~M.}\
  \bibnamefont {Jank}}, \bibinfo {author} {\bibfnamefont {E.}~\bibnamefont
  {Spiecker}},\ and\ \bibinfo {author} {\bibfnamefont {A.}~\bibnamefont
  {Hutzler}},\ }\href {https://doi.org/https://doi.org/10.1002/advs.202202803}
  {\bibfield  {journal} {\bibinfo  {journal} {Advanced Science}\ }\textbf
  {\bibinfo {volume} {9}},\ \bibinfo {pages} {2202803} (\bibinfo {year}
  {2022})}\BibitemShut {NoStop}%
\bibitem [{\citenamefont {Schneider}\ \emph {et~al.}(2014)\citenamefont
  {Schneider}, \citenamefont {Norton}, \citenamefont {Mendel}, \citenamefont
  {Grogan}, \citenamefont {Ross},\ and\ \citenamefont {Bau}}]{Schneider.2014}%
  \BibitemOpen
  \bibfield  {author} {\bibinfo {author} {\bibfnamefont {N.~M.}\ \bibnamefont
  {Schneider}}, \bibinfo {author} {\bibfnamefont {M.~M.}\ \bibnamefont
  {Norton}}, \bibinfo {author} {\bibfnamefont {B.~J.}\ \bibnamefont {Mendel}},
  \bibinfo {author} {\bibfnamefont {J.~M.}\ \bibnamefont {Grogan}}, \bibinfo
  {author} {\bibfnamefont {F.~M.}\ \bibnamefont {Ross}},\ and\ \bibinfo
  {author} {\bibfnamefont {H.~H.}\ \bibnamefont {Bau}},\ }\href
  {https://doi.org/10.1021/jp507400n} {\bibfield  {journal} {\bibinfo
  {journal} {The Journal of Physical Chemistry C}\ }\textbf {\bibinfo {volume}
  {118}},\ \bibinfo {pages} {22373} (\bibinfo {year} {2014})}\BibitemShut
  {NoStop}%
\bibitem [{\citenamefont {Gupta}\ \emph {et~al.}(2018)\citenamefont {Gupta},
  \citenamefont {Schneider}, \citenamefont {Park}, \citenamefont {Steingart},\
  and\ \citenamefont {Ross}}]{Gupta.2018}%
  \BibitemOpen
  \bibfield  {author} {\bibinfo {author} {\bibfnamefont {T.}~\bibnamefont
  {Gupta}}, \bibinfo {author} {\bibfnamefont {N.~M.}\ \bibnamefont
  {Schneider}}, \bibinfo {author} {\bibfnamefont {J.~H.}\ \bibnamefont {Park}},
  \bibinfo {author} {\bibfnamefont {D.}~\bibnamefont {Steingart}},\ and\
  \bibinfo {author} {\bibfnamefont {F.~M.}\ \bibnamefont {Ross}},\ }\href
  {https://doi.org/10.1039/c8nr01935e} {\bibfield  {journal} {\bibinfo
  {journal} {Nanoscale}\ }\textbf {\bibinfo {volume} {10}},\ \bibinfo {pages}
  {7702} (\bibinfo {year} {2018})},\ \Eprint {https://arxiv.org/abs/29651479}
  {29651479} \BibitemShut {NoStop}%
\bibitem [{\citenamefont {Abellan}\ \emph {et~al.}(2017)\citenamefont
  {Abellan}, \citenamefont {Moser}, \citenamefont {Lucas}, \citenamefont
  {Grate}, \citenamefont {Evans},\ and\ \citenamefont
  {Browning}}]{Abellan.2017}%
  \BibitemOpen
  \bibfield  {author} {\bibinfo {author} {\bibfnamefont {P.}~\bibnamefont
  {Abellan}}, \bibinfo {author} {\bibfnamefont {T.~H.}\ \bibnamefont {Moser}},
  \bibinfo {author} {\bibfnamefont {I.~T.}\ \bibnamefont {Lucas}}, \bibinfo
  {author} {\bibfnamefont {J.~W.}\ \bibnamefont {Grate}}, \bibinfo {author}
  {\bibfnamefont {J.~E.}\ \bibnamefont {Evans}},\ and\ \bibinfo {author}
  {\bibfnamefont {N.~D.}\ \bibnamefont {Browning}},\ }\href
  {https://doi.org/10.1039/c6ra27066b} {\bibfield  {journal} {\bibinfo
  {journal} {RSC Advances}\ }\textbf {\bibinfo {volume} {7}},\ \bibinfo {pages}
  {3831} (\bibinfo {year} {2017})}\BibitemShut {NoStop}%
\bibitem [{\citenamefont {Su}\ \emph {et~al.}(2019)\citenamefont {Su},
  \citenamefont {Mehdi}, \citenamefont {Patterson}, \citenamefont {Sommerdijk},
  \citenamefont {Browning},\ and\ \citenamefont {Friedrich}}]{Su.2019b}%
  \BibitemOpen
  \bibfield  {author} {\bibinfo {author} {\bibfnamefont {H.}~\bibnamefont
  {Su}}, \bibinfo {author} {\bibfnamefont {B.~L.}\ \bibnamefont {Mehdi}},
  \bibinfo {author} {\bibfnamefont {J.~P.}\ \bibnamefont {Patterson}}, \bibinfo
  {author} {\bibfnamefont {N.~A. J.~M.}\ \bibnamefont {Sommerdijk}}, \bibinfo
  {author} {\bibfnamefont {N.~D.}\ \bibnamefont {Browning}},\ and\ \bibinfo
  {author} {\bibfnamefont {H.}~\bibnamefont {Friedrich}},\ }\href
  {https://doi.org/10.1021/acs.jpcc.9b06078} {\bibfield  {journal} {\bibinfo
  {journal} {The Journal of Physical Chemistry C}\ }\textbf {\bibinfo {volume}
  {123}},\ \bibinfo {pages} {25448} (\bibinfo {year} {2019})}\BibitemShut
  {NoStop}%
\bibitem [{\citenamefont {Hutzler}\ \emph
  {et~al.}(2019{\natexlab{a}})\citenamefont {Hutzler}, \citenamefont {Fritsch},
  \citenamefont {Jank}, \citenamefont {Branscheid}, \citenamefont {Martens},
  \citenamefont {Spiecker},\ and\ \citenamefont {M{\"a}rz}}]{Hutzler.2019}%
  \BibitemOpen
  \bibfield  {author} {\bibinfo {author} {\bibfnamefont {A.}~\bibnamefont
  {Hutzler}}, \bibinfo {author} {\bibfnamefont {B.}~\bibnamefont {Fritsch}},
  \bibinfo {author} {\bibfnamefont {M.~P.~M.}\ \bibnamefont {Jank}}, \bibinfo
  {author} {\bibfnamefont {R.}~\bibnamefont {Branscheid}}, \bibinfo {author}
  {\bibfnamefont {R.~C.}\ \bibnamefont {Martens}}, \bibinfo {author}
  {\bibfnamefont {E.}~\bibnamefont {Spiecker}},\ and\ \bibinfo {author}
  {\bibfnamefont {M.}~\bibnamefont {M{\"a}rz}},\ }\href
  {https://doi.org/10.1002/admi.201901027} {\bibfield  {journal} {\bibinfo
  {journal} {Advanced Materials Interfaces}\ }\textbf {\bibinfo {volume}
  {345}},\ \bibinfo {pages} {1901027} (\bibinfo {year}
  {2019}{\natexlab{a}})}\BibitemShut {NoStop}%
\bibitem [{\citenamefont {Korpanty}\ \emph {et~al.}(2021)\citenamefont
  {Korpanty}, \citenamefont {Parent},\ and\ \citenamefont
  {Gianneschi}}]{Korpanty.2021}%
  \BibitemOpen
  \bibfield  {author} {\bibinfo {author} {\bibfnamefont {J.}~\bibnamefont
  {Korpanty}}, \bibinfo {author} {\bibfnamefont {L.~R.}\ \bibnamefont
  {Parent}},\ and\ \bibinfo {author} {\bibfnamefont {N.~C.}\ \bibnamefont
  {Gianneschi}},\ }\href {https://doi.org/10.1021/acs.nanolett.0c04636}
  {\bibfield  {journal} {\bibinfo  {journal} {Nano Letters}\ ,\ \bibinfo
  {pages} {1141}} (\bibinfo {year} {2021})},\ \Eprint
  {https://arxiv.org/abs/33448858} {33448858} \BibitemShut {NoStop}%
\bibitem [{\citenamefont {Aliyah}\ \emph {et~al.}(2020)\citenamefont {Aliyah},
  \citenamefont {Lyu}, \citenamefont {Goldmann}, \citenamefont {Bizien},
  \citenamefont {Hamon}, \citenamefont {Alloyeau},\ and\ \citenamefont
  {Constantin}}]{Aliyah.2020}%
  \BibitemOpen
  \bibfield  {author} {\bibinfo {author} {\bibfnamefont {K.}~\bibnamefont
  {Aliyah}}, \bibinfo {author} {\bibfnamefont {J.}~\bibnamefont {Lyu}},
  \bibinfo {author} {\bibfnamefont {C.}~\bibnamefont {Goldmann}}, \bibinfo
  {author} {\bibfnamefont {T.}~\bibnamefont {Bizien}}, \bibinfo {author}
  {\bibfnamefont {C.}~\bibnamefont {Hamon}}, \bibinfo {author} {\bibfnamefont
  {D.}~\bibnamefont {Alloyeau}},\ and\ \bibinfo {author} {\bibfnamefont
  {D.}~\bibnamefont {Constantin}},\ }\href
  {https://doi.org/10.1021/acs.jpclett.0c00121} {\bibfield  {journal} {\bibinfo
   {journal} {The Journal of Physical Chemistry letters}\ }\textbf {\bibinfo
  {volume} {11}},\ \bibinfo {pages} {2830} (\bibinfo {year} {2020})},\ \Eprint
  {https://arxiv.org/abs/32200632} {32200632} \BibitemShut {NoStop}%
\bibitem [{\citenamefont {Hermannsd{\"o}rfer}\ \emph
  {et~al.}(2015)\citenamefont {Hermannsd{\"o}rfer}, \citenamefont {{de
  Jonge}},\ and\ \citenamefont {Verch}}]{Hermannsdorfer.2015}%
  \BibitemOpen
  \bibfield  {author} {\bibinfo {author} {\bibfnamefont {J.}~\bibnamefont
  {Hermannsd{\"o}rfer}}, \bibinfo {author} {\bibfnamefont {N.}~\bibnamefont
  {{de Jonge}}},\ and\ \bibinfo {author} {\bibfnamefont {A.}~\bibnamefont
  {Verch}},\ }\href {https://doi.org/10.1039/C5CC06812F} {\bibfield  {journal}
  {\bibinfo  {journal} {Chemical communications (Cambridge, England)}\ }\textbf
  {\bibinfo {volume} {51}},\ \bibinfo {pages} {16393} (\bibinfo {year}
  {2015})},\ \Eprint {https://arxiv.org/abs/26412024} {26412024} \BibitemShut
  {NoStop}%
\bibitem [{\citenamefont {Dong}\ \emph {et~al.}(2021)\citenamefont {Dong},
  \citenamefont {Fu}, \citenamefont {Min}, \citenamefont {Zhang}, \citenamefont
  {Dong}, \citenamefont {Pan}, \citenamefont {Sun}, \citenamefont {Wei},
  \citenamefont {Qin}, \citenamefont {Zhu},\ and\ \citenamefont
  {Xu}}]{Dong.2021}%
  \BibitemOpen
  \bibfield  {author} {\bibinfo {author} {\bibfnamefont {M.}~\bibnamefont
  {Dong}}, \bibinfo {author} {\bibfnamefont {R.}~\bibnamefont {Fu}}, \bibinfo
  {author} {\bibfnamefont {H.}~\bibnamefont {Min}}, \bibinfo {author}
  {\bibfnamefont {Q.}~\bibnamefont {Zhang}}, \bibinfo {author} {\bibfnamefont
  {H.}~\bibnamefont {Dong}}, \bibinfo {author} {\bibfnamefont {Y.}~\bibnamefont
  {Pan}}, \bibinfo {author} {\bibfnamefont {L.}~\bibnamefont {Sun}}, \bibinfo
  {author} {\bibfnamefont {W.}~\bibnamefont {Wei}}, \bibinfo {author}
  {\bibfnamefont {M.}~\bibnamefont {Qin}}, \bibinfo {author} {\bibfnamefont
  {Z.}~\bibnamefont {Zhu}},\ and\ \bibinfo {author} {\bibfnamefont
  {F.}~\bibnamefont {Xu}},\ }\bibfield  {journal} {\bibinfo  {journal} {Crystal
  Growth {\&} Design}\ }\href {https://doi.org/10.1021/acs.cgd.0c01573}
  {10.1021/acs.cgd.0c01573} (\bibinfo {year} {2021})\BibitemShut {NoStop}%
\bibitem [{\citenamefont {Fritsch}\ \emph {et~al.}(2021)\citenamefont
  {Fritsch}, \citenamefont {Hutzler}, \citenamefont {Wu}, \citenamefont
  {Khadivianazar}, \citenamefont {{Vogl. Lilian}}, \citenamefont {Jank},
  \citenamefont {M{\"a}rz},\ and\ \citenamefont {Spiecker}}]{Fritsch.2021}%
  \BibitemOpen
  \bibfield  {author} {\bibinfo {author} {\bibfnamefont {B.}~\bibnamefont
  {Fritsch}}, \bibinfo {author} {\bibfnamefont {A.}~\bibnamefont {Hutzler}},
  \bibinfo {author} {\bibfnamefont {M.}~\bibnamefont {Wu}}, \bibinfo {author}
  {\bibfnamefont {S.}~\bibnamefont {Khadivianazar}}, \bibinfo {author}
  {\bibnamefont {{Vogl. Lilian}}}, \bibinfo {author} {\bibfnamefont {M.~P.~M.}\
  \bibnamefont {Jank}}, \bibinfo {author} {\bibfnamefont {M.}~\bibnamefont
  {M{\"a}rz}},\ and\ \bibinfo {author} {\bibfnamefont {E.}~\bibnamefont
  {Spiecker}},\ }\href {https://doi.org/10.1039/D0NA01027H} {\bibfield
  {journal} {\bibinfo  {journal} {Nanoscale Advances}\ }\textbf {\bibinfo
  {volume} {3}},\ \bibinfo {pages} {2466} (\bibinfo {year} {2021})}\BibitemShut
  {NoStop}%
\bibitem [{\citenamefont {Hutzler}\ \emph {et~al.}(2018)\citenamefont
  {Hutzler}, \citenamefont {Schmutzler}, \citenamefont {Jank}, \citenamefont
  {Branscheid}, \citenamefont {Unruh}, \citenamefont {Spiecker},\ and\
  \citenamefont {Frey}}]{Hutzler.2018}%
  \BibitemOpen
  \bibfield  {author} {\bibinfo {author} {\bibfnamefont {A.}~\bibnamefont
  {Hutzler}}, \bibinfo {author} {\bibfnamefont {T.}~\bibnamefont {Schmutzler}},
  \bibinfo {author} {\bibfnamefont {M.~P.~M.}\ \bibnamefont {Jank}}, \bibinfo
  {author} {\bibfnamefont {R.}~\bibnamefont {Branscheid}}, \bibinfo {author}
  {\bibfnamefont {T.}~\bibnamefont {Unruh}}, \bibinfo {author} {\bibfnamefont
  {E.}~\bibnamefont {Spiecker}},\ and\ \bibinfo {author} {\bibfnamefont
  {L.}~\bibnamefont {Frey}},\ }\href
  {https://doi.org/10.1021/acs.nanolett.8b03388} {\bibfield  {journal}
  {\bibinfo  {journal} {Nano Letters}\ }\textbf {\bibinfo {volume} {18}},\
  \bibinfo {pages} {7222} (\bibinfo {year} {2018})},\ \Eprint
  {https://arxiv.org/abs/30346790} {30346790} \BibitemShut {NoStop}%
\bibitem [{\citenamefont {Bae}\ \emph {et~al.}(2020)\citenamefont {Bae},
  \citenamefont {Lim}, \citenamefont {Kim}, \citenamefont {Kang}, \citenamefont
  {Kim}, \citenamefont {Kim}, \citenamefont {Kang}, \citenamefont {Jeon},
  \citenamefont {Cho}, \citenamefont {Lee}, \citenamefont {Lee},\ and\
  \citenamefont {Park}}]{Bae.2020}%
  \BibitemOpen
  \bibfield  {author} {\bibinfo {author} {\bibfnamefont {Y.}~\bibnamefont
  {Bae}}, \bibinfo {author} {\bibfnamefont {K.}~\bibnamefont {Lim}}, \bibinfo
  {author} {\bibfnamefont {S.}~\bibnamefont {Kim}}, \bibinfo {author}
  {\bibfnamefont {D.}~\bibnamefont {Kang}}, \bibinfo {author} {\bibfnamefont
  {B.~H.}\ \bibnamefont {Kim}}, \bibinfo {author} {\bibfnamefont
  {J.}~\bibnamefont {Kim}}, \bibinfo {author} {\bibfnamefont {S.}~\bibnamefont
  {Kang}}, \bibinfo {author} {\bibfnamefont {S.}~\bibnamefont {Jeon}}, \bibinfo
  {author} {\bibfnamefont {J.}~\bibnamefont {Cho}}, \bibinfo {author}
  {\bibfnamefont {W.~B.}\ \bibnamefont {Lee}}, \bibinfo {author} {\bibfnamefont
  {W.~C.}\ \bibnamefont {Lee}},\ and\ \bibinfo {author} {\bibfnamefont
  {J.}~\bibnamefont {Park}},\ }\bibfield  {journal} {\bibinfo  {journal} {Nano
  Letters}\ }\href {https://doi.org/10.1021/acs.nanolett.0c03517}
  {10.1021/acs.nanolett.0c03517} (\bibinfo {year} {2020}),\ \Eprint
  {https://arxiv.org/abs/33186041} {33186041} \BibitemShut {NoStop}%
\bibitem [{\citenamefont {Crook}\ \emph {et~al.}(2021)\citenamefont {Crook},
  \citenamefont {Laube}, \citenamefont {Moreno-Hernandez}, \citenamefont
  {Kahnt}, \citenamefont {Zahn}, \citenamefont {Ondry}, \citenamefont {Liu},\
  and\ \citenamefont {Alivisatos}}]{Crook.2021}%
  \BibitemOpen
  \bibfield  {author} {\bibinfo {author} {\bibfnamefont {M.~F.}\ \bibnamefont
  {Crook}}, \bibinfo {author} {\bibfnamefont {C.}~\bibnamefont {Laube}},
  \bibinfo {author} {\bibfnamefont {I.~A.}\ \bibnamefont {Moreno-Hernandez}},
  \bibinfo {author} {\bibfnamefont {A.}~\bibnamefont {Kahnt}}, \bibinfo
  {author} {\bibfnamefont {S.}~\bibnamefont {Zahn}}, \bibinfo {author}
  {\bibfnamefont {J.~C.}\ \bibnamefont {Ondry}}, \bibinfo {author}
  {\bibfnamefont {A.}~\bibnamefont {Liu}},\ and\ \bibinfo {author}
  {\bibfnamefont {A.~P.}\ \bibnamefont {Alivisatos}},\ }\bibfield  {journal}
  {\bibinfo  {journal} {Journal of the American Chemical Society}\ }\href
  {https://doi.org/10.1021/jacs.1c05099} {10.1021/jacs.1c05099} (\bibinfo
  {year} {2021}),\ \Eprint {https://arxiv.org/abs/34292703} {34292703}
  \BibitemShut {NoStop}%
\bibitem [{\citenamefont {Dang}\ \emph {et~al.}(2021)\citenamefont {Dang},
  \citenamefont {Manna},\ and\ \citenamefont {Baranov}}]{Dang.2021}%
  \BibitemOpen
  \bibfield  {author} {\bibinfo {author} {\bibfnamefont {Z.}~\bibnamefont
  {Dang}}, \bibinfo {author} {\bibfnamefont {L.}~\bibnamefont {Manna}},\ and\
  \bibinfo {author} {\bibfnamefont {D.}~\bibnamefont {Baranov}},\ }\bibfield
  {journal} {\bibinfo  {journal} {Nanoscale}\ }\href
  {https://doi.org/10.1039/D0NR08584G} {10.1039/D0NR08584G} (\bibinfo {year}
  {2021}),\ \Eprint {https://arxiv.org/abs/33459324} {33459324} \BibitemShut
  {NoStop}%
\bibitem [{\citenamefont {Loh}\ \emph {et~al.}(2017)\citenamefont {Loh},
  \citenamefont {Sen}, \citenamefont {Bosman}, \citenamefont {Tan},
  \citenamefont {Zhong}, \citenamefont {Nijhuis}, \citenamefont {Kr{\'a}l},
  \citenamefont {Matsudaira},\ and\ \citenamefont {Mirsaidov}}]{Loh.2017}%
  \BibitemOpen
  \bibfield  {author} {\bibinfo {author} {\bibfnamefont {N.~D.}\ \bibnamefont
  {Loh}}, \bibinfo {author} {\bibfnamefont {S.}~\bibnamefont {Sen}}, \bibinfo
  {author} {\bibfnamefont {M.}~\bibnamefont {Bosman}}, \bibinfo {author}
  {\bibfnamefont {S.~F.}\ \bibnamefont {Tan}}, \bibinfo {author} {\bibfnamefont
  {J.}~\bibnamefont {Zhong}}, \bibinfo {author} {\bibfnamefont {C.~A.}\
  \bibnamefont {Nijhuis}}, \bibinfo {author} {\bibfnamefont {P.}~\bibnamefont
  {Kr{\'a}l}}, \bibinfo {author} {\bibfnamefont {P.}~\bibnamefont
  {Matsudaira}},\ and\ \bibinfo {author} {\bibfnamefont {U.}~\bibnamefont
  {Mirsaidov}},\ }\href {https://doi.org/10.1038/nchem.2618} {\bibfield
  {journal} {\bibinfo  {journal} {Nature Chemistry}\ }\textbf {\bibinfo
  {volume} {9}},\ \bibinfo {pages} {77} (\bibinfo {year} {2017})}\BibitemShut
  {NoStop}%
\bibitem [{\citenamefont {Woehl}\ \emph {et~al.}(2012)\citenamefont {Woehl},
  \citenamefont {Evans}, \citenamefont {Arslan}, \citenamefont {Ristenpart},\
  and\ \citenamefont {Browning}}]{Woehl.2012}%
  \BibitemOpen
  \bibfield  {author} {\bibinfo {author} {\bibfnamefont {T.~J.}\ \bibnamefont
  {Woehl}}, \bibinfo {author} {\bibfnamefont {J.~E.}\ \bibnamefont {Evans}},
  \bibinfo {author} {\bibfnamefont {I.}~\bibnamefont {Arslan}}, \bibinfo
  {author} {\bibfnamefont {W.~D.}\ \bibnamefont {Ristenpart}},\ and\ \bibinfo
  {author} {\bibfnamefont {N.~D.}\ \bibnamefont {Browning}},\ }\href
  {https://doi.org/10.1021/nn303371y} {\bibfield  {journal} {\bibinfo
  {journal} {ACS nano}\ }\textbf {\bibinfo {volume} {6}},\ \bibinfo {pages}
  {8599} (\bibinfo {year} {2012})},\ \Eprint {https://arxiv.org/abs/22957797}
  {22957797} \BibitemShut {NoStop}%
\bibitem [{\citenamefont {Wang}\ \emph {et~al.}(2018)\citenamefont {Wang},
  \citenamefont {Park},\ and\ \citenamefont {Woehl}}]{Wang.2018}%
  \BibitemOpen
  \bibfield  {author} {\bibinfo {author} {\bibfnamefont {M.}~\bibnamefont
  {Wang}}, \bibinfo {author} {\bibfnamefont {C.}~\bibnamefont {Park}},\ and\
  \bibinfo {author} {\bibfnamefont {T.~J.}\ \bibnamefont {Woehl}},\ }\href
  {https://doi.org/10.1021/acs.chemmater.8b03050} {\bibfield  {journal}
  {\bibinfo  {journal} {Chemistry of Materials}\ }\textbf {\bibinfo {volume}
  {30}},\ \bibinfo {pages} {7727} (\bibinfo {year} {2018})}\BibitemShut
  {NoStop}%
\bibitem [{\citenamefont {Abellan}\ \emph {et~al.}(2014)\citenamefont
  {Abellan}, \citenamefont {Woehl}, \citenamefont {Parent}, \citenamefont
  {Browning}, \citenamefont {Evans},\ and\ \citenamefont
  {Arslan}}]{Abellan.2014}%
  \BibitemOpen
  \bibfield  {author} {\bibinfo {author} {\bibfnamefont {P.}~\bibnamefont
  {Abellan}}, \bibinfo {author} {\bibfnamefont {T.~J.}\ \bibnamefont {Woehl}},
  \bibinfo {author} {\bibfnamefont {L.~R.}\ \bibnamefont {Parent}}, \bibinfo
  {author} {\bibfnamefont {N.~D.}\ \bibnamefont {Browning}}, \bibinfo {author}
  {\bibfnamefont {J.~E.}\ \bibnamefont {Evans}},\ and\ \bibinfo {author}
  {\bibfnamefont {I.}~\bibnamefont {Arslan}},\ }\href
  {https://doi.org/10.1039/c3cc48479c} {\bibfield  {journal} {\bibinfo
  {journal} {Chemical Communications}\ }\textbf {\bibinfo {volume} {50}},\
  \bibinfo {pages} {4873} (\bibinfo {year} {2014})},\ \Eprint
  {https://arxiv.org/abs/24643324} {24643324} \BibitemShut {NoStop}%
\bibitem [{\citenamefont {Dong}\ \emph {et~al.}(2019)\citenamefont {Dong},
  \citenamefont {Wang}, \citenamefont {Wei}, \citenamefont {Hu}, \citenamefont
  {Qin}, \citenamefont {Zhang}, \citenamefont {Sun},\ and\ \citenamefont
  {Xu}}]{Dong.2019}%
  \BibitemOpen
  \bibfield  {author} {\bibinfo {author} {\bibfnamefont {M.}~\bibnamefont
  {Dong}}, \bibinfo {author} {\bibfnamefont {W.}~\bibnamefont {Wang}}, \bibinfo
  {author} {\bibfnamefont {W.}~\bibnamefont {Wei}}, \bibinfo {author}
  {\bibfnamefont {X.}~\bibnamefont {Hu}}, \bibinfo {author} {\bibfnamefont
  {M.}~\bibnamefont {Qin}}, \bibinfo {author} {\bibfnamefont {Q.}~\bibnamefont
  {Zhang}}, \bibinfo {author} {\bibfnamefont {L.}~\bibnamefont {Sun}},\ and\
  \bibinfo {author} {\bibfnamefont {F.}~\bibnamefont {Xu}},\ }\bibfield
  {journal} {\bibinfo  {journal} {The Journal of Physical Chemistry C}\ }\href
  {https://doi.org/10.1021/acs.jpcc.9b05267} {10.1021/acs.jpcc.9b05267}
  (\bibinfo {year} {2019})\BibitemShut {NoStop}%
\bibitem [{\citenamefont {{Le Ca{\"e}r}}(2011)}]{LeCaer.2011}%
  \BibitemOpen
  \bibfield  {author} {\bibinfo {author} {\bibfnamefont {S.}~\bibnamefont {{Le
  Ca{\"e}r}}},\ }\href {https://doi.org/10.3390/w3010235} {\bibfield  {journal}
  {\bibinfo  {journal} {Water}\ }\textbf {\bibinfo {volume} {3}},\ \bibinfo
  {pages} {235} (\bibinfo {year} {2011})},\ \bibinfo {note} {pII:
  w3010235}\BibitemShut {NoStop}%
\bibitem [{\citenamefont {Rumble}(2022)}]{Rumble.2022}%
  \BibitemOpen
  \bibinfo {editor} {\bibfnamefont {J.~R.}\ \bibnamefont {Rumble}},\ ed.,\
  \href@noop {} {\emph {\bibinfo {title} {CRC handbook of chemistry and
  physics}}},\ \bibinfo {edition} {103rd}\ ed.\ (\bibinfo  {publisher} {{CRC
  Press}},\ \bibinfo {year} {2022})\ \bibinfo {note} {rumble, John R.,
  (editor.)}\BibitemShut {NoStop}%
\bibitem [{\citenamefont {Horne}\ \emph {et~al.}(2016)\citenamefont {Horne},
  \citenamefont {Donoclift}, \citenamefont {Sims}, \citenamefont {Orr},\ and\
  \citenamefont {Pimblott}}]{Horne.2016}%
  \BibitemOpen
  \bibfield  {author} {\bibinfo {author} {\bibfnamefont {G.~P.}\ \bibnamefont
  {Horne}}, \bibinfo {author} {\bibfnamefont {T.~A.}\ \bibnamefont
  {Donoclift}}, \bibinfo {author} {\bibfnamefont {H.~E.}\ \bibnamefont {Sims}},
  \bibinfo {author} {\bibfnamefont {R.~M.}\ \bibnamefont {Orr}},\ and\ \bibinfo
  {author} {\bibfnamefont {S.~M.}\ \bibnamefont {Pimblott}},\ }\href
  {https://doi.org/10.1021/acs.jpcb.6b06862} {\bibfield  {journal} {\bibinfo
  {journal} {The Journal of Physical Chemistry. B}\ }\textbf {\bibinfo {volume}
  {120}},\ \bibinfo {pages} {11781} (\bibinfo {year} {2016})},\ \Eprint
  {https://arxiv.org/abs/27779879} {27779879} \BibitemShut {NoStop}%
\bibitem [{\citenamefont {Kelm}\ and\ \citenamefont
  {Bohnert}(2004)}]{Kelm.2004}%
  \BibitemOpen
  \bibfield  {author} {\bibinfo {author} {\bibfnamefont {M.}~\bibnamefont
  {Kelm}}\ and\ \bibinfo {author} {\bibfnamefont {E.}~\bibnamefont {Bohnert}},\
  }\href {https://doi.org/10.5445/IR/270057730} {\bibinfo {title} {A kinetic
  model for the radiolysis of chloride brine, its sensitivity against model
  parameters and a comparison with experiments}} (\bibinfo {year}
  {2004})\BibitemShut {NoStop}%
\bibitem [{\citenamefont {Williams}\ \emph {et~al.}(2002)\citenamefont
  {Williams}, \citenamefont {Dentener},\ and\ \citenamefont {{van den
  Berg}}}]{Williams.2002}%
  \BibitemOpen
  \bibfield  {author} {\bibinfo {author} {\bibfnamefont {J.~E.}\ \bibnamefont
  {Williams}}, \bibinfo {author} {\bibfnamefont {F.~J.}\ \bibnamefont
  {Dentener}},\ and\ \bibinfo {author} {\bibfnamefont {A.~R.}\ \bibnamefont
  {{van den Berg}}},\ }\href {https://doi.org/10.5194/acp-2-39-2002} {\bibfield
   {journal} {\bibinfo  {journal} {Atmospheric Chemistry and Physics}\ }\textbf
  {\bibinfo {volume} {2}},\ \bibinfo {pages} {39} (\bibinfo {year}
  {2002})}\BibitemShut {NoStop}%
\bibitem [{\citenamefont {Armstrong}\ \emph {et~al.}(2015)\citenamefont
  {Armstrong}, \citenamefont {Huie}, \citenamefont {Koppenol}, \citenamefont
  {Lymar}, \citenamefont {Mer{\'e}nyi}, \citenamefont {Neta}, \citenamefont
  {Ruscic}, \citenamefont {Stanbury}, \citenamefont {Steenken},\ and\
  \citenamefont {Wardman}}]{Armstrong.2015}%
  \BibitemOpen
  \bibfield  {author} {\bibinfo {author} {\bibfnamefont {D.~A.}\ \bibnamefont
  {Armstrong}}, \bibinfo {author} {\bibfnamefont {R.~E.}\ \bibnamefont {Huie}},
  \bibinfo {author} {\bibfnamefont {W.~H.}\ \bibnamefont {Koppenol}}, \bibinfo
  {author} {\bibfnamefont {S.~V.}\ \bibnamefont {Lymar}}, \bibinfo {author}
  {\bibfnamefont {G.}~\bibnamefont {Mer{\'e}nyi}}, \bibinfo {author}
  {\bibfnamefont {P.}~\bibnamefont {Neta}}, \bibinfo {author} {\bibfnamefont
  {B.}~\bibnamefont {Ruscic}}, \bibinfo {author} {\bibfnamefont {D.~M.}\
  \bibnamefont {Stanbury}}, \bibinfo {author} {\bibfnamefont {S.}~\bibnamefont
  {Steenken}},\ and\ \bibinfo {author} {\bibfnamefont {P.}~\bibnamefont
  {Wardman}},\ }\href {https://doi.org/10.1515/pac-2014-0502} {\bibfield
  {journal} {\bibinfo  {journal} {Pure and Applied Chemistry}\ }\textbf
  {\bibinfo {volume} {87}},\ \bibinfo {pages} {1139} (\bibinfo {year}
  {2015})}\BibitemShut {NoStop}%
\bibitem [{\citenamefont {Yuk}\ \emph {et~al.}(2012)\citenamefont {Yuk},
  \citenamefont {Park}, \citenamefont {Ercius}, \citenamefont {Kim},
  \citenamefont {Hellebusch}, \citenamefont {Crommie}, \citenamefont {Lee},
  \citenamefont {Zettl},\ and\ \citenamefont {Alivisatos}}]{Yuk.2012}%
  \BibitemOpen
  \bibfield  {author} {\bibinfo {author} {\bibfnamefont {J.~M.}\ \bibnamefont
  {Yuk}}, \bibinfo {author} {\bibfnamefont {J.}~\bibnamefont {Park}}, \bibinfo
  {author} {\bibfnamefont {P.}~\bibnamefont {Ercius}}, \bibinfo {author}
  {\bibfnamefont {K.}~\bibnamefont {Kim}}, \bibinfo {author} {\bibfnamefont
  {D.~J.}\ \bibnamefont {Hellebusch}}, \bibinfo {author} {\bibfnamefont
  {M.~F.}\ \bibnamefont {Crommie}}, \bibinfo {author} {\bibfnamefont {J.~Y.}\
  \bibnamefont {Lee}}, \bibinfo {author} {\bibfnamefont {A.}~\bibnamefont
  {Zettl}},\ and\ \bibinfo {author} {\bibfnamefont {A.~P.}\ \bibnamefont
  {Alivisatos}},\ }\href {https://doi.org/10.1126/science.1217654} {\bibfield
  {journal} {\bibinfo  {journal} {Science}\ }\textbf {\bibinfo {volume}
  {336}},\ \bibinfo {pages} {61} (\bibinfo {year} {2012})},\ \Eprint
  {https://arxiv.org/abs/22491849} {22491849} \BibitemShut {NoStop}%
\bibitem [{\citenamefont {Yin}\ \emph {et~al.}(2019)\citenamefont {Yin},
  \citenamefont {Betzler}, \citenamefont {Sheng}, \citenamefont {Zhang},
  \citenamefont {Peng}, \citenamefont {Shangguan}, \citenamefont {Bustillo},
  \citenamefont {Li}, \citenamefont {Sun},\ and\ \citenamefont
  {Zheng}}]{Yin.2019}%
  \BibitemOpen
  \bibfield  {author} {\bibinfo {author} {\bibfnamefont {Z.-W.}\ \bibnamefont
  {Yin}}, \bibinfo {author} {\bibfnamefont {S.~B.}\ \bibnamefont {Betzler}},
  \bibinfo {author} {\bibfnamefont {T.}~\bibnamefont {Sheng}}, \bibinfo
  {author} {\bibfnamefont {Q.}~\bibnamefont {Zhang}}, \bibinfo {author}
  {\bibfnamefont {X.}~\bibnamefont {Peng}}, \bibinfo {author} {\bibfnamefont
  {J.}~\bibnamefont {Shangguan}}, \bibinfo {author} {\bibfnamefont {K.~C.}\
  \bibnamefont {Bustillo}}, \bibinfo {author} {\bibfnamefont {J.-T.}\
  \bibnamefont {Li}}, \bibinfo {author} {\bibfnamefont {S.-G.}\ \bibnamefont
  {Sun}},\ and\ \bibinfo {author} {\bibfnamefont {H.}~\bibnamefont {Zheng}},\
  }\href {https://doi.org/10.1016/j.nanoen.2019.05.068} {\bibfield  {journal}
  {\bibinfo  {journal} {Nano Energy}\ }\textbf {\bibinfo {volume} {62}},\
  \bibinfo {pages} {507} (\bibinfo {year} {2019})}\BibitemShut {NoStop}%
\bibitem [{\citenamefont {Hutzler}\ \emph
  {et~al.}(2019{\natexlab{b}})\citenamefont {Hutzler}, \citenamefont {Fritsch},
  \citenamefont {Jank}, \citenamefont {Branscheid}, \citenamefont {Spiecker},\
  and\ \citenamefont {M{\"a}rz}}]{Hutzler.2019b}%
  \BibitemOpen
  \bibfield  {author} {\bibinfo {author} {\bibfnamefont {A.}~\bibnamefont
  {Hutzler}}, \bibinfo {author} {\bibfnamefont {B.}~\bibnamefont {Fritsch}},
  \bibinfo {author} {\bibfnamefont {M.~P.~M.}\ \bibnamefont {Jank}}, \bibinfo
  {author} {\bibfnamefont {R.}~\bibnamefont {Branscheid}}, \bibinfo {author}
  {\bibfnamefont {E.}~\bibnamefont {Spiecker}},\ and\ \bibinfo {author}
  {\bibfnamefont {M.}~\bibnamefont {M{\"a}rz}},\ }\href
  {https://doi.org/10.3791/59751} {\bibfield  {journal} {\bibinfo  {journal}
  {Journal of visualized experiments: JoVE}\ ,\ \bibinfo {pages} {e59751}}
  (\bibinfo {year} {2019}{\natexlab{b}})}\BibitemShut {NoStop}%
\bibitem [{\citenamefont {Lim}\ \emph {et~al.}(2020)\citenamefont {Lim},
  \citenamefont {Bae}, \citenamefont {Jeon}, \citenamefont {Kim}, \citenamefont
  {Kim}, \citenamefont {Kim}, \citenamefont {Kang}, \citenamefont {Heo},
  \citenamefont {Park},\ and\ \citenamefont {Lee}}]{Lim.2020}%
  \BibitemOpen
  \bibfield  {author} {\bibinfo {author} {\bibfnamefont {K.}~\bibnamefont
  {Lim}}, \bibinfo {author} {\bibfnamefont {Y.}~\bibnamefont {Bae}}, \bibinfo
  {author} {\bibfnamefont {S.}~\bibnamefont {Jeon}}, \bibinfo {author}
  {\bibfnamefont {K.}~\bibnamefont {Kim}}, \bibinfo {author} {\bibfnamefont
  {B.~H.}\ \bibnamefont {Kim}}, \bibinfo {author} {\bibfnamefont
  {J.}~\bibnamefont {Kim}}, \bibinfo {author} {\bibfnamefont {S.}~\bibnamefont
  {Kang}}, \bibinfo {author} {\bibfnamefont {T.}~\bibnamefont {Heo}}, \bibinfo
  {author} {\bibfnamefont {J.}~\bibnamefont {Park}},\ and\ \bibinfo {author}
  {\bibfnamefont {W.~C.}\ \bibnamefont {Lee}},\ }\href
  {https://doi.org/10.1002/adma.202002889} {\bibfield  {journal} {\bibinfo
  {journal} {Advanced materials (Deerfield Beach, Fla.)}\ ,\ \bibinfo {pages}
  {e2002889}} (\bibinfo {year} {2020})},\ \Eprint
  {https://arxiv.org/abs/32844520} {32844520} \BibitemShut {NoStop}%
\bibitem [{\citenamefont {Kelly}\ \emph {et~al.}(2018)\citenamefont {Kelly},
  \citenamefont {Zhou}, \citenamefont {Clark}, \citenamefont {Hamer},
  \citenamefont {Lewis}, \citenamefont {Rakowski}, \citenamefont {Haigh},\ and\
  \citenamefont {Gorbachev}}]{Kelly.2018}%
  \BibitemOpen
  \bibfield  {author} {\bibinfo {author} {\bibfnamefont {D.~J.}\ \bibnamefont
  {Kelly}}, \bibinfo {author} {\bibfnamefont {M.}~\bibnamefont {Zhou}},
  \bibinfo {author} {\bibfnamefont {N.}~\bibnamefont {Clark}}, \bibinfo
  {author} {\bibfnamefont {M.~J.}\ \bibnamefont {Hamer}}, \bibinfo {author}
  {\bibfnamefont {E.~A.}\ \bibnamefont {Lewis}}, \bibinfo {author}
  {\bibfnamefont {A.~M.}\ \bibnamefont {Rakowski}}, \bibinfo {author}
  {\bibfnamefont {S.~J.}\ \bibnamefont {Haigh}},\ and\ \bibinfo {author}
  {\bibfnamefont {R.~V.}\ \bibnamefont {Gorbachev}},\ }\href
  {https://doi.org/10.1021/acs.nanolett.7b04713} {\bibfield  {journal}
  {\bibinfo  {journal} {Nano Letters}\ }\textbf {\bibinfo {volume} {18}},\
  \bibinfo {pages} {1168} (\bibinfo {year} {2018})},\ \Eprint
  {https://arxiv.org/abs/29323499} {29323499} \BibitemShut {NoStop}%
\bibitem [{\citenamefont {Liu}\ \emph {et~al.}(2021)\citenamefont {Liu},
  \citenamefont {Cao}, \citenamefont {He}, \citenamefont {Zhang}, \citenamefont
  {Ge},\ and\ \citenamefont {Chen}}]{Liu.2021}%
  \BibitemOpen
  \bibfield  {author} {\bibinfo {author} {\bibfnamefont {Z.}~\bibnamefont
  {Liu}}, \bibinfo {author} {\bibfnamefont {Z.}~\bibnamefont {Cao}}, \bibinfo
  {author} {\bibfnamefont {J.}~\bibnamefont {He}}, \bibinfo {author}
  {\bibfnamefont {H.}~\bibnamefont {Zhang}}, \bibinfo {author} {\bibfnamefont
  {Y.}~\bibnamefont {Ge}},\ and\ \bibinfo {author} {\bibfnamefont
  {B.}~\bibnamefont {Chen}},\ }\href
  {https://doi.org/10.1021/acs.nanolett.1c01901} {\bibfield  {journal}
  {\bibinfo  {journal} {Nano Letters}\ }\textbf {\bibinfo {volume} {21}},\
  \bibinfo {pages} {6882} (\bibinfo {year} {2021})},\ \Eprint
  {https://arxiv.org/abs/34387492} {34387492} \BibitemShut {NoStop}%
\bibitem [{\citenamefont {Yang}\ \emph {et~al.}(2019)\citenamefont {Yang},
  \citenamefont {Choi}, \citenamefont {Sheng}, \citenamefont {Jung},
  \citenamefont {Bustillo}, \citenamefont {Chen}, \citenamefont {Lee},
  \citenamefont {Ercius}, \citenamefont {Kim}, \citenamefont {Warner},
  \citenamefont {Chan},\ and\ \citenamefont {Zheng}}]{Yang.2019d}%
  \BibitemOpen
  \bibfield  {author} {\bibinfo {author} {\bibfnamefont {J.}~\bibnamefont
  {Yang}}, \bibinfo {author} {\bibfnamefont {M.~K.}\ \bibnamefont {Choi}},
  \bibinfo {author} {\bibfnamefont {Y.}~\bibnamefont {Sheng}}, \bibinfo
  {author} {\bibfnamefont {J.}~\bibnamefont {Jung}}, \bibinfo {author}
  {\bibfnamefont {K.~C.}\ \bibnamefont {Bustillo}}, \bibinfo {author}
  {\bibfnamefont {T.}~\bibnamefont {Chen}}, \bibinfo {author} {\bibfnamefont
  {S.-W.}\ \bibnamefont {Lee}}, \bibinfo {author} {\bibfnamefont
  {P.}~\bibnamefont {Ercius}}, \bibinfo {author} {\bibfnamefont {J.~H.}\
  \bibnamefont {Kim}}, \bibinfo {author} {\bibfnamefont {J.~H.}\ \bibnamefont
  {Warner}}, \bibinfo {author} {\bibfnamefont {E.~M.}\ \bibnamefont {Chan}},\
  and\ \bibinfo {author} {\bibfnamefont {H.}~\bibnamefont {Zheng}},\ }\bibfield
   {journal} {\bibinfo  {journal} {Nano Letters}\ }\href
  {https://doi.org/10.1021/acs.nanolett.8b04821} {10.1021/acs.nanolett.8b04821}
  (\bibinfo {year} {2019}),\ \Eprint {https://arxiv.org/abs/30741548}
  {30741548} \BibitemShut {NoStop}%
\bibitem [{\citenamefont {Shingledecker}\ and\ \citenamefont
  {Herbst}(2018)}]{Shingledecker.2018}%
  \BibitemOpen
  \bibfield  {author} {\bibinfo {author} {\bibfnamefont {C.~N.}\ \bibnamefont
  {Shingledecker}}\ and\ \bibinfo {author} {\bibfnamefont {E.}~\bibnamefont
  {Herbst}},\ }\href {https://doi.org/10.1039/C7CP05901A} {\bibfield  {journal}
  {\bibinfo  {journal} {Phys. Chem. Chem. Phys.}\ }\textbf {\bibinfo {volume}
  {20}},\ \bibinfo {pages} {5359} (\bibinfo {year} {2018})}\BibitemShut
  {NoStop}%
\bibitem [{\citenamefont {Arumainayagam}\ \emph {et~al.}(2019)\citenamefont
  {Arumainayagam}, \citenamefont {Garrod}, \citenamefont {Boyer}, \citenamefont
  {Hay}, \citenamefont {Bao}, \citenamefont {Campbell}, \citenamefont {Wang},
  \citenamefont {Nowak}, \citenamefont {Arumainayagam},\ and\ \citenamefont
  {Hodge}}]{Arumainayagam.2019}%
  \BibitemOpen
  \bibfield  {author} {\bibinfo {author} {\bibfnamefont {C.~R.}\ \bibnamefont
  {Arumainayagam}}, \bibinfo {author} {\bibfnamefont {R.~T.}\ \bibnamefont
  {Garrod}}, \bibinfo {author} {\bibfnamefont {M.~C.}\ \bibnamefont {Boyer}},
  \bibinfo {author} {\bibfnamefont {A.~K.}\ \bibnamefont {Hay}}, \bibinfo
  {author} {\bibfnamefont {S.~T.}\ \bibnamefont {Bao}}, \bibinfo {author}
  {\bibfnamefont {J.~S.}\ \bibnamefont {Campbell}}, \bibinfo {author}
  {\bibfnamefont {J.}~\bibnamefont {Wang}}, \bibinfo {author} {\bibfnamefont
  {C.~M.}\ \bibnamefont {Nowak}}, \bibinfo {author} {\bibfnamefont {M.~R.}\
  \bibnamefont {Arumainayagam}},\ and\ \bibinfo {author} {\bibfnamefont
  {P.~J.}\ \bibnamefont {Hodge}},\ }\href {https://doi.org/10.1039/C7CS00443E}
  {\bibfield  {journal} {\bibinfo  {journal} {Chem. Soc. Rev.}\ }\textbf
  {\bibinfo {volume} {48}},\ \bibinfo {pages} {2293} (\bibinfo {year}
  {2019})}\BibitemShut {NoStop}%
\bibitem [{\citenamefont {Hill}\ and\ \citenamefont {Smith}(1994)}]{Hill.1994}%
  \BibitemOpen
  \bibfield  {author} {\bibinfo {author} {\bibfnamefont {M.~A.}\ \bibnamefont
  {Hill}}\ and\ \bibinfo {author} {\bibfnamefont {F.~A.}\ \bibnamefont
  {Smith}},\ }\href {https://doi.org/10.1016/0969-806X(94)90190-2} {\bibfield
  {journal} {\bibinfo  {journal} {Radiation Physics and Chemistry}\ }\textbf
  {\bibinfo {volume} {43}},\ \bibinfo {pages} {265} (\bibinfo {year} {1994})},\
  \bibinfo {note} {pII: 0969806X94901902}\BibitemShut {NoStop}%
\bibitem [{\citenamefont {Pastina}\ and\ \citenamefont
  {LaVerne}(2001)}]{Pastina.2001}%
  \BibitemOpen
  \bibfield  {author} {\bibinfo {author} {\bibfnamefont {B.}~\bibnamefont
  {Pastina}}\ and\ \bibinfo {author} {\bibfnamefont {J.~A.}\ \bibnamefont
  {LaVerne}},\ }\href {https://doi.org/10.1021/jp012245j} {\bibfield  {journal}
  {\bibinfo  {journal} {The Journal of Physical Chemistry A}\ }\textbf
  {\bibinfo {volume} {105}},\ \bibinfo {pages} {9316} (\bibinfo {year}
  {2001})}\BibitemShut {NoStop}%
\bibitem [{\citenamefont {Holmes}\ \emph {et~al.}(2021)\citenamefont {Holmes},
  \citenamefont {Rothman},\ and\ \citenamefont {Zimmerman}}]{Holmes.2021}%
  \BibitemOpen
  \bibfield  {author} {\bibinfo {author} {\bibfnamefont {T.~D.}\ \bibnamefont
  {Holmes}}, \bibinfo {author} {\bibfnamefont {R.~H.}\ \bibnamefont
  {Rothman}},\ and\ \bibinfo {author} {\bibfnamefont {W.~B.}\ \bibnamefont
  {Zimmerman}},\ }\href {https://doi.org/10.1007/s11090-021-10152-z} {\bibfield
   {journal} {\bibinfo  {journal} {Plasma Chemistry and Plasma Processing}\
  }\textbf {\bibinfo {volume} {41}},\ \bibinfo {pages} {531} (\bibinfo {year}
  {2021})},\ \bibinfo {note} {pII: 10152}\BibitemShut {NoStop}%
\bibitem [{\citenamefont {{El Omar}}\ \emph {et~al.}(2013)\citenamefont {{El
  Omar}}, \citenamefont {Schmidhammer}, \citenamefont {Balcerzyk},
  \citenamefont {LaVerne},\ and\ \citenamefont {Mostafavi}}]{ElOmar.2013}%
  \BibitemOpen
  \bibfield  {author} {\bibinfo {author} {\bibfnamefont {A.~K.}\ \bibnamefont
  {{El Omar}}}, \bibinfo {author} {\bibfnamefont {U.}~\bibnamefont
  {Schmidhammer}}, \bibinfo {author} {\bibfnamefont {A.}~\bibnamefont
  {Balcerzyk}}, \bibinfo {author} {\bibfnamefont {J.}~\bibnamefont {LaVerne}},\
  and\ \bibinfo {author} {\bibfnamefont {M.}~\bibnamefont {Mostafavi}},\ }\href
  {https://doi.org/10.1021/jp312023r} {\bibfield  {journal} {\bibinfo
  {journal} {The Journal of Physical Chemistry A}\ }\textbf {\bibinfo {volume}
  {117}},\ \bibinfo {pages} {2287} (\bibinfo {year} {2013})},\ \Eprint
  {https://arxiv.org/abs/23441977} {23441977} \BibitemShut {NoStop}%
\bibitem [{\citenamefont {Yang}\ and\ \citenamefont
  {Pignatello}(2017)}]{Yang.2017}%
  \BibitemOpen
  \bibfield  {author} {\bibinfo {author} {\bibfnamefont {Y.}~\bibnamefont
  {Yang}}\ and\ \bibinfo {author} {\bibfnamefont {J.~J.}\ \bibnamefont
  {Pignatello}},\ }\bibfield  {journal} {\bibinfo  {journal} {Molecules (Basel,
  Switzerland)}\ }\textbf {\bibinfo {volume} {22}},\ \href
  {https://doi.org/10.3390/molecules22101684} {10.3390/molecules22101684}
  (\bibinfo {year} {2017}),\ \Eprint {https://arxiv.org/abs/29027977}
  {29027977} \BibitemShut {NoStop}%
\bibitem [{\citenamefont {Schwarz}\ and\ \citenamefont
  {Gill}(1977)}]{Schwarz.1977}%
  \BibitemOpen
  \bibfield  {author} {\bibinfo {author} {\bibfnamefont {H.~A.}\ \bibnamefont
  {Schwarz}}\ and\ \bibinfo {author} {\bibfnamefont {P.~S.}\ \bibnamefont
  {Gill}},\ }\href {https://doi.org/10.1021/j100516a006} {\bibfield  {journal}
  {\bibinfo  {journal} {The Journal of Physical Chemistry}\ }\textbf {\bibinfo
  {volume} {81}},\ \bibinfo {pages} {22} (\bibinfo {year} {1977})}\BibitemShut
  {NoStop}%
\bibitem [{\citenamefont {Huie}(2003)}]{Huie.2003}%
  \BibitemOpen
  \bibfield  {author} {\bibinfo {author} {\bibfnamefont {R.~E.}\ \bibnamefont
  {Huie}},\ }\href {https://kinetics.nist.gov/solution/} {\bibinfo {title}
  {Ndrl/nist solution kinetics database on the web}} (\bibinfo {year}
  {2003})\BibitemShut {NoStop}%
\bibitem [{\citenamefont {Haag}\ and\ \citenamefont
  {Hoigne}(1983)}]{Haag.1983}%
  \BibitemOpen
  \bibfield  {author} {\bibinfo {author} {\bibfnamefont {W.~R.}\ \bibnamefont
  {Haag}}\ and\ \bibinfo {author} {\bibfnamefont {J.}~\bibnamefont {Hoigne}},\
  }\href {https://doi.org/10.1021/es00111a004} {\bibfield  {journal} {\bibinfo
  {journal} {Environmental Science {\&} Technology}\ }\textbf {\bibinfo
  {volume} {17}},\ \bibinfo {pages} {261} (\bibinfo {year} {1983})}\BibitemShut
  {NoStop}%
\bibitem [{\citenamefont {Kl{\"a}ning}\ and\ \citenamefont
  {Wolff}(1985)}]{Klaning.1985b}%
  \BibitemOpen
  \bibfield  {author} {\bibinfo {author} {\bibfnamefont {U.~K.}\ \bibnamefont
  {Kl{\"a}ning}}\ and\ \bibinfo {author} {\bibfnamefont {T.}~\bibnamefont
  {Wolff}},\ }\href {https://doi.org/10.1002/bbpc.19850890309} {\bibfield
  {journal} {\bibinfo  {journal} {Berichte der Bunsengesellschaft f{\"u}r
  physikalische Chemie}\ }\textbf {\bibinfo {volume} {89}},\ \bibinfo {pages}
  {243} (\bibinfo {year} {1985})}\BibitemShut {NoStop}%
\bibitem [{\citenamefont {Buxton}\ and\ \citenamefont
  {Dainton}(1968)}]{Buxton.1968}%
  \BibitemOpen
  \bibfield  {author} {\bibinfo {author} {\bibfnamefont {G.~V.}\ \bibnamefont
  {Buxton}}\ and\ \bibinfo {author} {\bibfnamefont {F.~S.}\ \bibnamefont
  {Dainton}},\ }\href {http://www.jstor.org/stable/2416051} {\bibfield
  {journal} {\bibinfo  {journal} {Proceedings of the Royal Society of London.
  Series A, Mathematical and Physical Sciences}\ ,\ \bibinfo {pages} {427}}
  (\bibinfo {year} {1968})}\BibitemShut {NoStop}%
\bibitem [{\citenamefont {Buxton}\ \emph {et~al.}(1988)\citenamefont {Buxton},
  \citenamefont {Greenstock}, \citenamefont {Helman},\ and\ \citenamefont
  {Ross}}]{Buxton.1988}%
  \BibitemOpen
  \bibfield  {author} {\bibinfo {author} {\bibfnamefont {G.~V.}\ \bibnamefont
  {Buxton}}, \bibinfo {author} {\bibfnamefont {C.~L.}\ \bibnamefont
  {Greenstock}}, \bibinfo {author} {\bibfnamefont {W.~P.}\ \bibnamefont
  {Helman}},\ and\ \bibinfo {author} {\bibfnamefont {A.~B.}\ \bibnamefont
  {Ross}},\ }\href {https://doi.org/10.1063/1.555805} {\bibfield  {journal}
  {\bibinfo  {journal} {Journal of Physical and Chemical Reference Data}\
  }\textbf {\bibinfo {volume} {17}},\ \bibinfo {pages} {513} (\bibinfo {year}
  {1988})}\BibitemShut {NoStop}%
\bibitem [{\citenamefont {Mezyk}\ and\ \citenamefont
  {Bartels}(1997)}]{Mezyk.1997}%
  \BibitemOpen
  \bibfield  {author} {\bibinfo {author} {\bibfnamefont {S.~P.}\ \bibnamefont
  {Mezyk}}\ and\ \bibinfo {author} {\bibfnamefont {D.~M.}\ \bibnamefont
  {Bartels}},\ }\href {https://doi.org/10.1021/jp970934i} {\bibfield  {journal}
  {\bibinfo  {journal} {The Journal of Physical Chemistry A}\ }\textbf
  {\bibinfo {volume} {101}},\ \bibinfo {pages} {6233} (\bibinfo {year}
  {1997})}\BibitemShut {NoStop}%
\bibitem [{\citenamefont {McKenzie}\ \emph {et~al.}(2016)\citenamefont
  {McKenzie}, \citenamefont {MacDonald-Taylor}, \citenamefont {McLachlan},
  \citenamefont {Orr},\ and\ \citenamefont {Woodhead}}]{McKenzie.2016}%
  \BibitemOpen
  \bibfield  {author} {\bibinfo {author} {\bibfnamefont {H.}~\bibnamefont
  {McKenzie}}, \bibinfo {author} {\bibfnamefont {J.}~\bibnamefont
  {MacDonald-Taylor}}, \bibinfo {author} {\bibfnamefont {F.}~\bibnamefont
  {McLachlan}}, \bibinfo {author} {\bibfnamefont {R.}~\bibnamefont {Orr}},\
  and\ \bibinfo {author} {\bibfnamefont {D.}~\bibnamefont {Woodhead}},\ }\href
  {https://doi.org/10.1016/j.proche.2016.10.067} {\bibfield  {journal}
  {\bibinfo  {journal} {Procedia Chemistry}\ }\textbf {\bibinfo {volume}
  {21}},\ \bibinfo {pages} {481} (\bibinfo {year} {2016})},\ \bibinfo {note}
  {pII: S1876619616301097}\BibitemShut {NoStop}%
\bibitem [{\citenamefont {Mikhailova}\ and\ \citenamefont
  {Ershov}(1993)}]{Mikhailova.1993}%
  \BibitemOpen
  \bibfield  {author} {\bibinfo {author} {\bibfnamefont {T.~L.}\ \bibnamefont
  {Mikhailova}}\ and\ \bibinfo {author} {\bibfnamefont {B.~G.}\ \bibnamefont
  {Ershov}},\ }\href@noop {} {\bibfield  {journal} {\bibinfo  {journal}
  {Russian Chemical Bulletin}\ }\textbf {\bibinfo {volume} {42}},\ \bibinfo
  {pages} {235} (\bibinfo {year} {1993})}\BibitemShut {NoStop}%
\bibitem [{\citenamefont {Halpern}\ and\ \citenamefont
  {Rabani}(1966)}]{Halpern.1966}%
  \BibitemOpen
  \bibfield  {author} {\bibinfo {author} {\bibfnamefont {J.}~\bibnamefont
  {Halpern}}\ and\ \bibinfo {author} {\bibfnamefont {J.}~\bibnamefont
  {Rabani}},\ }\href {https://doi.org/10.1021/ja00956a015} {\bibfield
  {journal} {\bibinfo  {journal} {Journal of the American Chemical Society}\
  }\textbf {\bibinfo {volume} {88}},\ \bibinfo {pages} {699} (\bibinfo {year}
  {1966})}\BibitemShut {NoStop}%
\bibitem [{\citenamefont {Leriche}(2003)}]{Leriche.2003}%
  \BibitemOpen
  \bibfield  {author} {\bibinfo {author} {\bibfnamefont {M.}~\bibnamefont
  {Leriche}},\ }\bibfield  {journal} {\bibinfo  {journal} {Journal of
  Geophysical Research}\ }\textbf {\bibinfo {volume} {108}},\ \href
  {https://doi.org/10.1029/2002JD002950} {10.1029/2002JD002950} (\bibinfo
  {year} {2003})\BibitemShut {NoStop}%
\bibitem [{\citenamefont {Herrmann}\ \emph {et~al.}(2000)\citenamefont
  {Herrmann}, \citenamefont {Ervens}, \citenamefont {Jacobi}, \citenamefont
  {Wolke}, \citenamefont {Nowacki},\ and\ \citenamefont
  {Zellner}}]{Herrmann.2000}%
  \BibitemOpen
  \bibfield  {author} {\bibinfo {author} {\bibfnamefont {H.}~\bibnamefont
  {Herrmann}}, \bibinfo {author} {\bibfnamefont {B.}~\bibnamefont {Ervens}},
  \bibinfo {author} {\bibfnamefont {H.-W.}\ \bibnamefont {Jacobi}}, \bibinfo
  {author} {\bibfnamefont {R.}~\bibnamefont {Wolke}}, \bibinfo {author}
  {\bibfnamefont {P.}~\bibnamefont {Nowacki}},\ and\ \bibinfo {author}
  {\bibfnamefont {R.}~\bibnamefont {Zellner}},\ }\href
  {https://doi.org/10.1023/A:1006318622743} {\bibfield  {journal} {\bibinfo
  {journal} {Journal of Atmospheric Chemistry}\ }\textbf {\bibinfo {volume}
  {36}},\ \bibinfo {pages} {231} (\bibinfo {year} {2000})},\ \bibinfo {note}
  {pII: 244887}\BibitemShut {NoStop}%
\bibitem [{\citenamefont {Hoign{\'e}}\ \emph {et~al.}(1985)\citenamefont
  {Hoign{\'e}}, \citenamefont {Bader}, \citenamefont {Haag},\ and\
  \citenamefont {Staehelin}}]{Hoigne.1985}%
  \BibitemOpen
  \bibfield  {author} {\bibinfo {author} {\bibfnamefont {J.}~\bibnamefont
  {Hoign{\'e}}}, \bibinfo {author} {\bibfnamefont {H.}~\bibnamefont {Bader}},
  \bibinfo {author} {\bibfnamefont {W.}~\bibnamefont {Haag}},\ and\ \bibinfo
  {author} {\bibfnamefont {J.}~\bibnamefont {Staehelin}},\ }\href
  {https://doi.org/10.1016/0043-1354(85)90368-9} {\bibfield  {journal}
  {\bibinfo  {journal} {Water Research}\ }\textbf {\bibinfo {volume} {19}},\
  \bibinfo {pages} {993} (\bibinfo {year} {1985})},\ \bibinfo {note} {pII:
  0043135485903689}\BibitemShut {NoStop}%
\bibitem [{\citenamefont {Sehested}\ \emph {et~al.}(1992)\citenamefont
  {Sehested}, \citenamefont {Corfitzen}, \citenamefont {Holcman},\ and\
  \citenamefont {Hart}}]{Sehested.1992}%
  \BibitemOpen
  \bibfield  {author} {\bibinfo {author} {\bibfnamefont {K.}~\bibnamefont
  {Sehested}}, \bibinfo {author} {\bibfnamefont {H.}~\bibnamefont {Corfitzen}},
  \bibinfo {author} {\bibfnamefont {J.}~\bibnamefont {Holcman}},\ and\ \bibinfo
  {author} {\bibfnamefont {E.~J.}\ \bibnamefont {Hart}},\ }\href
  {https://doi.org/10.1021/j100181a084} {\bibfield  {journal} {\bibinfo
  {journal} {The Journal of Physical Chemistry C}\ }\textbf {\bibinfo {volume}
  {96}},\ \bibinfo {pages} {1005} (\bibinfo {year} {1992})}\BibitemShut
  {NoStop}%
\bibitem [{\citenamefont {Levanov}\ and\ \citenamefont
  {Isaikina}(2020)}]{Levanov.2020}%
  \BibitemOpen
  \bibfield  {author} {\bibinfo {author} {\bibfnamefont {A.~V.}\ \bibnamefont
  {Levanov}}\ and\ \bibinfo {author} {\bibfnamefont {O.~Y.}\ \bibnamefont
  {Isaikina}},\ }\href {https://doi.org/10.1021/acs.iecr.0c02770} {\bibfield
  {journal} {\bibinfo  {journal} {Industrial {\&} Engineering Chemistry
  Research}\ }\textbf {\bibinfo {volume} {59}},\ \bibinfo {pages} {14278}
  (\bibinfo {year} {2020})}\BibitemShut {NoStop}%
\bibitem [{\citenamefont {Kl{\"a}ning}\ \emph {et~al.}(1984)\citenamefont
  {Kl{\"a}ning}, \citenamefont {Sehested},\ and\ \citenamefont
  {Wolff}}]{Klaning.1984}%
  \BibitemOpen
  \bibfield  {author} {\bibinfo {author} {\bibfnamefont {U.~K.}\ \bibnamefont
  {Kl{\"a}ning}}, \bibinfo {author} {\bibfnamefont {K.}~\bibnamefont
  {Sehested}},\ and\ \bibinfo {author} {\bibfnamefont {T.}~\bibnamefont
  {Wolff}},\ }\href {https://doi.org/10.1039/f19848002969} {\bibfield
  {journal} {\bibinfo  {journal} {Journal of the Chemical Society, Faraday
  Transactions 1: Physical Chemistry in Condensed Phases}\ }\textbf {\bibinfo
  {volume} {80}},\ \bibinfo {pages} {2969} (\bibinfo {year}
  {1984})}\BibitemShut {NoStop}%
\bibitem [{\citenamefont {Sauer}\ \emph {et~al.}(1984)\citenamefont {Sauer},
  \citenamefont {Brown},\ and\ \citenamefont {Hart}}]{Sauer.1984}%
  \BibitemOpen
  \bibfield  {author} {\bibinfo {author} {\bibfnamefont {M.~C.}\ \bibnamefont
  {Sauer}}, \bibinfo {author} {\bibfnamefont {W.~G.}\ \bibnamefont {Brown}},\
  and\ \bibinfo {author} {\bibfnamefont {E.~J.}\ \bibnamefont {Hart}},\ }\href
  {https://doi.org/10.1021/j150651a033} {\bibfield  {journal} {\bibinfo
  {journal} {The Journal of Physical Chemistry}\ }\textbf {\bibinfo {volume}
  {88}},\ \bibinfo {pages} {1398} (\bibinfo {year} {1984})}\BibitemShut
  {NoStop}%
\bibitem [{\citenamefont {Sunder}\ and\ \citenamefont
  {Christensen}(1993)}]{Sunder.1993}%
  \BibitemOpen
  \bibfield  {author} {\bibinfo {author} {\bibfnamefont {S.}~\bibnamefont
  {Sunder}}\ and\ \bibinfo {author} {\bibfnamefont {H.}~\bibnamefont
  {Christensen}},\ }\href {https://doi.org/10.13182/NT93-A34900} {\bibfield
  {journal} {\bibinfo  {journal} {Nuclear Technology}\ }\textbf {\bibinfo
  {volume} {104}},\ \bibinfo {pages} {403} (\bibinfo {year}
  {1993})}\BibitemShut {NoStop}%
\bibitem [{\citenamefont {Trummal}\ \emph {et~al.}(2016)\citenamefont
  {Trummal}, \citenamefont {Lipping}, \citenamefont {Kaljurand}, \citenamefont
  {Koppel},\ and\ \citenamefont {Leito}}]{Trummal.2016}%
  \BibitemOpen
  \bibfield  {author} {\bibinfo {author} {\bibfnamefont {A.}~\bibnamefont
  {Trummal}}, \bibinfo {author} {\bibfnamefont {L.}~\bibnamefont {Lipping}},
  \bibinfo {author} {\bibfnamefont {I.}~\bibnamefont {Kaljurand}}, \bibinfo
  {author} {\bibfnamefont {I.~A.}\ \bibnamefont {Koppel}},\ and\ \bibinfo
  {author} {\bibfnamefont {I.}~\bibnamefont {Leito}},\ }\href
  {https://doi.org/10.1021/acs.jpca.6b02253} {\bibfield  {journal} {\bibinfo
  {journal} {The Journal of Physical Chemistry A}\ }\textbf {\bibinfo {volume}
  {120}},\ \bibinfo {pages} {3663} (\bibinfo {year} {2016})},\ \Eprint
  {https://arxiv.org/abs/27115918} {27115918} \BibitemShut {NoStop}%
\bibitem [{\citenamefont {Wu}\ \emph {et~al.}(1980)\citenamefont {Wu},
  \citenamefont {Wong},\ and\ \citenamefont {{Di Bartolo}}}]{Wu.1980}%
  \BibitemOpen
  \bibfield  {author} {\bibinfo {author} {\bibfnamefont {D.}~\bibnamefont
  {Wu}}, \bibinfo {author} {\bibfnamefont {D.}~\bibnamefont {Wong}},\ and\
  \bibinfo {author} {\bibfnamefont {B.}~\bibnamefont {{Di Bartolo}}},\ }\href
  {https://doi.org/10.1016/0047-2670(80)85102-1} {\bibfield  {journal}
  {\bibinfo  {journal} {Journal of Photochemistry}\ }\textbf {\bibinfo {volume}
  {14}},\ \bibinfo {pages} {303} (\bibinfo {year} {1980})},\ \bibinfo {note}
  {pII: 0047267080851021}\BibitemShut {NoStop}%
\bibitem [{\citenamefont {Kl{\"a}ning}\ \emph {et~al.}(1985)\citenamefont
  {Kl{\"a}ning}, \citenamefont {Sehested},\ and\ \citenamefont
  {Holcman}}]{Klaning.1985}%
  \BibitemOpen
  \bibfield  {author} {\bibinfo {author} {\bibfnamefont {U.~K.}\ \bibnamefont
  {Kl{\"a}ning}}, \bibinfo {author} {\bibfnamefont {K.}~\bibnamefont
  {Sehested}},\ and\ \bibinfo {author} {\bibfnamefont {J.}~\bibnamefont
  {Holcman}},\ }\href {https://doi.org/10.1021/j100251a008} {\bibfield
  {journal} {\bibinfo  {journal} {The Journal of Physical Chemistry}\ }\textbf
  {\bibinfo {volume} {89}},\ \bibinfo {pages} {760} (\bibinfo {year}
  {1985})}\BibitemShut {NoStop}%
\bibitem [{\citenamefont {Dunn}\ and\ \citenamefont {Simon}(1992)}]{Dunn.1992}%
  \BibitemOpen
  \bibfield  {author} {\bibinfo {author} {\bibfnamefont {R.~C.}\ \bibnamefont
  {Dunn}}\ and\ \bibinfo {author} {\bibfnamefont {J.~D.}\ \bibnamefont
  {Simon}},\ }\href {https://doi.org/10.1021/ja00038a060} {\bibfield  {journal}
  {\bibinfo  {journal} {Journal of the American Chemical Society}\ }\textbf
  {\bibinfo {volume} {114}},\ \bibinfo {pages} {4856} (\bibinfo {year}
  {1992})}\BibitemShut {NoStop}%
\end{thebibliography}%

\clearpage

\begin{widetext}
\begin{center}
\large
\textbf{Tailoring the Acidity of Liquid Media with Ionizing Radiation \\
– Rethinking the Acid-Base Correlation Beyond pH} \\ \vspace{0.2cm}
\normalsize
Birk Fritsch, Andreas Körner, Tha\"{i}s Couasnon, Roberts Blukis, Mehran Taherkhani, Liane G. Benning, Michael P. M. Jank, Erdmann Spieker, Andreas Hutzler
\end{center}

\end{widetext}

\appendix
\setcounter{page}{1}

\section{Supporting Information}
\setcounter{figure}{0}
\renewcommand{\figurename}{FIG.}
\renewcommand{\thefigure}{S\arabic{figure}}

\setcounter{table}{0}
\renewcommand{\tablename}{TABLE}
\renewcommand{\thetable}{S\arabic{table}}

\begin{figure*}[t!]
\includegraphics[width=1\textwidth]{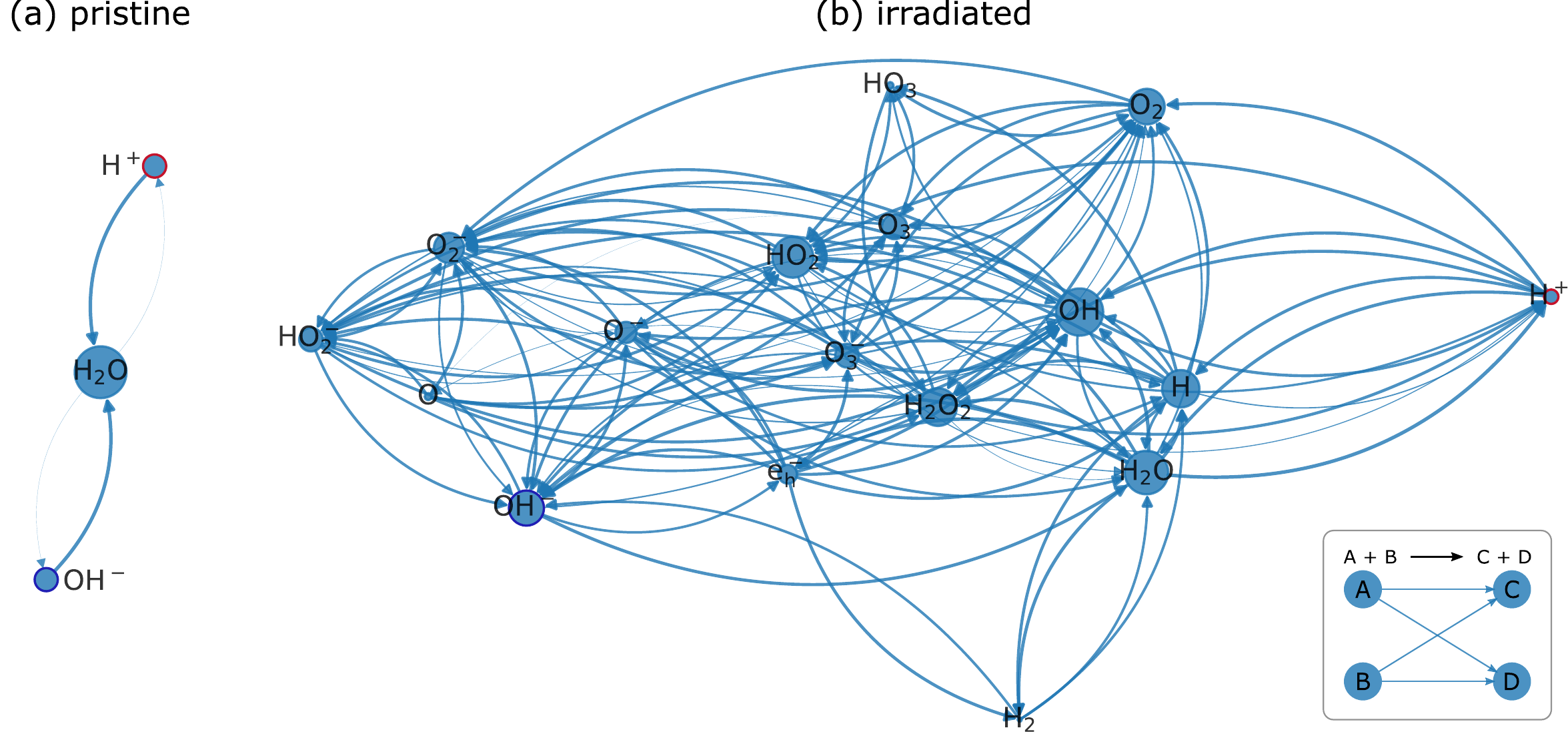}
\caption{\label{fig:Network1}Graph representation of the reaction interplay impacting $\rm H^+$ and $\rm OH^-$ concentrations in (a) pristine and (b) radiolytic water. Tabular representation is found in Table\,\ref{tab:model_Water_S1}.}
\end{figure*}

\subsection{\label{app:pistarelaborations}Discussion on $\pi^*$}

\subsubsection{General considerations}
During irradiation, the acidity is characterized by two instead of one quantity, namely $c$(H$^+$) and $c$(OH$^-$).
In order to access the general outcome for radiation chemistry, a 2x2 equation system (eq. (\ref{eq:KW*}) and eq. (\ref{eq:pi*}) is used including four variables ($c$(H$^+$), $c$(OH$^-$), $K_{\rm W}^*$, and $\pi^*$) to characterize the situation.
To unambiguously describe this system, two of these parameters need to be known.

Using $\pi^*$ allows a direct judgement of the acidity in irradiated solutions, as it would be obtained from pH in non-irradiated solutions.
Moreover, if the reactivity of both, $c(\mathrm{H}^+$), and $c(\mathrm{OH}^-$) within the desired solution is known and kinetically-driven, alternative reaction pathways are negligible, $\pi^*$ solely describes the acid-base equilibrium even without knowledge of $K_{\rm W}^*$.

As here, dynamic steady states are regarded in which forward and backward reactions constantly occur but without changing the net concentrations of the reactants.
Hence, the product distribution will only depend on the ratio of forward and backward reaction rate.
If both ions are equally reactive within the environment of interest, $\pi^*$ is easily interpretable as the net acidity of the solution.
Yet, for scenarios where this assumption would not hold, it would be reasonable to fall back on using two of these four parameters and solve eq. (\ref{eq:Kw}) and eq. (\ref{eq:pi*}) accordingly.

\subsubsection{Relation to pH in non-irradiated solutions}

For non-irradiated solutions, the decadic logarithm of the concentration ratio of $\rm H^+$ and $\rm OH^-$ 
is coupled and depends on pH. The following equation is designed to be applied in non-irradiated solutions only, therefore it is avoided to call it $\pi^*$ here.
\begin{align}
\begin{split}
\frac{\lg(c(\mathrm{H^+} ))}{\lg (c(\mathrm{OH^-}))} = &\lg(c(\mathrm{H^+})) - \lg(c(\mathrm{ OH^-}))\\
= &\lg(c(\mathrm{ H^+} )) - \lg(c(\mathrm{ OH^-})) \\
& + \lg(c(\mathrm{ H^+})) -\lg(c(\mathrm{ H^+} )) \\
= &~2 \lg(c(\mathrm{ H^+} ))  -\lg(c(\mathrm{ OH^-}))\\
  &- \lg(c(\mathrm{ H^+} ))\\
= &-2 \left[-\lg(c(\mathrm{ H^+} ))\right] \\
& - \left[\lg(c(\mathrm{ OH^-})) + \lg(c(\mathrm{ H^+}))\right]\\
= &-2\left[-\lg(c(\mathrm{ H^+ }))\right] \\
  &-\lg(c(\mathrm{H}^+ ) \cdot c(\mathrm{ OH^-}))\\
\end{split}
\label{eq:pistar_S1}
\end{align} 

\begin{figure}
\includegraphics[scale=0.60]{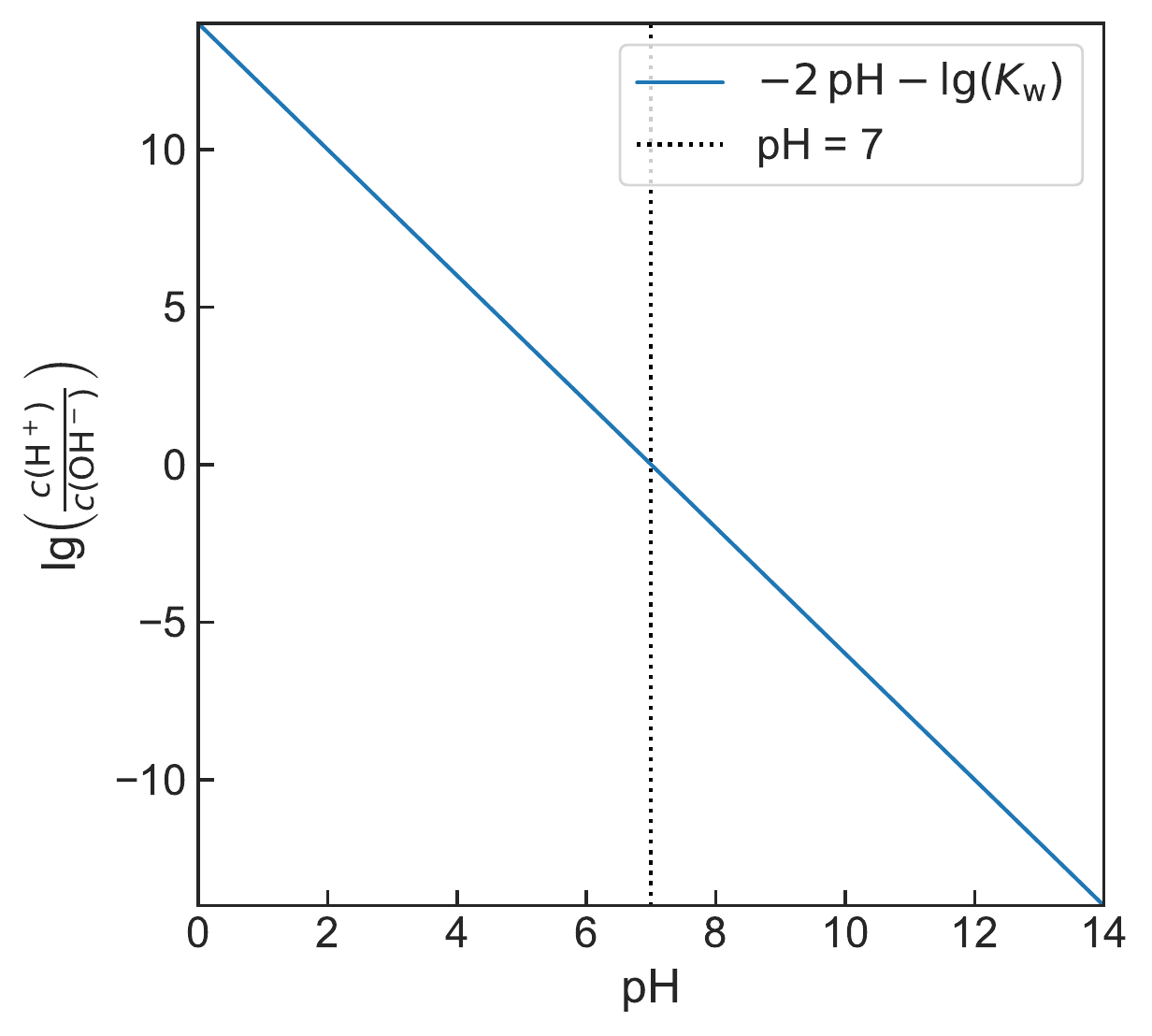}
\caption{\label{fig:pistar_dep_doserate}Linear relation of the decadic logarithm of the $\rm H^+$ and $\rm OH^-$ concentrations at a given pH value for non-irradiated solutions after eq.\,(\ref{eq:pistar_from_pHandKw_S2}).}
\end{figure}

\begin{figure}
\includegraphics[scale=0.60]{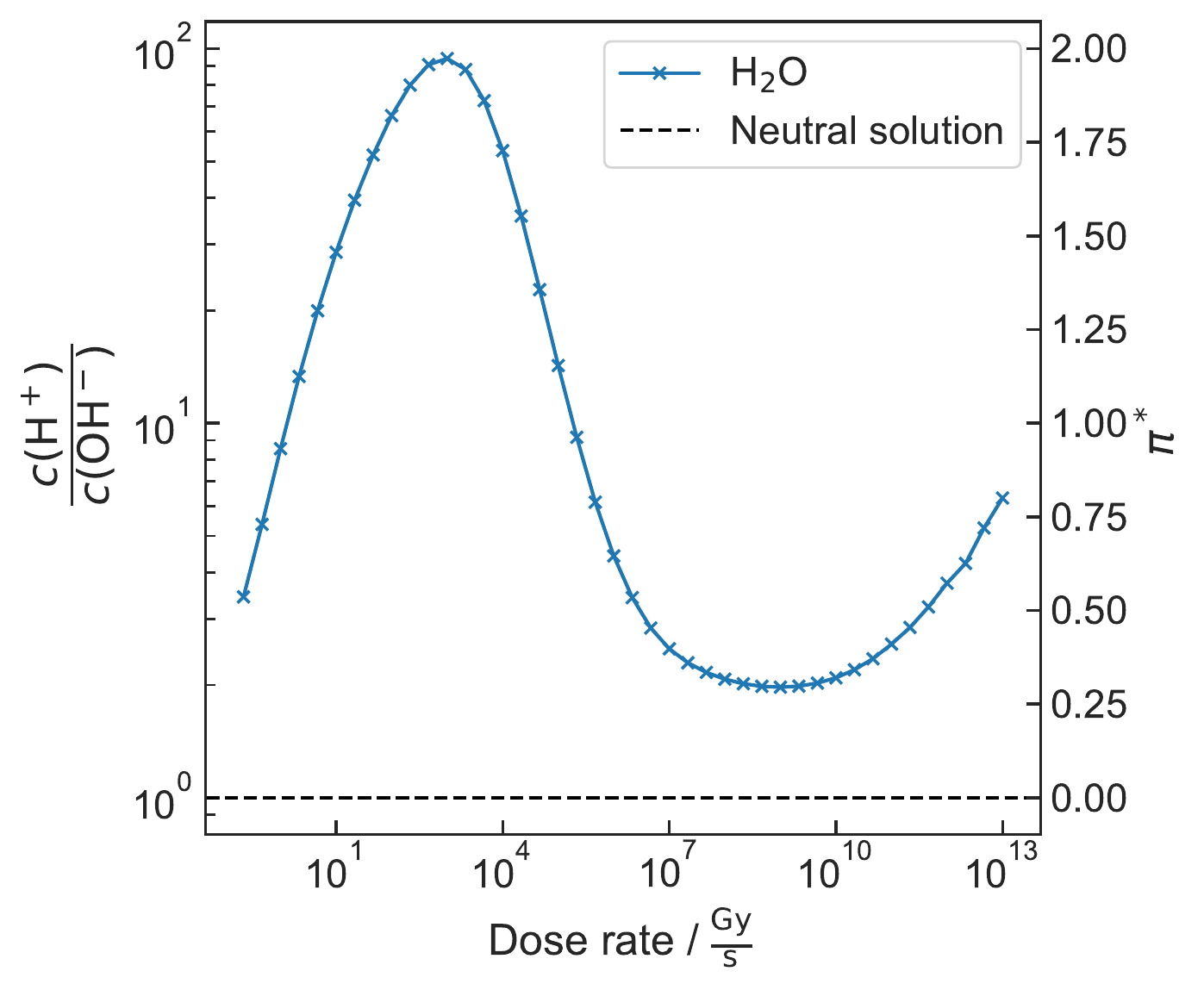}
\caption{\label{fig:excessratioH2O}Dependence the decadic logarithm of the $\rm H^+$ and $\rm OH^-$ concentrations at pH of seven for irradiated solutions. The data shown here is a zoom of the data presented for pure water in Figure\,\ref{fig:ClBrNO3}.}
\end{figure}

Inserting equations (\ref{eq:Kw}) and (\ref{eq:DecadicLog}) into (\ref{eq:pistar_S1}) yields:
\begin{eqnarray}
    \lg\left(\frac{c(\rm H^+ )}{c(\rm OH^-)}\right) = -2 {\rm pH} - \lg(K_{\rm W})
    \label{eq:pistar_from_pHandKw_S2}
\end{eqnarray} 
This linear relationship is depicted in Figure\,\ref{fig:pistar_dep_doserate}.
The simulation for neat, aerated water under standard conditions (no radiation, $\rm 25^\circ C$, $\rm pH=7$) is shown in Figure \ref{fig:excessratioH2O}.
In this case, $\pi^*$ remains at positive values between 0.25 and 2 for a dose rate regime between $0.1\rm\,Gy\cdot\,s^{-1}$ and $1\rm\,PGy\cdot\,s^{-1}$. 
This can be compared to non-irradiated solutions with pH of 6 – 6.875. A $\pi^*$ of unity can be considered neutral condition (i.e. the ratio of $c(\rm H^+)$ and $c(\rm OH^-)$ of almost unity).
A peak appears at $1 \rm\,kGy\cdot\,s^{-1}$ with $\pi^* \approx 2$ yielding an acidic environment that can be compared to $\rm pH = 6$ in a non-irradiated environment.

\subsection{\label{app:Anions}Different additives}

\begin{figure*}[t!]
\includegraphics[scale=0.550]{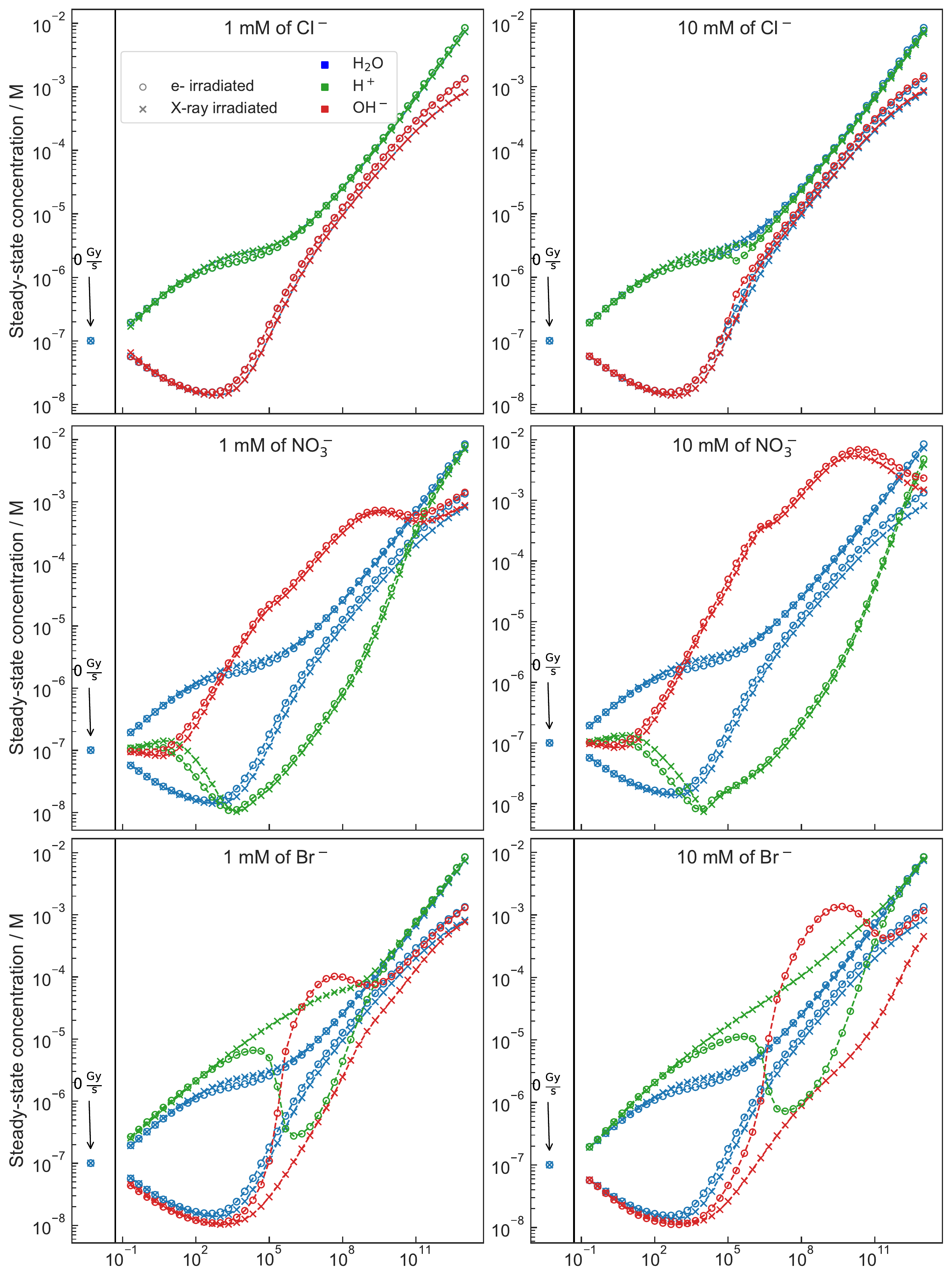}
\caption{\label{fig:SteadyStates4}Steady-state concentrations of $\rm H^+$ and $\rm OH^-$ in both, pure, aerated water, and aqueous solutions containing either $1\,\rm mM$ (left) or $10\,\rm mM$ (right) $\rm Cl^-$, $\rm Br^-$ or $\rm NO_3^-$ ions as functions of the dose rate. 
}
\end{figure*}

In Figure\,\ref{fig:SteadyStates4} the concentrations of $\rm H^+$ and $\rm OH^-$ for initial concentrations of the anions $\rm Cl^-$, $\rm NO_3^-$ and $\rm Br^-$ of $1\,\rm mM$ as well as $10\,\rm mM$ are compared against pure water (blue).

\begin{figure*}
\includegraphics[scale=0.60]{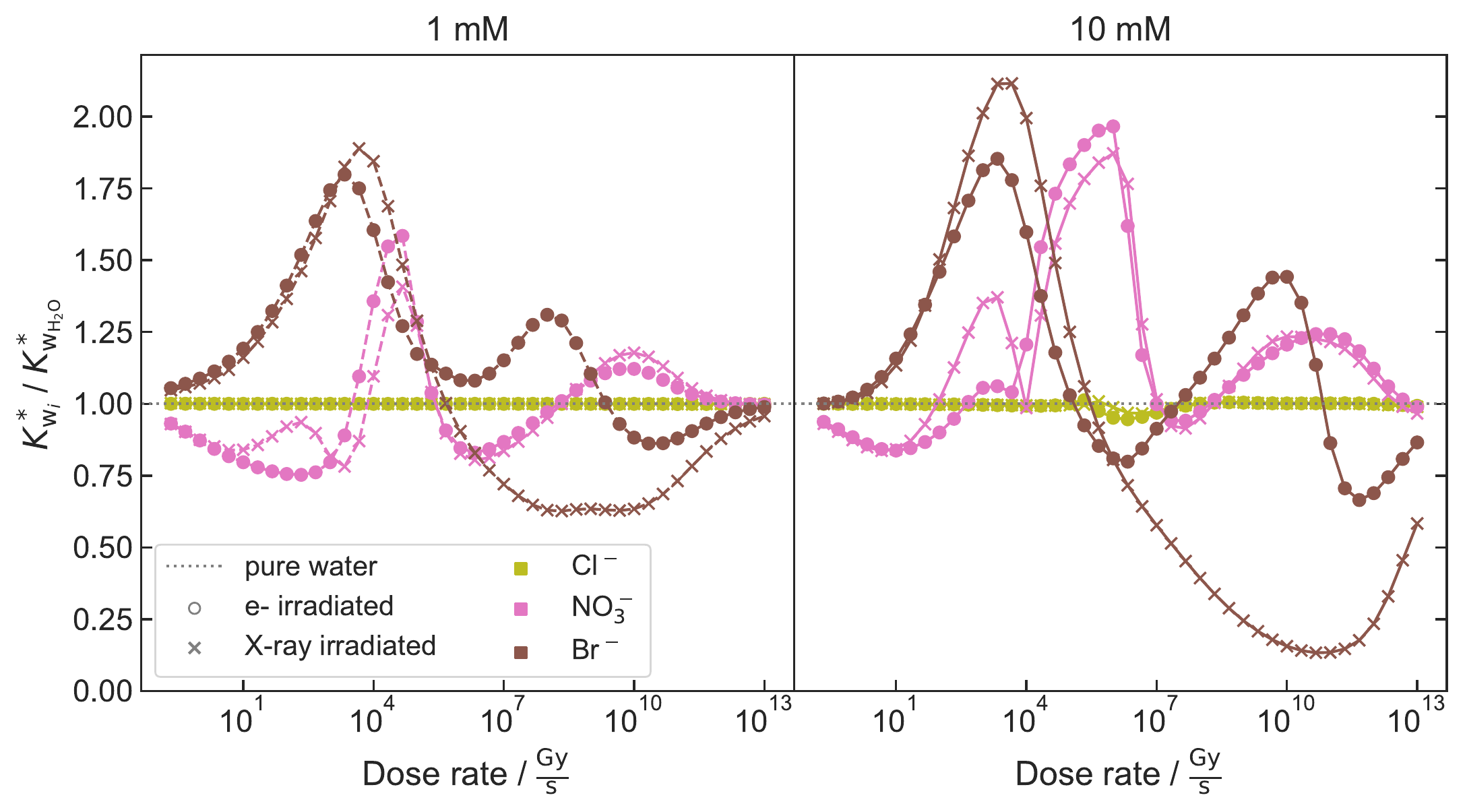}
\caption{\label{fig:AnionsKW_Relation}Relation of the ion product $K_{\rm W}^*$ to $K^*_{\rm W_{H_2O}}$ of areated, non-irradiated water for aqueous solutions of $1\,\rm mM$ (left) and $10\,\rm mM$ (right) $\rm Cl^-$, $\rm Br^-$ or $\rm NO_3^-$ ions. `o'-markers correspond to the simulation performed based on $G$-values of electrons. `X'-markers correspond to the simulation performed based on $G$-values of X-rays.}
\end{figure*}

From Figure \ref{fig:AnionsKW_Relation}, the relation of $K_{\rm W} ^*$ to $K_{\rm W, H_2O}^*$ appears to scale with the initial concentration of $\rm Br^-$ or $\rm NO_3^-$. At $10\,\rm mM$ $\rm NO_3^-$, a $\pi^*$ between -3 and -4 is established from $0.1\rm\,MGy\cdot\,s^{-1}$ until $10\rm\,GGy\cdot\,s^{-1}$. In non-irradiated solutions, such ratios of $c(\rm H^+)$ and $c(\rm OH^-)$ would correspond to moderately basic solutions (pH $8.5 - 9$).

\begin{figure*}
\includegraphics[
scale=0.66]{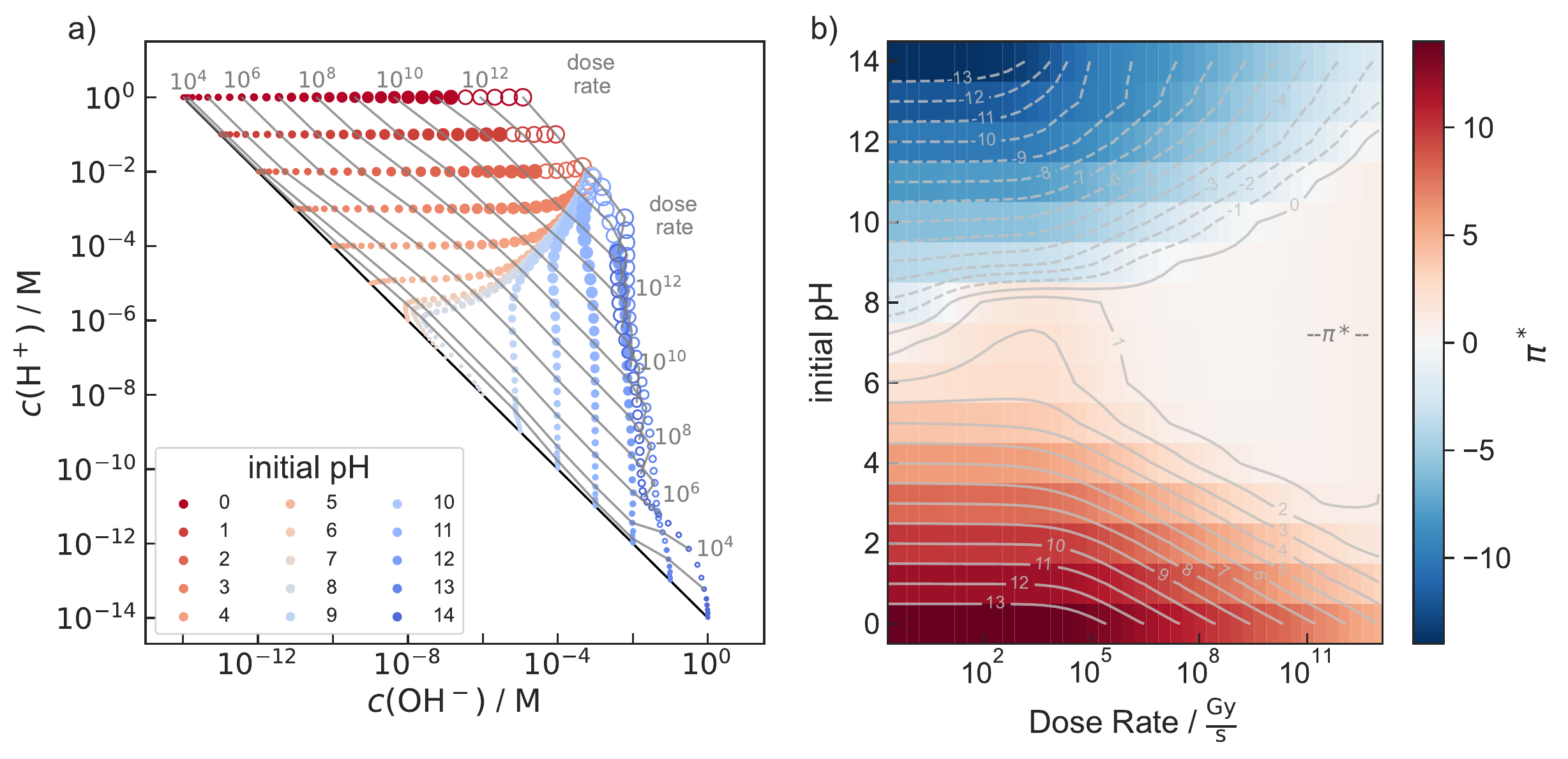}
\caption{\label{fig:pHMap_RadAcidity_Xray}Acid/base chemistry of of neat, aerated water as a function of dose rate of incident X-ray radiation and the initial pH value. (a) Concentrations of $\rm H^+$ and $\rm OH^-$ in the steady state. Each dot represents the steady state concentration of a simulation, while its size is a measure of the dose rate.  Dose rate (grey numbers) is given in $\rm Gy\cdot s^{-1}$ and indicated by contour lines. The black diagonal line corresponds to water under equilibrium conditions ($K_{\rm W} = 10^{-14}\,\rm M^2$) without irradiation. Empty dots represent steady states, where the concentration of water is below 99\% of the unirradiated solution. (b) $\pi^*$ (color map and grey contour lines) as function of initial pH and dose rate. The equivalent plots for electron beam irradiation are shown in Figure\,\ref{fig:pHMap_RadAcidity}.}
\end{figure*}

\FloatBarrier

\subsection{\label{app:SimulationTablesAndGraphs}Simulation tables and graphs}

\begin{table}[h]
\caption{\label{tab:GenerationValues}Generation values used in this work.}
\begin{ruledtabular}
\begin{tabular}{ccc}
& \multicolumn{2}{c}{G-value / (Molecules/100 eV)} \\
primary species	&	e-beam irrad. \cite{Schneider.2014} & X-ray irrad. \cite{Pastina.2001}	\\
$e_h^-$	&	3.47 & 2.60	\\
$\rm H^+$	&	4.42 & 3.10	\\
$\rm OH^-$	&	0.95 & 0.50	\\
$\rm H_2O_2$	&	0.47 & 0.70	\\
$\rm H$	&	1.00 & 0.66	\\
$\rm OH$	&	3.63 & 2.70 \\
$\rm HO_2$	&	0.08 & 0.02	\\
$\rm H_2$	&	0.17 & 0.45	\\
$\rm H_2O$	&	-5.68  & -4.64	\\	\\
\end{tabular}
\end{ruledtabular}
\end{table}

The following section comprises the reaction sets utilized for simulations shown in this work in tabular and graphical network \cite{Fritsch.2022b, Holmes.2021} format.
The latter emphasizes the fundamental difference between irradiated and non-irradiated solutions.
The equilibrium chemistry is fully described by Equation (\ref{eq:Chem_WaterAutoprotolysis}) and shown in Figure\,\ref{fig:Network1}, the reaction interplay is fully described in (a), while the generation of reactive species by irradiation (Eq.~(\ref{eq:ionizing_radiation})) triggers a reaction cascade comprising 83 reactions and 17 species (b). A tabular representation is shown in Table\,\ref{tab:model_Water_S1}.

\def\LT@LR@e{\LTleft\z@   \LTright\z@}%
\afterpage{\onecolumngrid
\renewcommand{\doublerulesep}{0pt}
\begin{longtable*}[e]{@{\extracolsep{\fill}}crclcc}\caption{\label{tab:model_Water_S1}Kinetic model for irradiation of neat, aerated water used in this work. $k$ describes the kinetic constant in units of $\text{mol}^{-n+1}\,\text{L}^{3(n-1)}\,\text{s}^{-1}$, where $n$ denotes the reaction order.}\\
\hline\hline
& \multicolumn{3}{c}{Reaction} & $k$ & Reference  \\
\hline
\endfirsthead
\hline\hline
& \multicolumn{3}{c}{Reaction} & $k$ & Reference  \\
\hline
\endhead
\hline\hline
\multicolumn{6}{c}{\textit{Continued on next page.}}\\
\endfoot
\hline\hline
\endlastfoot
1 & H$_{2}$O & $\longrightarrow$ & H$^+$ + OH$^-$ & $2.599\cdot10^{-5}$ & \cite{Kelm.2004} \\
2 & H$^+$ + OH$^-$ & $\longrightarrow$ & H$_{2}$O & $1.43\cdot10^{11}$ & \cite{Kelm.2004} \\
3 & H$_{2}$O$_{2}$ & $\longrightarrow$ & H$^+$ + HO$_{2}^-$ & $1.119\cdot10^{-1}$ & \cite{Schneider.2014} \\
4 & H$^+$ + HO$_{2}^-$ & $\longrightarrow$ & H$_{2}$O$_{2}$ & $5\cdot10^{10}$ & \cite{Schneider.2014} \\
5 & H$_{2}$O$_{2}$ + OH$^-$ & $\longrightarrow$ & HO$_{2}^-$ + H$_{2}$O & $1.3\cdot10^{10}$ & \cite{Schneider.2014} \\
6 & HO$_{2}^-$ + H$_{2}$O & $\longrightarrow$ & H$_{2}$O$_{2}$ + OH$^-$ & $5.82\cdot10^{7}$ & \cite{Schneider.2014} \\
7 & e$_\mathrm{h}^-$ + H$_{2}$O & $\longrightarrow$ & H + OH$^-$ & $1.9\cdot10^{1}$ & \cite{Schneider.2014} \\
8 & H + OH$^-$ & $\longrightarrow$ & e$_\mathrm{h}^-$ + H$_{2}$O & $2.2\cdot10^{7}$ & \cite{Schneider.2014} \\
9 & H & $\longrightarrow$ & e$_\mathrm{h}^-$ + H$^+$ & $3.9\cdot10^{0}$ & \cite{Schneider.2014} \\
10 & e$_\mathrm{h}^-$ + H$^+$ & $\longrightarrow$ & H & $2.3\cdot10^{10}$ & \cite{Schneider.2014} \\
11 & OH + OH$^-$ & $\longrightarrow$ & O$^-$ + H$_{2}$O & $1.3\cdot10^{10}$ & \cite{Schneider.2014} \\
12 & O$^-$ + H$_{2}$O & $\longrightarrow$ & OH + OH$^-$ & $1\cdot10^{8}$ & \cite{Schneider.2014} \\
13 & OH & $\longrightarrow$ & O$^-$ + H$^+$ & $1.259\cdot10^{-1}$ & \cite{Schneider.2014} \\
14 & O$^-$ + H$^+$ & $\longrightarrow$ & OH & $1\cdot10^{11}$ & \cite{Schneider.2014} \\
15 & HO$_{2}$ & $\longrightarrow$ & O$_{2}^-$ + H$^+$ & $1.346\cdot10^{6}$ & \cite{Schneider.2014} \\
16 & O$_{2}^-$ + H$^+$ & $\longrightarrow$ & HO$_{2}$ & $5\cdot10^{10}$ & \cite{Schneider.2014} \\
17 & HO$_{2}$ + OH$^-$ & $\longrightarrow$ & O$_{2}^-$ + H$_{2}$O & $5\cdot10^{10}$ & \cite{Schneider.2014} \\
18 & O$_{2}^-$ + H$_{2}$O & $\longrightarrow$ & HO$_{2}$ + OH$^-$ & $1.862\cdot10^{1}$ & \cite{Schneider.2014} \\
19 & e$_\mathrm{h}^-$ + OH & $\longrightarrow$ & OH$^-$ & $3\cdot10^{10}$ & \cite{Schneider.2014} \\
20 & e$_\mathrm{h}^-$ + H$_{2}$O$_{2}$ & $\longrightarrow$ & OH + OH$^-$ & $1.1\cdot10^{10}$ & \cite{Schneider.2014} \\
21 & e$_\mathrm{h}^-$ + O$_{2}^-$ + H$_{2}$O & $\longrightarrow$ & HO$_{2}^-$ + OH$^-$ & $1.3\cdot10^{10}$ & \cite{Schneider.2014} \\
22 & e$_\mathrm{h}^-$ + HO$_{2}$ & $\longrightarrow$ & HO$_{2}^-$ & $2\cdot10^{10}$ & \cite{Schneider.2014} \\
23 & e$_\mathrm{h}^-$ + O$_{2}$ & $\longrightarrow$ & O$_{2}^-$ & $1.9\cdot10^{10}$ & \cite{Schneider.2014} \\
24 & 2 e$_\mathrm{h}^-$ + 2 H$_{2}$O & $\longrightarrow$ & H$_{2}$ + 2 OH$^-$ & $5.5\cdot10^{9}$ & \cite{Schneider.2014} \\
25 & e$_\mathrm{h}^-$ + H + H$_{2}$O & $\longrightarrow$ & H$_{2}$ + OH$^-$ & $2.5\cdot10^{10}$ & \cite{Schneider.2014} \\
26 & e$_\mathrm{h}^-$ + HO$_{2}^-$ & $\longrightarrow$ & O$^-$ + OH$^-$ & $3.5\cdot10^{9}$ & \cite{Schneider.2014} \\
27 & e$_\mathrm{h}^-$ + O$^-$ + H$_{2}$O & $\longrightarrow$ & OH$^-$ + OH$^-$ & $2.2\cdot10^{10}$ & \cite{Schneider.2014} \\
28 & e$_\mathrm{h}^-$ + O$_{3}^-$ + H$_{2}$O & $\longrightarrow$ & O$_{2}$ + OH$^-$ + OH$^-$ & $1.6\cdot10^{10}$ & \cite{Schneider.2014} \\
29 & e$_\mathrm{h}^-$ + O$_{3}$ & $\longrightarrow$ & O$_{3}^-$ & $3.6\cdot10^{10}$ & \cite{Schneider.2014} \\
30 & H + H$_{2}$O & $\longrightarrow$ & H$_{2}$ + OH & $1.1\cdot10^{1}$ & \cite{Schneider.2014} \\
31 & H + O$^-$ & $\longrightarrow$ & OH$^-$ & $1\cdot10^{10}$ & \cite{Schneider.2014} \\
32 & H + HO$_{2}^-$ & $\longrightarrow$ & OH + OH$^-$ & $9\cdot10^{7}$ & \cite{Schneider.2014} \\
33 & H + O$_{3}^-$ & $\longrightarrow$ & OH$^-$ + O$_{2}$ & $1\cdot10^{10}$ & \cite{Schneider.2014} \\
34 & 2 H & $\longrightarrow$ & H$_{2}$ & $7.8\cdot10^{9}$ & \cite{Schneider.2014} \\
35 & H + OH & $\longrightarrow$ & H$_{2}$O & $7\cdot10^{9}$ & \cite{Schneider.2014} \\
36 & H + H$_{2}$O$_{2}$ & $\longrightarrow$ & OH + H$_{2}$O & $9\cdot10^{7}$ & \cite{Schneider.2014} \\
37 & H + O$_{2}$ & $\longrightarrow$ & HO$_{2}$ & $2.1\cdot10^{10}$ & \cite{Schneider.2014} \\
38 & H + HO$_{2}$ & $\longrightarrow$ & H$_{2}$O$_{2}$ & $1.8\cdot10^{10}$ & \cite{Schneider.2014} \\
39 & H + O$_{2}^-$ & $\longrightarrow$ & HO$_{2}^-$ & $1.8\cdot10^{10}$ & \cite{Schneider.2014} \\
40 & H + O$_{3}$ & $\longrightarrow$ & HO$_{3}$ & $3.8\cdot10^{10}$ & \cite{Schneider.2014} \\
41 & 2 OH & $\longrightarrow$ & H$_{2}$O$_{2}$ & $3.6\cdot10^{9}$ & \cite{Schneider.2014} \\
42 & OH + HO$_{2}$ & $\longrightarrow$ & H$_{2}$O + O$_{2}$ & $6\cdot10^{9}$ & \cite{Schneider.2014} \\
43 & OH + O$_{2}^-$ & $\longrightarrow$ & OH$^-$ + O$_{2}$ & $8.2\cdot10^{9}$ & \cite{Schneider.2014} \\
44 & OH + H$_{2}$ & $\longrightarrow$ & H + H$_{2}$O & $4.3\cdot10^{7}$ & \cite{Schneider.2014} \\
45 & OH + H$_{2}$O$_{2}$ & $\longrightarrow$ & HO$_{2}$ + H$_{2}$O & $2.7\cdot10^{7}$ & \cite{Schneider.2014} \\
46 & OH + O$^-$ & $\longrightarrow$ & HO$_{2}^-$ & $2.5\cdot10^{10}$ & \cite{Schneider.2014} \\
47 & OH + HO$_{2}^-$ & $\longrightarrow$ & HO$_{2}$ + OH$^-$ & $7.5\cdot10^{9}$ & \cite{Schneider.2014} \\
48 & OH + O$_{3}^-$ & $\longrightarrow$ & O$_{3}$ + OH$^-$ & $2.6\cdot10^{9}$ & \cite{Schneider.2014} \\
49 & OH + O$_{3}^-$ & $\longrightarrow$ & 2 O$_{2}^-$ + H$^+$ & $6\cdot10^{9}$ & \cite{Schneider.2014} \\
50 & OH + O$_{3}$ & $\longrightarrow$ & HO$_{2}$ + O$_{2}$ & $1.1\cdot10^{8}$ & \cite{Schneider.2014} \\
51 & HO$_{2}$ + O$_{2}^-$ & $\longrightarrow$ & HO$_{2}^-$ + O$_{2}$ & $8\cdot10^{7}$ & \cite{Schneider.2014} \\
52 & HO$_{2}$ + HO$_{2}$ & $\longrightarrow$ & H$_{2}$O$_{2}$ + O$_{2}$ & $7\cdot10^{5}$ & \cite{Schneider.2014} \\
53 & HO$_{2}$ + O$^-$ & $\longrightarrow$ & O$_{2}$ + OH$^-$ & $6\cdot10^{9}$ & \cite{Schneider.2014} \\
54 & HO$_{2}$ + H$_{2}$O$_{2}$ & $\longrightarrow$ & OH + O$_{2}$ + H$_{2}$O & $5\cdot10^{-1}$ & \cite{Schneider.2014} \\
55 & HO$_{2}$ + HO$_{2}^-$ & $\longrightarrow$ & OH + O$_{2}$ + OH$^-$ & $5\cdot10^{-1}$ & \cite{Schneider.2014} \\
56 & HO$_{2}$ + O$_{3}^-$ & $\longrightarrow$ & O$_{2}$ + O$_{2}$ + OH$^-$ & $6\cdot10^{9}$ & \cite{Schneider.2014} \\
57 & HO$_{2}$ + O$_{3}$ & $\longrightarrow$ & HO$_{3}$ + O$_{2}$ & $5\cdot10^{8}$ & \cite{Schneider.2014} \\
58 & 2 O$_{2}^-$ + 2 H$_{2}$O & $\longrightarrow$ & H$_{2}$O$_{2}$ + O$_{2}$ + 2 OH$^-$ & $1\cdot10^{2}$ & \cite{Schneider.2014} \\
59 & O$_{2}^-$ + O$^-$ + H$_{2}$O & $\longrightarrow$ & O$_{2}$ + 2 OH$^-$ & $6\cdot10^{8}$ & \cite{Schneider.2014} \\
60 & O$_{2}^-$ + H$_{2}$O$_{2}$ & $\longrightarrow$ & OH + O$_{2}$ + OH$^-$ & $1.3\cdot10^{-1}$ & \cite{Schneider.2014} \\
61 & O$_{2}^-$ + HO$_{2}^-$ & $\longrightarrow$ & O$^-$ + O$_{2}$ + OH$^-$ & $1.3\cdot10^{-1}$ & \cite{Schneider.2014} \\
62 & O$_{2}^-$ + O$_{3}^-$ + H$_{2}$O & $\longrightarrow$ & O$_{2}$ + O$_{2}$ + 2 OH$^-$ & $1\cdot10^{4}$ & \cite{Schneider.2014} \\
63 & O$_{2}^-$ + O$_{3}$ & $\longrightarrow$ & O$_{3}^-$ + O$_{2}$ & $1.5\cdot10^{9}$ & \cite{Schneider.2014} \\
64 & 2 O$^-$ + H$_{2}$O & $\longrightarrow$ & HO$_{2}^-$ + OH$^-$ & $1\cdot10^{9}$ & \cite{Schneider.2014} \\
65 & O$^-$ + O$_{2}$ & $\longrightarrow$ & O$_{3}^-$ & $3.6\cdot10^{9}$ & \cite{Schneider.2014} \\
66 & O$^-$ + H$_{2}$ & $\longrightarrow$ & H + OH$^-$ & $8\cdot10^{7}$ & \cite{Schneider.2014} \\
67 & O$^-$ + H$_{2}$O$_{2}$ & $\longrightarrow$ & O$_{2}^-$ + H$_{2}$O & $5\cdot10^{8}$ & \cite{Schneider.2014} \\
68 & O$^-$ + HO$_{2}^-$ & $\longrightarrow$ & O$_{2}^-$ + OH$^-$ & $4\cdot10^{8}$ & \cite{Schneider.2014} \\
69 & O$^-$ + O$_{3}^-$ & $\longrightarrow$ & O$_{2}^-$ + O$_{2}^-$ & $7\cdot10^{8}$ & \cite{Schneider.2014} \\
70 & O$^-$ + O$_{3}$ & $\longrightarrow$ & O$_{2}^-$ + O$_{2}$ & $5\cdot10^{9}$ & \cite{Schneider.2014} \\
71 & O$_{3}^-$ & $\longrightarrow$ & O$_{2}$ + O$^-$ & $3.3\cdot10^{3}$ & \cite{Schneider.2014} \\
72 & O$_{3}^-$ + H$^+$ & $\longrightarrow$ & O$_{2}$ + OH & $9\cdot10^{10}$ & \cite{Schneider.2014} \\
73 & HO$_{3}$ & $\longrightarrow$ & O$_{2}$ + OH & $1.1\cdot10^{5}$ & \cite{Schneider.2014} \\
74 & H$_{2}$O$_{2}$ & $\longrightarrow$ & H$_{2}$O + O & $1\cdot10^{-3}$ & \cite{Kelm.2004} \\
75 & 2 O & $\longrightarrow$ & O$_{2}$ & $1\cdot10^{9}$ & \cite{Kelm.2004} \\
76 & O$_{3}$ & $\longrightarrow$ & O$_{2}$ + O & $3\cdot10^{-6}$ & \cite{Sehested.1992} \\
77 & 2 O$_{3}^-$ + H$_{2}$O & $\longrightarrow$ & OH$^-$ + HO$_{2}^-$ + 2 O$_{2}$ & $1\cdot10^{4}$ & \cite{Levanov.2020} \\
78 & 2 HO$_{3}$ & $\longrightarrow$ & H$_{2}$O$_{2}$ + 2 O$_{2}$ & $5\cdot10^{9}$ & \cite{Levanov.2020} \\
79 & O$_{3}$ + OH$^-$ & $\longrightarrow$ & HO$_{2}^-$ + O$_{2}$ & $1\cdot10^{2}$ & \cite{Levanov.2020} \\
80 & O$_{2}$ + O & $\longrightarrow$ & O$_{3}$ & $4\cdot10^{9}$ & \cite{Klaning.1984} \\
81 & H$_{2}$O$_{2}$ + O & $\longrightarrow$ & OH + HO$_{2}$ & $1.6\cdot10^{9}$ & \cite{Sauer.1984} \\
82 & O + HO$_{2}^-$ & $\longrightarrow$ & OH + O$_{2}^-$ & $5.3\cdot10^{9}$ & \cite{Sauer.1984} \\
83 & O + OH$^-$ & $\longrightarrow$ & HO$_{2}^-$ & $4.2\cdot10^{8}$ & \cite{Sauer.1984}\\
\end{longtable*}
\twocolumngrid
}

\begin{figure*}[t!]
\includegraphics[width=0.8\textwidth]{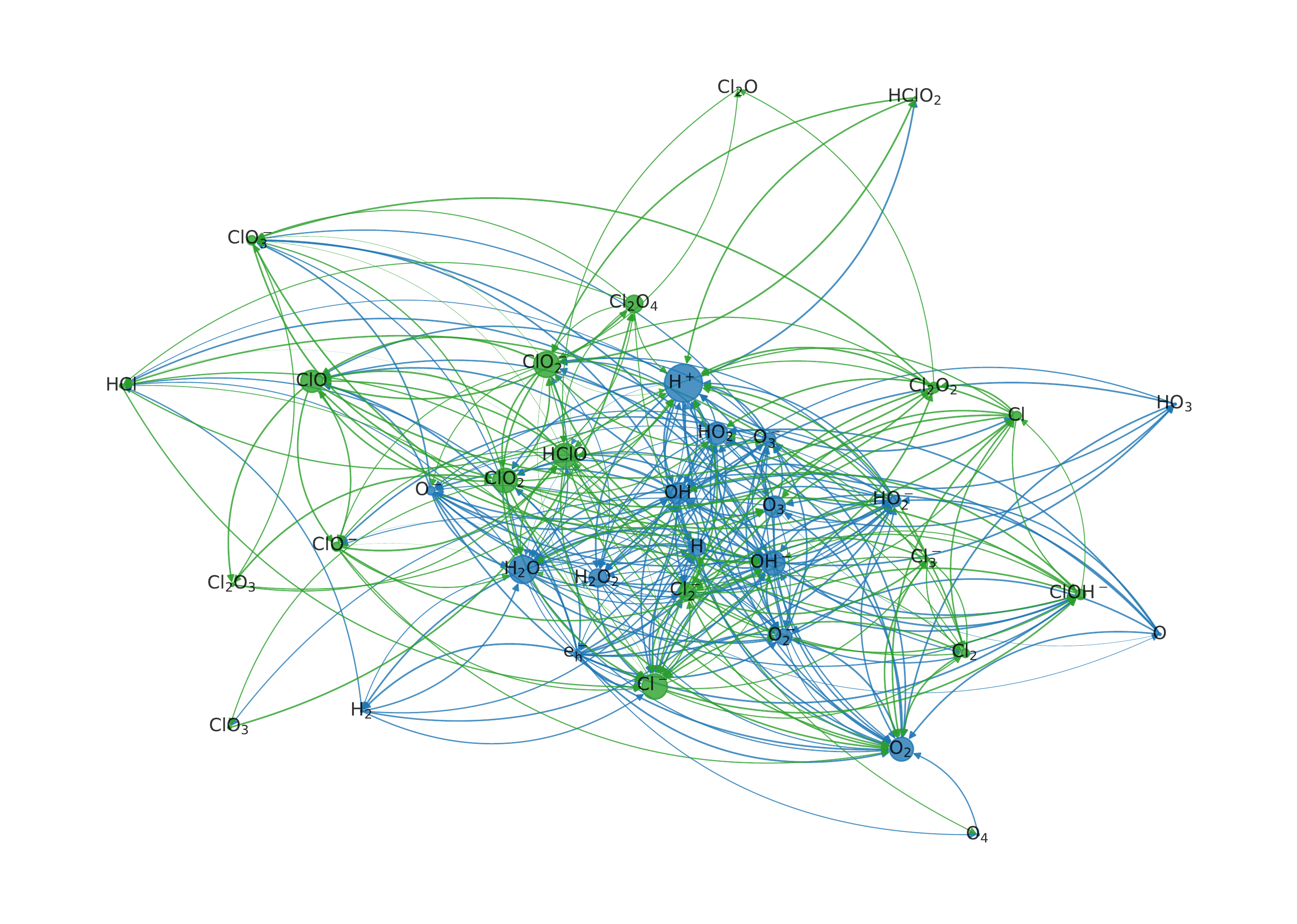}
\caption{\label{fig:Graph_Cl_S3}Graph representation of the kinetic model of $\rm Cl^-$-containing aqueous solutions. Tabular representation is found in Table\,\ref{tab:model_Cl-_S4}.}
\end{figure*}

In addition, the chlorine set comprises 89 reactions and 19 new species (Table\,\ref{tab:model_Cl-_S4}, Figure\,\ref{fig:Graph_Cl_S3}). It is a subset of the reaction set for aqueous $\rm HAuCl_4$ solutions introduced earlier \cite{Fritsch.2022b}.

\afterpage{\onecolumngrid
\renewcommand{\doublerulesep}{0pt}
\begin{longtable*}[e]{@{\extracolsep{\fill}}crclcc}\caption{\label{tab:model_Cl-_S4}Kinetic model used to describe the radiolysis of $\rm Cl^-$-containing aqueous solutions. Here, $k$ denotes the respective kinetic constant in units of $\text{mol}^{-n+1}\,\text{L}^{3(n-1)}\,\text{s}^{-1}$, where $n$ denotes the reaction order. Please refer to Supporting Table\,\ref{tab:model_Water_S1} for the first 83 reactions.}\\
\hline\hline
& \multicolumn{3}{c}{Reaction} & $k$ & Reference  \\
\hline
\endfirsthead
\hline\hline
& \multicolumn{3}{c}{Reaction} & $k$ & Reference  \\
\hline
\endhead
\hline\hline
\multicolumn{6}{c}{\textit{Continued on next page.}}\\
\endfoot
\hline\hline
\endlastfoot
84 & OH + Cl$^-$ & $\longrightarrow$ & ClOH$^-$ & $4.3\cdot10^{9}$ & \cite{Kelm.2004} \\
85 & OH + HClO & $\longrightarrow$ & ClO + H$_{2}$O & $9\cdot10^{9}$ & \cite{Kelm.2004} \\
86 & OH + ClO$_{2}^-$ + H$^+$ & $\longrightarrow$ & ClO$_{2}$ + H$_{2}$O & $6.3\cdot10^{9}$ & \cite{Kelm.2004} \\
87 & e$_\mathrm{h}^-$ + Cl & $\longrightarrow$ & Cl$^-$ & $1\cdot10^{10}$ & \cite{Kelm.2004} \\
88 & e$_\mathrm{h}^-$ + Cl$_{2}^-$ & $\longrightarrow$ & 2 Cl$^-$ & $1\cdot10^{10}$ & \cite{Kelm.2004} \\
89 & e$_\mathrm{h}^-$ + ClOH$^-$ & $\longrightarrow$ & Cl$^-$ + OH$^-$ & $1\cdot10^{10}$ & \cite{Kelm.2004} \\
90 & e$_\mathrm{h}^-$ + HClO & $\longrightarrow$ & ClOH$^-$ & $5.3\cdot10^{10}$ & \cite{Kelm.2004} \\
91 & e$_\mathrm{h}^-$ + Cl$_{2}$ & $\longrightarrow$ & Cl$_{2}^-$ & $1\cdot10^{10}$ & \cite{Kelm.2004} \\
92 & e$_\mathrm{h}^-$ + Cl$_{3}^-$ & $\longrightarrow$ & Cl$_{2}^-$ + Cl$^-$ & $1\cdot10^{10}$ & \cite{Kelm.2004} \\
93 & e$_\mathrm{h}^-$ + ClO$_{2}^-$ + H$^+$ & $\longrightarrow$ & ClO + OH$^-$ & $4.5\cdot10^{10}$ & \cite{Kelm.2004} \\
94 & e$_\mathrm{h}^-$ + ClO$_{3}^-$ + H$^+$ & $\longrightarrow$ & ClO$_{2}$ + OH$^-$ & $1\cdot10^{10}$ & \cite{Sunder.1993} \\
95 & H + Cl & $\longrightarrow$ & Cl$^-$ + H$^+$ & $1\cdot10^{10}$ & \cite{Kelm.2004} \\
96 & H + Cl$_{2}^-$ & $\longrightarrow$ & 2 Cl$^-$ + H$^+$ & $8\cdot10^{9}$ & \cite{Kelm.2004} \\
97 & H + ClOH$^-$ & $\longrightarrow$ & Cl$^-$ + H$_{2}$O & $1\cdot10^{10}$ & \cite{Kelm.2004} \\
98 & H + Cl$_{2}$ & $\longrightarrow$ & Cl$_{2}^-$ + H$^+$ & $7\cdot10^{9}$ & \cite{Kelm.2004} \\
99 & H + HClO & $\longrightarrow$ & ClOH$^-$ + H$^+$ & $1\cdot10^{10}$ & \cite{Kelm.2004} \\
100 & H + Cl$_{3}^-$ & $\longrightarrow$ & Cl$_{2}^-$ + Cl$^-$ + H$^+$ & $1\cdot10^{10}$ & \cite{Kelm.2004} \\
101 & HO$_{2}$ + Cl$_{2}^-$ & $\longrightarrow$ & Cl$^-$ + HCl + O$_{2}$ & $4\cdot10^{9}$ & \cite{Kelm.2004} \\
102 & HCl & $\longrightarrow$ & Cl$^-$ + H$^+$ & $5\cdot10^{5}$ & \cite{Kelm.2004} \\
103 & Cl$^-$ + H$^+$ & $\longrightarrow$ & HCl & $6.29\cdot10^{-1}$ & \cite{Kelm.2004, Trummal.2016} \\
104 & HO$_{2}$ + Cl$_{2}$ & $\longrightarrow$ & Cl$_{2}^-$ + O$_{2}$ + H$^+$ & $1\cdot10^{9}$ & \cite{Kelm.2004} \\
105 & HO$_{2}$ + Cl$_{3}^-$ & $\longrightarrow$ & Cl$_{2}^-$ + HCl + O$_{2}$ & $1\cdot10^{9}$ & \cite{Kelm.2004} \\
106 & O$_{2}^-$ + Cl$_{2}^-$ & $\longrightarrow$ & 2 Cl$^-$ + O$_{2}$ & $1.2\cdot10^{10}$ & \cite{Kelm.2004} \\
107 & O$_{2}^-$ + HClO & $\longrightarrow$ & ClOH$^-$ + O$_{2}$ & $7.5\cdot10^{6}$ & \cite{Kelm.2004} \\
108 & H$_{2}$O$_{2}$ + Cl$_{2}^-$ & $\longrightarrow$ & 2 HCl + O$_{2}^-$ & $1.4\cdot10^{5}$ & \cite{Kelm.2004} \\
109 & H$_{2}$O$_{2}$ + Cl$_{2}$ & $\longrightarrow$ & HO$_{2}$ + Cl$_{2}^-$ + H$^+$ & $1.9\cdot10^{2}$ & \cite{Kelm.2004} \\
110 & H$_{2}$O$_{2}$ + HClO & $\longrightarrow$ & HCl + H$_{2}$O + O$_{2}$ & $1.7\cdot10^{5}$ & \cite{Kelm.2004} \\
111 & OH$^-$ + Cl$_{2}^-$ & $\longrightarrow$ & ClOH$^-$ + Cl$^-$ & $7.3\cdot10^{6}$ & \cite{Kelm.2004} \\
112 & OH$^-$ + Cl$_{2}$ & $\longrightarrow$ & HClO + Cl$^-$ & $6\cdot10^{8}$ & \cite{Kelm.2004} \\
113 & H$^+$ + ClOH$^-$ & $\longrightarrow$ & Cl + H$_{2}$O & $2.1\cdot10^{10}$ & \cite{Kelm.2004} \\
114 & H$_{2}$O + Cl$_{2}$O$_{2}$ & $\longrightarrow$ & HClO + ClO$_{2}^-$ + H$^+$ & $1\cdot10^{4}$ & \cite{Levanov.2020} \\
115 & H$_{2}$O + Cl$_{2}$O & $\longrightarrow$ & 2 HClO & $1\cdot10^{2}$ & \cite{Kelm.2004} \\
116 & H$_{2}$O + Cl$_{2}$O$_{4}$ & $\longrightarrow$ & ClO$_{2}^-$ + ClO$_{3}^-$ + 2 H$^+$ & $1\cdot10^{2}$ & \cite{Kelm.2004} \\
117 & H$_{2}$O + Cl$_{2}$O$_{4}$ & $\longrightarrow$ & HClO + HCl + O$_{4}$ & $1\cdot10^{2}$ & \cite{Kelm.2004} \\
118 & O$_{4}$ & $\longrightarrow$ & 2 O$_{2}$ & $1\cdot10^{5}$ & \cite{Kelm.2004} \\
119 & Cl$^-$ + Cl & $\longrightarrow$ & Cl$_{2}^-$ & $2.1\cdot10^{10}$ & \cite{Kelm.2004} \\
120 & Cl$^-$ + ClOH$^-$ & $\longrightarrow$ & Cl$_{2}^-$ + OH$^-$ & $9\cdot10^{4}$ & \cite{Kelm.2004} \\
121 & Cl$^-$ + HClO & $\longrightarrow$ & Cl$_{2}$ + OH$^-$ & $1\cdot10^{1}$ & \cite{Sunder.1993} \\
122 & Cl$^-$ + Cl$_{2}$ & $\longrightarrow$ & Cl$_{3}^-$ & $1\cdot10^{4}$ & \cite{Kelm.2004} \\
123 & ClOH$^-$ & $\longrightarrow$ & OH + Cl$^-$ & $6.1\cdot10^{9}$ & \cite{Kelm.2004} \\
124 & Cl$_{2}^-$ & $\longrightarrow$ & Cl + Cl$^-$ & $1.1\cdot10^{5}$ & \cite{Kelm.2004} \\
125 & 2 Cl$_{2}^-$ & $\longrightarrow$ & Cl$_{3}^-$ + Cl$^-$ & $7\cdot10^{9}$ & \cite{Kelm.2004} \\
126 & Cl$_{3}^-$ & $\longrightarrow$ & Cl$_{2}$ + Cl$^-$ & $5\cdot10^{4}$ & \cite{Kelm.2004} \\
127 & 2 ClO & $\longrightarrow$ & Cl$_{2}$O$_{2}$ & $1.5\cdot10^{10}$ & \cite{Kelm.2004} \\
128 & 2 ClO$_{2}$ & $\longrightarrow$ & Cl$_{2}$O$_{4}$ & $1\cdot10^{2}$ & \cite{Kelm.2004} \\
129 & Cl$_{2}$O$_{2}$ + ClO$_{2}^-$ & $\longrightarrow$ & ClO$_{3}^-$ + Cl$_{2}$O & $1\cdot10^{2}$ & \cite{Kelm.2004} \\
130 & 2 HClO & $\longrightarrow$ & Cl$^-$ + ClO$_{2}^-$ + 2 H$^+$ & $6\cdot10^{-9}$ & \cite{Kelm.2004} \\
131 & ClO$_{2}^-$ + HClO & $\longrightarrow$ & Cl$^-$ + ClO$_{3}^-$ + H$^+$ & $9\cdot10^{-7}$ & \cite{Kelm.2004} \\
132 & 2 HClO & $\longrightarrow$ & O$_{2}$ + 2 HCl & $3\cdot10^{-10}$ & \cite{Kelm.2004} \\
133 & HClO + Cl$^-$ + H$^+$ & $\longrightarrow$ & Cl$_{2}$ + H$_{2}$O & $9\cdot10^{3}$ & \cite{Kelm.2004} \\
134 & Cl$_{2}$ + H$_{2}$O & $\longrightarrow$ & HClO + Cl$^-$ + H$^+$ & $1.5\cdot10^{1}$ & \cite{Kelm.2004} \\
135 & Cl$_{2}^-$ + H$_{2}$ & $\longrightarrow$ & H + HCl + Cl$^-$ & $4.3\cdot10^{5}$ & \cite{Kelm.2004} \\
136 & 2 Cl & $\longrightarrow$ & Cl$_{2}$ & $8.8\cdot10^{7}$ & \cite{Wu.1980} \\
137 & ClO$_{2}$ + O$_{3}$ & $\longrightarrow$ & O$_{2}$ + ClO$_{3}$ & $1.1\cdot10^{3}$ & \cite{Klaning.1985} \\
138 & ClO$_{2}$ + OH & $\longrightarrow$ & ClO$_{3}^-$ + H$^+$ & $4\cdot10^{9}$ & \cite{Klaning.1985} \\
139 & ClO$_{2}$ + O$^-$ & $\longrightarrow$ & ClO$_{3}^-$ & $2.7\cdot10^{9}$ & \cite{Klaning.1985} \\
140 & ClO$_{2}$ + O$_{3}^-$ & $\longrightarrow$ & O$_{2}$ + ClO$_{3}^-$ & $1.8\cdot10^{5}$ & \cite{Klaning.1985} \\
141 & ClO$_{2}$ + O$_{3}^-$ & $\longrightarrow$ & O$_{3}$ + ClO$_{2}^-$ & $1.8\cdot10^{5}$ & \cite{Klaning.1985} \\
142 & ClO$_{2}^-$ + O$_{3}$ & $\longrightarrow$ & O$_{3}^-$ + ClO$_{2}$ & $4\cdot10^{6}$ & \cite{Klaning.1985} \\
143 & ClO$_{2}$ & $\longrightarrow$ & O$_{2}$ + Cl & $6.7\cdot10^{9}$ & \cite{Dunn.1992} \\
144 & HClO & $\longrightarrow$ & H$^+$ + ClO$^-$ & $2\cdot10^{3}$ & \cite{Levanov.2020} \\
145 & H$^+$ + ClO$^-$ & $\longrightarrow$ & HClO & $5\cdot10^{10}$ & \cite{Levanov.2020} \\
146 & HClO$_{2}$ & $\longrightarrow$ & H$^+$ + ClO$_{2}^-$ & $9.53\cdot10^{8}$ & \cite{Levanov.2020} \\
147 & H$^+$ + ClO$_{2}^-$ & $\longrightarrow$ & HClO$_{2}$ & $5\cdot10^{10}$ & \cite{Levanov.2020} \\
148 & Cl + O$_{3}^-$ & $\longrightarrow$ & Cl$^-$ + O$_{3}$ & $1\cdot10^{9}$ & \cite{Levanov.2020} \\
149 & ClO + O$_{3}^-$ & $\longrightarrow$ & ClO$^-$ + O$_{3}$ & $1\cdot10^{9}$ & \cite{Levanov.2020} \\
150 & Cl$_{2}^-$ + ClO$_{2}$ & $\longrightarrow$ & Cl$_{2}$O$_{2}$ + Cl$^-$ & $1\cdot10^{9}$ & \cite{Levanov.2020} \\
151 & Cl + ClO$_{2}$ & $\longrightarrow$ & Cl$_{2}$O$_{2}$ & $1\cdot10^{9}$ & \cite{Levanov.2020} \\
152 & ClO + ClO$_{2}^-$ & $\longrightarrow$ & ClO$^-$ + ClO$_{2}$ & $9.4\cdot10^{8}$ & \cite{Levanov.2020} \\
153 & ClO$^-$ + O$^-$ + H$^+$ & $\longrightarrow$ & ClO + OH$^-$ & $2.3\cdot10^{8}$ & \cite{Levanov.2020} \\
154 & Cl$^-$ + H$_{2}$O$_{2}$ & $\longrightarrow$ & ClO$^-$ + H$_{2}$O & $1.8\cdot10^{-9}$ & \cite{Levanov.2020} \\
155 & Cl$^-$ + H$_{2}$O$_{2}$ + H$^+$ & $\longrightarrow$ & HClO + H$_{2}$O & $8.3\cdot10^{-7}$ & \cite{Levanov.2020} \\
156 & ClO$^-$ + H$_{2}$O$_{2}$ & $\longrightarrow$ & Cl$^-$ + O$_{2}$ + H$_{2}$O & $3.4\cdot10^{3}$ & \cite{Levanov.2020} \\
157 & HClO + HO$_{2}^-$ & $\longrightarrow$ & Cl$^-$ + O$_{2}$ + H$_{2}$O & $4.4\cdot10^{7}$ & \cite{Levanov.2020} \\
158 & Cl$_{2}$ + HO$_{2}^-$ & $\longrightarrow$ & 2 Cl$^-$ + O$_{2}$ + H$^+$ & $1.1\cdot10^{8}$ & \cite{Levanov.2020} \\
159 & Cl + H$_{2}$O$_{2}$ & $\longrightarrow$ & Cl$^-$ + H$^+$ + HO$_{2}$ & $2\cdot10^{9}$ & \cite{Levanov.2020} \\
160 & Cl + HO$_{2}$ & $\longrightarrow$ & Cl$^-$ + H$^+$ + O$_{2}$ & $3.1\cdot10^{9}$ & \cite{Levanov.2020} \\
161 & Cl + OH$^-$ & $\longrightarrow$ & ClOH$^-$ & $1.8\cdot10^{10}$ & \cite{Levanov.2020} \\
162 & ClO$_{2}$ + H$_{2}$O$_{2}$ & $\longrightarrow$ & ClO$_{2}^-$ + H$^+$ + HO$_{2}$ & $4\cdot10^{0}$ & \cite{Levanov.2020} \\
163 & ClO$_{2}$ + HO$_{2}^-$ & $\longrightarrow$ & ClO$_{2}^-$ + HO$_{2}$ & $1.3\cdot10^{5}$ & \cite{Levanov.2020} \\
164 & ClO$_{2}$ + HO$_{2}$ & $\longrightarrow$ & ClO$_{2}^-$ + H$^+$ + O$_{2}$ & $1\cdot10^{6}$ & \cite{Levanov.2020} \\
165 & ClO$_{2}$ + O$_{2}^-$ & $\longrightarrow$ & ClO$_{2}^-$ + O$_{2}$ & $3\cdot10^{9}$ & \cite{Levanov.2020} \\
166 & ClO$_{2}^-$ + O$_{2}^-$ & $\longrightarrow$ & ClO$^-$ + O$^-$ + O$_{2}$ & $4\cdot10^{1}$ & \cite{Levanov.2020} \\
167 & ClO + ClO$_{2}$ & $\longrightarrow$ & Cl$_{2}$O$_{3}$ & $7.4\cdot10^{9}$ & \cite{Levanov.2020} \\
168 & ClO + ClO$_{3}$ & $\longrightarrow$ & Cl$_{2}$O$_{4}$ & $7.4\cdot10^{9}$ & \cite{Levanov.2020} \\
169 & Cl$_{2}$O$_{2}$ + OH$^-$ & $\longrightarrow$ & Cl$^-$ + ClO$_{3}^-$ + H$^+$ & $1\cdot10^{10}$ & \cite{Levanov.2020} \\
170 & Cl$_{2}$O$_{3}$ + H$_{2}$O & $\longrightarrow$ & HClO + ClO$_{3}^-$ + H$^+$ & $1\cdot10^{4}$ & \cite{Levanov.2020} \\
171 & ClOH$^-$ & $\longrightarrow$ & Cl + OH$^-$ & $2.3\cdot10^{1}$ & \cite{Levanov.2020} \\
172 & Cl + H$_{2}$O & $\longrightarrow$ & ClOH$^-$ + H$^+$ & $1.8\cdot10^{5}$ & \cite{Levanov.2020} \\
173 & Cl$_{2}^-$ + O$_{3}$ & $\longrightarrow$ & ClO + Cl$^-$ + O$_{2}$ & $9\cdot10^{7}$ & \cite{Levanov.2020} \\
\end{longtable*}
\twocolumngrid
}

\begin{figure*}
\includegraphics[width=0.8\textwidth]{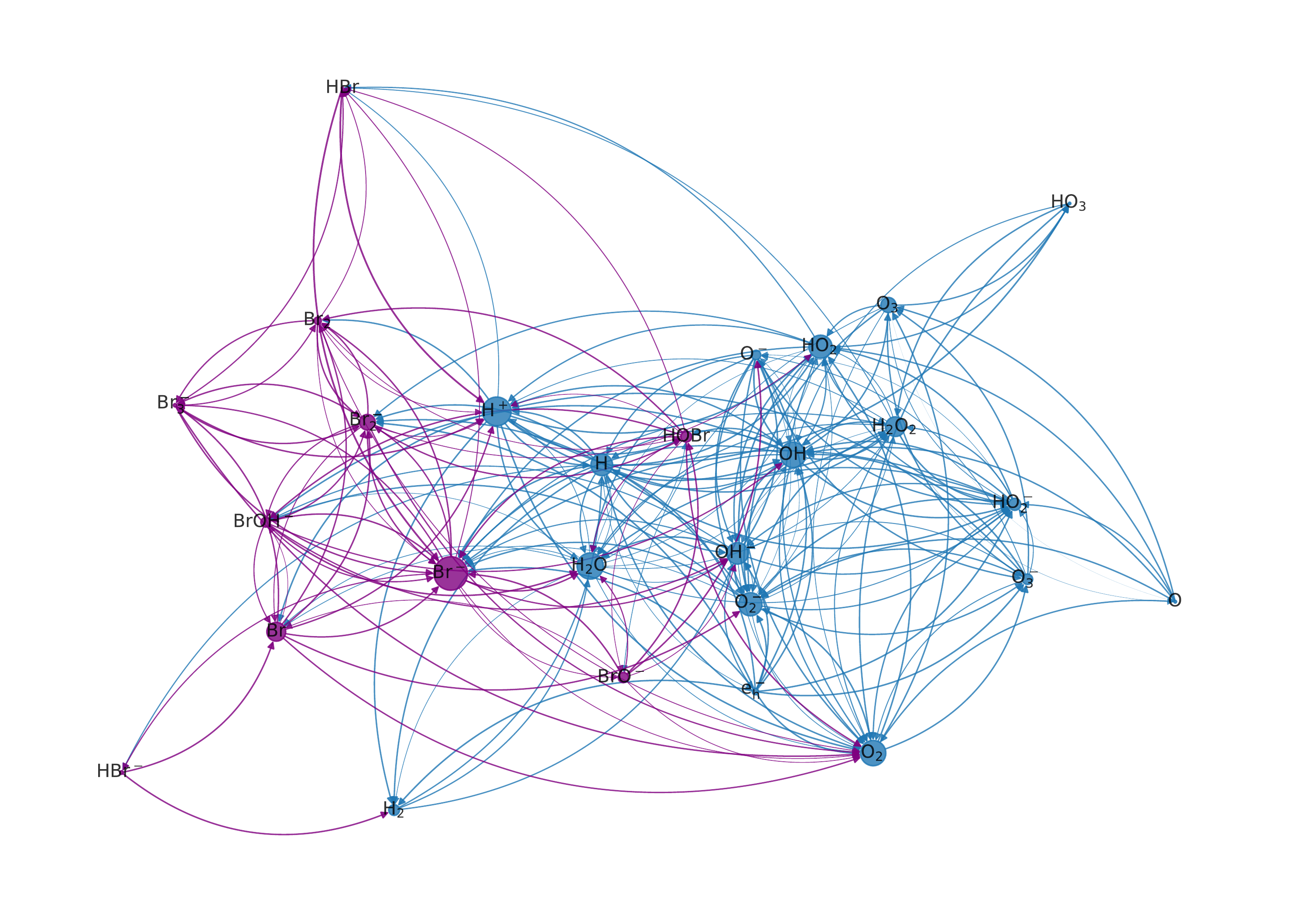}
\caption{\label{fig:Graph_Br_S2}Graph representation of the kinetic model of $\rm Br^-$-containing aqueous solutions. Tabular representation is found in Table\,\ref{tab:model_Br_S2}.}
\end{figure*}

$\rm Br^-$-containing solutions were described by 52 additional reactions and 10 additional species (Table\,\ref{tab:model_Br_S2}, Figure\,\ref{fig:Graph_Br_S2}) \cite{ElOmar.2013, Williams.2002, Yang.2017, Schwarz.1977, Huie.2003, Haag.1983, Klaning.1985b, Buxton.1968}. 

\afterpage{\onecolumngrid
\renewcommand{\doublerulesep}{0pt}
\begin{longtable*}[e]{@{\extracolsep{\fill}}crclcc}\caption{\label{tab:model_Br_S2}Kinetic model used to describe the radiolysis of $\rm Br^-$-containing aqueous solutions. Here, $k$ denotes the respective kinetic constant in units of $\text{mol}^{-n+1}\,\text{L}^{3(n-1)}\,\text{s}^{-1}$, where $n$ denotes the reaction order. Please refer to Supporting Table\,\ref{tab:model_Water_S1} for the first 83 reactions.}\\
\hline\hline
& \multicolumn{3}{c}{Reaction} & $k$ & Reference  \\
\hline
\endfirsthead
\hline\hline
& \multicolumn{3}{c}{Reaction} & $k$ & Reference  \\
\hline
\endhead
\hline\hline
\multicolumn{6}{c}{\textit{Continued on next page.}}\\
\endfoot
\hline\hline
\endlastfoot
 84  &                     Br$^-$ + OH &  $\longrightarrow$ &                       BrOH$^-$ &  $1.1\cdot10^{10}$ &             \cite{ElOmar.2013} \\
85  &                        BrOH$^-$ &  $\longrightarrow$ &                    Br$^-$ + OH &   $3.3\cdot10^{7}$ &             \cite{ElOmar.2013} \\
86  &                BrOH$^-$ + H$^+$ &  $\longrightarrow$ &                  Br + H$_{2}$O &  $4.4\cdot10^{10}$ &             \cite{ElOmar.2013} \\
87  &                        BrOH$^-$ &  $\longrightarrow$ &                    Br + OH$^-$ &   $4.2\cdot10^{6}$ &             \cite{ElOmar.2013} \\
88  &                     Br + OH$^-$ &  $\longrightarrow$ &                       BrOH$^-$ &  $1.3\cdot10^{10}$ &             \cite{ElOmar.2013} \\
89  &                     Br + Br$^-$ &  $\longrightarrow$ &                     Br$_{2}^-$ &  $1.2\cdot10^{10}$ &             \cite{ElOmar.2013} \\
90  &                      Br$_{2}^-$ &  $\longrightarrow$ &                    Br + Br$^-$ &   $1.9\cdot10^{4}$ &             \cite{ElOmar.2013} \\
91  &                    2 Br$_{2}^-$ &  $\longrightarrow$ &            Br$_{3}^-$ + Br$^-$ &   $2.4\cdot10^{9}$ &             \cite{ElOmar.2013} \\
92  &                 Br + Br$_{2}^-$ &  $\longrightarrow$ &                     Br$_{3}^-$ &     $5\cdot10^{9}$ &             \cite{ElOmar.2013} \\
93  &               Br$_{2}$ + Br$^-$ &  $\longrightarrow$ &                     Br$_{3}^-$ &   $1.6\cdot10^{8}$ &             \cite{ElOmar.2013} \\
94  &                      Br$_{3}^-$ &  $\longrightarrow$ &              Br$_{2}$ + Br$^-$ &     $1\cdot10^{7}$ &             \cite{ElOmar.2013} \\
95  &                            2 Br &  $\longrightarrow$ &                       Br$_{2}$ &     $5\cdot10^{9}$ &             \cite{ElOmar.2013} \\
96  &           Br + e$_\mathrm{h}^-$ &  $\longrightarrow$ &                         Br$^-$ &    $1\cdot10^{10}$ &             \cite{ElOmar.2013} \\
97  &   Br$_{2}^-$ + e$_\mathrm{h}^-$ &  $\longrightarrow$ &                       2 Br$^-$ &  $1.3\cdot10^{10}$ &             \cite{ElOmar.2013} \\
98  &   Br$_{3}^-$ + e$_\mathrm{h}^-$ &  $\longrightarrow$ &            Br$_{2}^-$ + Br$^-$ &  $2.7\cdot10^{10}$ &             \cite{ElOmar.2013} \\
99  &                          H + Br &  $\longrightarrow$ &                 H$^+$ + Br$^-$ &    $1\cdot10^{10}$ &             \cite{ElOmar.2013} \\
100 &                  Br$_{2}^-$ + H &  $\longrightarrow$ &               2 Br$^-$ + H$^+$ &  $1.4\cdot10^{10}$ &             \cite{ElOmar.2013} \\
101 &                  Br$_{3}^-$ + H &  $\longrightarrow$ &    Br$_{2}^-$ + Br$^-$ + H$^+$ &  $1.2\cdot10^{10}$ &             \cite{ElOmar.2013} \\
102 &           Br$_{2}^-$ + HO$_{2}$ &  $\longrightarrow$ &     O$_{2}$ + H$^+$ + 2 Br$^-$ &     $1\cdot10^{8}$ &             \cite{ElOmar.2013} \\
103 &           Br$_{3}^-$ + HO$_{2}$ &  $\longrightarrow$ &     Br$_{2}^-$ + HBr + O$_{2}$ &     $1\cdot10^{7}$ &             \cite{ElOmar.2013} \\
104 &               BrOH$^-$ + Br$^-$ &  $\longrightarrow$ &            Br$_{2}^-$ + OH$^-$ &   $1.9\cdot10^{8}$ &             \cite{ElOmar.2013} \\
105 &             Br$_{2}^-$ + OH$^-$ &  $\longrightarrow$ &              BrOH$^-$ + Br$^-$ &   $2.7\cdot10^{6}$ &             \cite{ElOmar.2013} \\
106 &                      Br$^-$ + H &  $\longrightarrow$ &                        HBr$^-$ &   $1.7\cdot10^{6}$ &             \cite{ElOmar.2013} \\
107 &                 HBr$^-$ + H$^+$ &  $\longrightarrow$ &                   H$_{2}$ + Br &  $1.1\cdot10^{10}$ &             \cite{ElOmar.2013} \\
108 &                  H$^+$ + Br$^-$ &  $\longrightarrow$ &                            HBr &     $1\cdot10^{4}$ &           \cite{Williams.2002} \\
109 &                             HBr &  $\longrightarrow$ &                 H$^+$ + Br$^-$ &    $1\cdot10^{13}$ &           \cite{Williams.2002} \\
110 &                    2 Br$_{2}^-$ &  $\longrightarrow$ &            Br$_{2}$ + 2 Br$^-$ &   $1.9\cdot10^{9}$ &               \cite{Yang.2017} \\
111 &                 Br + Br$_{2}^-$ &  $\longrightarrow$ &              Br$_{2}$ + Br$^-$ &     $2\cdot10^{9}$ &               \cite{Yang.2017} \\
112 &     Br$_{2}$ + e$_\mathrm{h}^-$ &  $\longrightarrow$ &                     Br$_{2}^-$ &  $5.3\cdot10^{10}$ &            \cite{Schwarz.1977} \\
113 &                    Br$_{2}$ + H &  $\longrightarrow$ &             Br$_{2}^-$ + H$^+$ &    $1\cdot10^{10}$ &               \cite{Huie.2003} \\
114 &                Br$^-$ + O$_{3}$ &  $\longrightarrow$ &              BrO$^-$ + O$_{2}$ &   $1.6\cdot10^{2}$ &               \cite{Haag.1983} \\
115 &                   Br + H$_{2}$O &  $\longrightarrow$ &               BrOH$^-$ + H$^+$ &  $1.36\cdot10^{0}$ &           \cite{Klaning.1985b} \\
116 &             Br + H$_{2}$O$_{2}$ &  $\longrightarrow$ &   O$_{2}^-$ + Br$^-$ + 2 H$^+$ &     $4\cdot10^{9}$ &               \cite{Yang.2017} \\
117 &                   Br + HO$_{2}$ &  $\longrightarrow$ &       H$^+$ + O$_{2}$ + Br$^-$ &     $1\cdot10^{9}$ &               \cite{Yang.2017} \\
118 &                 Br$_{2}^-$ + Br &  $\longrightarrow$ &              Br$_{2}$ + Br$^-$ &     $2\cdot10^{9}$ &               \cite{Yang.2017} \\
119 &     Br$_{2}^-$ + H$_{2}$O$_{2}$ &  $\longrightarrow$ &    HO$_{2}$ + 2 Br$^-$ + H$^+$ &     $5\cdot10^{2}$ &               \cite{Yang.2017} \\
120 &          Br$_{2}^-$ + O$_{2}^-$ &  $\longrightarrow$ &             O$_{2}$ + 2 Br$^-$ &   $1.7\cdot10^{8}$ &               \cite{Yang.2017} \\
121 &             Br$_{2}$ + HO$_{2}$ &  $\longrightarrow$ &   Br$_{2}^-$ + O$_{2}$ + H$^+$ &   $1.1\cdot10^{8}$ &               \cite{Yang.2017} \\
122 &            Br$_{2}$ + O$_{2}^-$ &  $\longrightarrow$ &           Br$_{2}^-$ + O$_{2}$ &   $5.6\cdot10^{9}$ &               \cite{Yang.2017} \\
123 &       Br$_{2}$ + H$_{2}$O$_{2}$ &  $\longrightarrow$ &                2 HBr + O$_{2}$ &   $1.3\cdot10^{3}$ &               \cite{Yang.2017} \\
124 &             Br$_{2}$ + H$_{2}$O &  $\longrightarrow$ &          HOBr + Br$^-$ + H$^+$ &   $9.7\cdot10^{1}$ &               \cite{Yang.2017} \\
125 &          Br$_{3}^-$ + O$_{2}^-$ &  $\longrightarrow$ &  Br$_{2}^-$ + Br$^-$ + O$_{2}$ &   $3.8\cdot10^{9}$ &               \cite{Yang.2017} \\
126 &                 BrO$^-$ + H$^+$ &  $\longrightarrow$ &                           HOBr &    $1\cdot10^{10}$ &           \cite{Williams.2002} \\
127 &                            HOBr &  $\longrightarrow$ &                H$^+$ + BrO$^-$ &   $2.3\cdot10^{1}$ &           \cite{Williams.2002} \\
128 &                 Br$_{2}^-$ + OH &  $\longrightarrow$ &                  HOBr + Br$^-$ &     $1\cdot10^{9}$ &               \cite{Yang.2017} \\
129 &           HOBr + Br$^-$ + H$^+$ &  $\longrightarrow$ &            Br$_{2}$ + H$_{2}$O &     $5\cdot10^{9}$ &               \cite{Yang.2017} \\
130 &               HOBr + HO$_{2}^-$ &  $\longrightarrow$ &    Br$^-$ + H$_{2}$O + O$_{2}$ &   $7.6\cdot10^{8}$ &               \cite{Yang.2017} \\
131 &           HOBr + H$_{2}$O$_{2}$ &  $\longrightarrow$ &       HBr + H$_{2}$O + O$_{2}$ &   $1.5\cdot10^{4}$ &               \cite{Yang.2017} \\
132 &                HOBr + O$_{2}^-$ &  $\longrightarrow$ &             BrOH$^-$ + O$_{2}$ &   $3.5\cdot10^{9}$ &               \cite{Yang.2017} \\
133 &        BrO$^-$ + H$_{2}$O$_{2}$ &  $\longrightarrow$ &    Br$^-$ + H$_{2}$O + O$_{2}$ &   $1.2\cdot10^{6}$ &               \cite{Yang.2017} \\
134 &  BrO$^-$ + O$_{2}^-$ + H$_{2}$O &  $\longrightarrow$ &        Br + 2 OH$^-$ + O$_{2}$ &     $1\cdot10^{2}$ &               \cite{Yang.2017} \\
135 &      BrO$^-$ + e$_\mathrm{h}^-$ &  $\longrightarrow$ &                 Br$^-$ + O$^-$ &  $1.5\cdot10^{10}$ &  \cite{Buxton.1968, Huie.2003} \\
\end{longtable*}
\twocolumngrid
}

\begin{figure*}
\includegraphics[width=0.8\textwidth]{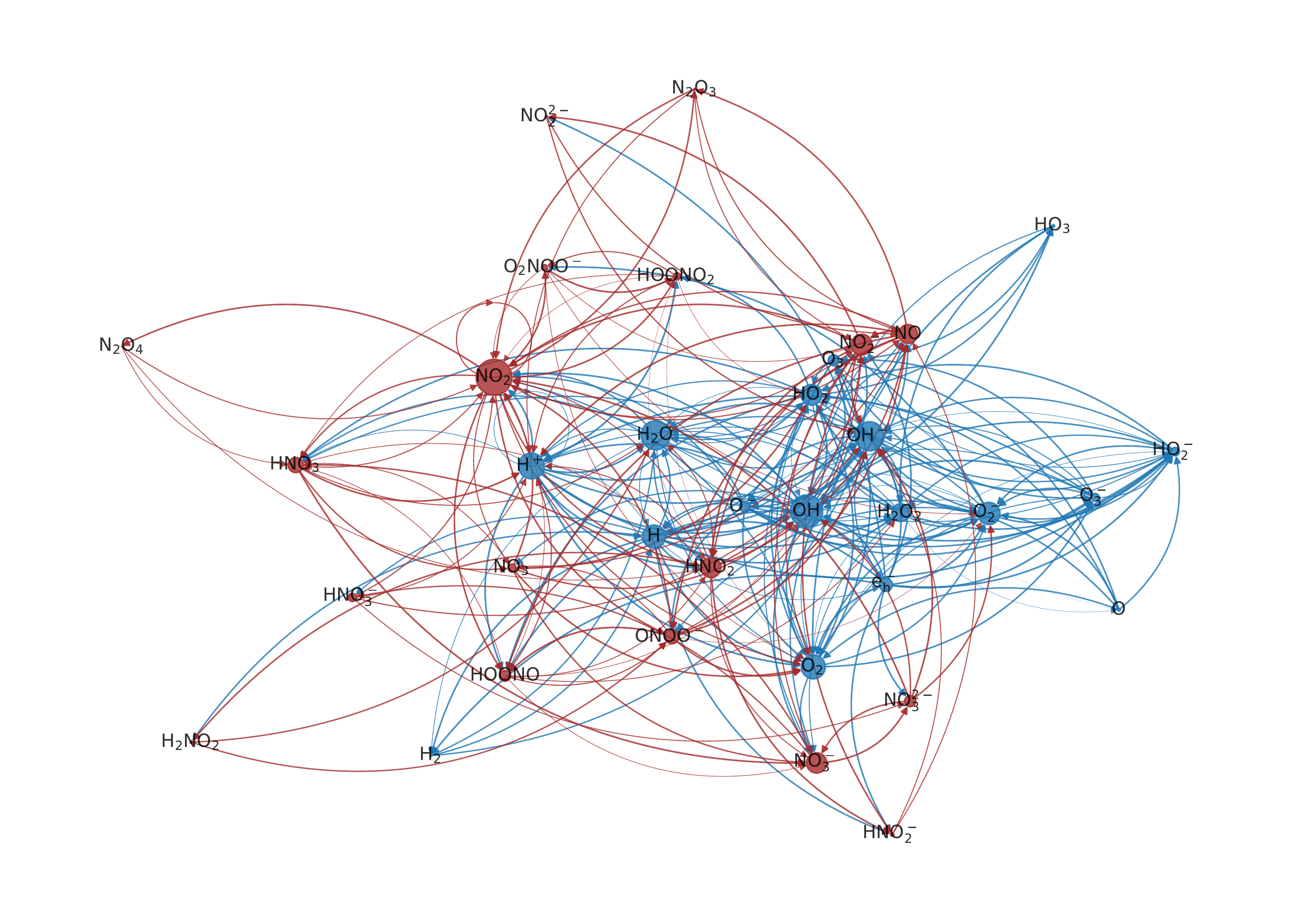}
\caption{\label{fig:Graph_NO3_S4}Graph representation of the kinetic model of $\rm NO_3^-$-containing aqueous solutions. Tabular representation is found in Table\,\ref{tab:model_NO3_S4}.}
\end{figure*}

$\rm NO_3^-$-solutions were simulated using a reaction set of 18 additional species distributed over 73 reactions (Table\,\ref{tab:model_NO3_S4}, Figure\,\ref{fig:Graph_NO3_S4}) \cite{Buxton.1988,Horne.2016,Mezyk.1997,McKenzie.2016, Huie.2003, Mikhailova.1993, Halpern.1966, Leriche.2003, Herrmann.2000, Hoigne.1985}.

\afterpage{\onecolumngrid
\renewcommand{\doublerulesep}{0pt}
\begin{longtable*}[e]{@{\extracolsep{\fill}}crclcc}\caption{\label{tab:model_NO3_S4}Kinetic model used to describe the radiolysis of $\rm NO_3^-$-containing aqueous solutions. Here, $k$ denotes the respective kinetic constant in units of $\text{mol}^{-n+1}\,\text{L}^{3(n-1)}\,\text{s}^{-1}$, where $n$ denotes the reaction order. Please refer to Supporting Table\,\ref{tab:model_Water_S1} for the first 83 reactions.}\\
\hline\hline
& \multicolumn{3}{c}{Reaction} & $k$ & Reference  \\
\hline
\endfirsthead
\hline\hline
& \multicolumn{3}{c}{Reaction} & $k$ & Reference  \\
\hline
\endhead
\hline\hline
\multicolumn{6}{c}{\textit{Continued on next page.}}\\
\endfoot
\hline\hline
\endlastfoot
84  &   NO$_{3}^-$ + e$_\mathrm{h}^-$ &  $\longrightarrow$ &                  NO$_{3}^{2-}$ &    $9.7\cdot10^{9}$ &                                                                                    \cite{Buxton.1988} \\
85  &                  NO$_{3}^-$ + H &  $\longrightarrow$ &                    HNO$_{3}^-$ &    $5.6\cdot10^{6}$ &                                                                                     \cite{Horne.2016} \\
86  &              NO$_{3}^-$ + H$^+$ &  $\longrightarrow$ &                      HNO$_{3}$ &      $6\cdot10^{8}$ &                                                                                     \cite{Horne.2016} \\
87  &                       HNO$_{3}$ &  $\longrightarrow$ &             H$^+$ + NO$_{3}^-$ &  $1.46\cdot10^{10}$ &                                                                                     \cite{Horne.2016} \\
88  &                  HNO$_{3}$ + OH &  $\longrightarrow$ &            NO$_{3}$ + H$_{2}$O &    $1.9\cdot10^{7}$ &                                                                                     \cite{Horne.2016} \\
89  &              NO$_{3}^{2-}$ + OH &  $\longrightarrow$ &            NO$_{3}^-$ + OH$^-$ &      $3\cdot10^{9}$ &                                                                                     \cite{Horne.2016} \\
90  &  NO$_{3}^{2-}$ + H$_{2}$O$_{2}$ &  $\longrightarrow$ &       NO$_{3}^-$ + OH + OH$^-$ &    $1.6\cdot10^{8}$ &                                                                                     \cite{Horne.2016} \\
91  &         NO$_{3}^{2-}$ + O$_{2}$ &  $\longrightarrow$ &         NO$_{3}^-$ + O$_{2}^-$ &    $2.4\cdot10^{8}$ &                                                                                     \cite{Horne.2016} \\
92  &        NO$_{3}^{2-}$ + H$_{2}$O &  $\longrightarrow$ &            NO$_{2}$ + 2 OH$^-$ &      $1\cdot10^{3}$ &                                                                                     \cite{Horne.2016} \\
93  &                     HNO$_{3}^-$ &  $\longrightarrow$ &          NO$_{3}^{2-}$ + H$^+$ &    $1.6\cdot10^{3}$ &                                                                                     \cite{Horne.2016} \\
94  &     NO$_{2}$ + e$_\mathrm{h}^-$ &  $\longrightarrow$ &                     NO$_{2}^-$ &     $1\cdot10^{10}$ &                                                                                     \cite{Horne.2016} \\
95  &                   NO$_{2}$ + OH &  $\longrightarrow$ &                          HOONO &    $4.5\cdot10^{9}$ &                                                                                     \cite{Horne.2016} \\
96  &             NO$_{2}$ + HO$_{2}$ &  $\longrightarrow$ &                    HOONO$_{2}$ &    $1.8\cdot10^{9}$ &                                                                                     \cite{Horne.2016} \\
97  &                    NO$_{2}$ + H &  $\longrightarrow$ &                      HNO$_{2}$ &     $1\cdot10^{10}$ &                                                                                     \cite{Horne.2016} \\
98  &            NO$_{2}$ + O$_{2}^-$ &  $\longrightarrow$ &                 O$_{2}$NOO$^-$ &    $4.5\cdot10^{9}$ &                                                                                     \cite{Horne.2016} \\
99  &                      2 NO$_{2}$ &  $\longrightarrow$ &                 N$_{2}$O$_{4}$ &    $4.5\cdot10^{9}$ &                                                                                     \cite{Horne.2016} \\
100 &             NO$_{2}$ + NO$_{3}$ &  $\longrightarrow$ &        NO + NO$_{2}$ + O$_{2}$ &   $2.41\cdot10^{5}$ &                                                                                     \cite{Horne.2016} \\
101 &                   NO$_{2}$ + NO &  $\longrightarrow$ &                 N$_{2}$O$_{3}$ &    $1.1\cdot10^{9}$ &                                                                                     \cite{Horne.2016} \\
102 &                NO$_{2}$ + O$^-$ &  $\longrightarrow$ &                       ONOO$^-$ &    $3.5\cdot10^{9}$ &                                                                                     \cite{Horne.2016} \\
103 &                  N$_{2}$O$_{4}$ &  $\longrightarrow$ &                     2 NO$_{2}$ &      $6\cdot10^{3}$ &                                                                                     \cite{Horne.2016} \\
104 &       N$_{2}$O$_{4}$ + H$_{2}$O &  $\longrightarrow$ &          HNO$_{2}$ + HNO$_{3}$ &    $1.8\cdot10^{1}$ &                                                                                     \cite{Horne.2016} \\
105 &                  HNO$_{2}$ + OH &  $\longrightarrow$ &            NO$_{2}$ + H$_{2}$O &      $2\cdot10^{9}$ &                                                                                     \cite{Horne.2016} \\
106 &                       HNO$_{2}$ &  $\longrightarrow$ &             NO$_{2}^-$ + H$^+$ &      $3\cdot10^{7}$ &                                                                                     \cite{Horne.2016} \\
107 &                     2 HNO$_{2}$ &  $\longrightarrow$ &       NO$_{2}$ + NO + H$_{2}$O &   $1.34\cdot10^{1}$ &                                                                                     \cite{Horne.2016} \\
108 &    HNO$_{2}$ + e$_\mathrm{h}^-$ &  $\longrightarrow$ &                    HNO$_{2}^-$ &      $4\cdot10^{9}$ &                                                                                     \cite{Horne.2016} \\
109 &                   HNO$_{2}$ + H &  $\longrightarrow$ &                H$_{2}$NO$_{2}$ &   $3.88\cdot10^{8}$ &                                                                                     \cite{Mezyk.1997} \\
110 &            HNO$_{2}$ + NO$_{3}$ &  $\longrightarrow$ &           NO$_{2}$ + HNO$_{3}$ &      $2\cdot10^{8}$ &                                                                                     \cite{Horne.2016} \\
111 &           HNO$_{2}$ + HNO$_{3}$ &  $\longrightarrow$ &          2 NO$_{2}$ + H$_{2}$O &   $6.62\cdot10^{3}$ &                                                                                     \cite{Horne.2016} \\
112 &              NO$_{2}^-$ + H$^+$ &  $\longrightarrow$ &                      HNO$_{2}$ &     $5\cdot10^{10}$ &                                                                                     \cite{Horne.2016} \\
113 &                 NO$_{2}^-$ + OH &  $\longrightarrow$ &              NO$_{2}$ + OH$^-$ &     $1\cdot10^{10}$ &                                                                                     \cite{Horne.2016} \\
114 &                  NO$_{2}^-$ + H &  $\longrightarrow$ &                    HNO$_{2}^-$ &   $1.64\cdot10^{9}$ &                                        \cite{Mezyk.1997} \\
115 &   NO$_{2}^-$ + O$^-$ + H$_{2}$O &  $\longrightarrow$ &            NO$_{2}$ + 2 OH$^-$ &    $3.1\cdot10^{8}$ &                                                                                     \cite{Horne.2016} \\
116 &   NO$_{2}^-$ + e$_\mathrm{h}^-$ &  $\longrightarrow$ &                  NO$_{2}^{2-}$ &    $4.1\cdot10^{9}$ &                                                                                     \cite{Horne.2016} \\
117 &           NO$_{2}^-$ + NO$_{3}$ &  $\longrightarrow$ &          NO$_{2}$ + NO$_{3}^-$ &    $4.4\cdot10^{9}$ &                                                                                     \cite{Horne.2016} \\
118 &                      2 NO$_{3}$ &  $\longrightarrow$ &           2 NO$_{2}$ + O$_{2}$ &    $1.3\cdot10^{5}$ &  \cite{McKenzie.2016} \\
119 &       NO$_{3}$ + H$_{2}$O$_{2}$ &  $\longrightarrow$ &           HNO$_{3}$ + HO$_{2}$ &    $7.1\cdot10^{6}$ &                                                                                     \cite{Horne.2016} \\
120 &                   NO$_{3}$ + OH &  $\longrightarrow$ &            NO$_{2}$ + HO$_{2}$ &     $1\cdot10^{10}$ &                                                                                     \cite{Horne.2016} \\
121 &             NO$_{3}$ + HO$_{2}$ &  $\longrightarrow$ &            HNO$_{3}$ + O$_{2}$ &      $3\cdot10^{9}$ &                                                                                     \cite{Horne.2016} \\
122 &             NO$_{3}$ + H$_{2}$O &  $\longrightarrow$ &                 HNO$_{3}$ + OH &      $3\cdot10^{2}$ &                                                                                     \cite{Horne.2016} \\
123 &               NO$_{3}$ + OH$^-$ &  $\longrightarrow$ &                NO$_{3}^-$ + OH &    $8.2\cdot10^{7}$ &                                                                                     \cite{Horne.2016} \\
124 &                           HOONO &  $\longrightarrow$ &             NO$_{3}^-$ + H$^+$ &     $9\cdot10^{-1}$ &                                                                                     \cite{Horne.2016} \\
125 &                           HOONO &  $\longrightarrow$ &                  NO$_{2}$ + OH &   $3.5\cdot10^{-1}$ &                                                                                     \cite{Horne.2016} \\
126 &                     HOONO$_{2}$ &  $\longrightarrow$ &            NO$_{2}$ + HO$_{2}$ &   $2.6\cdot10^{-2}$ &                                                                                     \cite{Horne.2016} \\
127 &                     HOONO$_{2}$ &  $\longrightarrow$ &            HNO$_{2}$ + O$_{2}$ &     $7\cdot10^{-4}$ &                                                                                     \cite{Horne.2016} \\
128 &                HOONO + H$_{2}$O &  $\longrightarrow$ &     HNO$_{2}$ + H$_{2}$O$_{2}$ &      $3\cdot10^{2}$ &                                                                                     \cite{Horne.2016} \\
129 &                     HOONO$_{2}$ &  $\longrightarrow$ &         O$_{2}$NOO$^-$ + H$^+$ &    $7.1\cdot10^{4}$ &                                                                                     \cite{Horne.2016} \\
130 &         HOONO$_{2}$ + HNO$_{2}$ &  $\longrightarrow$ &                    2 HNO$_{3}$ &    $1.2\cdot10^{1}$ &                                                                                     \cite{Horne.2016} \\
131 &                  O$_{2}$NOO$^-$ &  $\longrightarrow$ &           NO$_{2}^-$ + O$_{2}$ &   $1.35\cdot10^{0}$ &                                                                                     \cite{Horne.2016} \\
132 &                  O$_{2}$NOO$^-$ &  $\longrightarrow$ &           NO$_{2}$ + O$_{2}^-$ &      $1\cdot10^{0}$ &                                                                                     \cite{Horne.2016} \\
133 &          O$_{2}$NOO$^-$ + H$^+$ &  $\longrightarrow$ &                    HOONO$_{2}$ &     $5\cdot10^{10}$ &                                                                                     \cite{Horne.2016} \\
134 &                     HNO$_{2}^-$ &  $\longrightarrow$ &                    NO + OH$^-$ &      $5\cdot10^{3}$ &                                                                                     \cite{Horne.2016} \\
135 &        NO$_{2}^{2-}$ + H$_{2}$O &  $\longrightarrow$ &                  NO + 2 OH$^-$ &    $1.6\cdot10^{6}$ &                                                                                     \cite{Horne.2016} \\
136 &                  2 NO + O$_{2}$ &  $\longrightarrow$ &                     2 NO$_{2}$ &    $5.9\cdot10^{6}$ &                                                                                     \cite{Horne.2016} \\
137 &                         NO + OH &  $\longrightarrow$ &             NO$_{2}^-$ + H$^+$ &     $1\cdot10^{10}$ &                                                                                     \cite{Horne.2016} \\
138 &                   NO + HO$_{2}$ &  $\longrightarrow$ &                          HOONO &    $3.2\cdot10^{9}$ &                                                                                     \cite{Horne.2016} \\
139 &                  NO + O$_{2}^-$ &  $\longrightarrow$ &                       ONOO$^-$ &      $5\cdot10^{9}$ &                                                                                     \cite{Horne.2016} \\
140 &                        ONOO$^-$ &  $\longrightarrow$ &                 NO + O$_{2}^-$ &     $2\cdot10^{-2}$ &                                                                                     \cite{Horne.2016} \\
141 &                   HOONO + H$^+$ &  $\longrightarrow$ &              HNO$_{3}$ + H$^+$ &    $4.3\cdot10^{0}$ &                                                                                     \cite{Horne.2016} \\
142 &                   ONOO$^-$ + OH &  $\longrightarrow$ &          NO + O$_{2}$ + OH$^-$ &    $4.8\cdot10^{9}$ &                                                                                     \cite{Horne.2016} \\
143 &                  N$_{2}$O$_{3}$ &  $\longrightarrow$ &                  NO + NO$_{2}$ &    $8.4\cdot10^{4}$ &                                                                                     \cite{Horne.2016} \\
144 &       ONOO$^-$ + N$_{2}$O$_{3}$ &  $\longrightarrow$ &        2 NO$_{2}$ + NO$_{2}^-$ &    $3.1\cdot10^{8}$ &                                                                                     \cite{Horne.2016} \\
145 &       N$_{2}$O$_{3}$ + H$_{2}$O &  $\longrightarrow$ &         2 NO$_{2}^-$ + 2 H$^+$ &      $2\cdot10^{3}$ &                                                                                     \cite{Horne.2016} \\
146 &                ONOO$^-$ + H$^+$ &  $\longrightarrow$ &                          HOONO &     $5\cdot10^{10}$ &                                                                                     \cite{Horne.2016} \\
147 &             ONOO$^-$ + NO$_{2}$ &  $\longrightarrow$ &          NO$_{2}^-$ + NO$_{3}$ &    $2.4\cdot10^{4}$ &                                                                                     \cite{Horne.2016} \\
148 &     H$_{2}$NO$_{2}$ + O$_{2}^-$ &  $\longrightarrow$ &            ONOO$^-$ + H$_{2}$O &    $2.3\cdot10^{7}$ &                                                                     \cite{Mikhailova.1993, Huie.2003} \\
149 &           NO$_{3}^{2-}$ + H$^+$ &  $\longrightarrow$ &              NO$_{2}$ + OH$^-$ &     $2\cdot10^{10}$ &                                                                                      \cite{Huie.2003} \\
150 &                     HNO$_{3}^-$ &  $\longrightarrow$ &              NO$_{2}$ + OH$^-$ &      $2\cdot10^{5}$ &                                                                                      \cite{Huie.2003} \\
151 &                   HNO$_{2}$ + H &  $\longrightarrow$ &                  H$_{2}$O + NO &    $4.5\cdot10^{8}$ &                                                                                   \cite{Halpern.1966} \\
152 &      HNO$_{2}$ + H$_{2}$O$_{2}$ &  $\longrightarrow$ &  NO$_{3}^-$ + H$^+$ + H$_{2}$O &    $4.6\cdot10^{3}$ &                                                                                   \cite{Leriche.2003} \\
153 &        NO + NO$_{2}$ + H$_{2}$O &  $\longrightarrow$ &                    2 HNO$_{2}$ &   $1.58\cdot10^{8}$ &                                                                                  \cite{McKenzie.2016} \\
154 &           2 NO$_{2}$ + H$_{2}$O &  $\longrightarrow$ &          HNO$_{2}$ + HNO$_{3}$ &    $4.8\cdot10^{7}$ &                                                                                  \cite{McKenzie.2016} \\
155 &                           HOONO &  $\longrightarrow$ &               ONOO$^-$ + H$^+$ &      $5\cdot10^{4}$ &                                                                                  \cite{Herrmann.2000} \\
156 &            NO$_{2}^-$ + O$_{3}$ &  $\longrightarrow$ &           O$_{2}$ + NO$_{3}^-$ &    $3.7\cdot10^{5}$ &                                                                                    \cite{Hoigne.1985} \\
\end{longtable*}
\twocolumngrid
}

\end{document}